\documentclass[oldversion]{aa}  
\usepackage{graphicx}
\usepackage{txfonts}
\usepackage{epsfig}
\usepackage{natbib}
\begin{document}
   \title{Supermassive black holes in the Sbc spiral galaxies NGC 3310, NGC 4303 and NGC 4258\thanks{Based on observations made with the NASA/ESA Hubble Space Telescope, obtained at the Space Telescope Science Institute, which is operated by the Association of Universities for Research in Astronomy, Inc., under NASA contract NAS 5-26555. These observations are associated with proposal 8228.}}

   \author{Guia Pastorini\inst{1},
                  Alessandro Marconi\inst{1,2},
                  Alessandro Capetti\inst{3},
                  David J. Axon\inst{4},
                  Almudena Alonso-Herrero\inst{5},
                  John Atkinson\inst{6},
                  Dan Batcheldor\inst{4},
                  C.\ Marcella Carollo\inst{7},
                  James Collett\inst{6},
                  Linda Dressel\inst{8},
                  Mark A. Hughes\inst{6},
                  Duccio Macchetto\inst{8,9},
                  Witold Maciejewski\inst{10},
                  William Sparks\inst{8},
                  \and Roeland van der Marel\inst{8}		
		  }         

   \offprints{Guia Pastorini}

   \institute{Dipartimento di Astronomia e Scienza dello Spazio, Universit\`a degli Studi di Firenze, Largo E.Fermi 2, I-50125 Firenze, Italy
   \and INAF-Osservatorio Astrofisico di Arcetri, Largo E.Fermi 5, I-50125 Firenze, Italy
   \and INAF-Osservatorio Astrofisico di Torino, Strada Osservatorio 20, I-10025 Pino Torinese, Italy
   \and Department of Physics, Rochester Institute of Technology, 85 Lomb Memorial Drive, Rochester, NY 14623, USA
  \and Damir, Instituto de Estructura de la Materia, CSIC, Serrano 121, 28006 Madrid, Spain
   \and School of Physics, Astronomy and Mathematics, University of Hertfordshire, College Lane, Hatfield, AL10 9AB, UK
   \and Institute of Astronomy, Physics Department, ETH, Zurich, Switzerland
    \and Space Telescope Science Institute, 3700 San Martin Drive, Baltimore, MD 21218, USA
   \and ESA Research and Space Science Department
    \and Department of Astrophysics, University of Oxford, Denys Wilkinson Building, Keble Road, Oxford OX1 3RH, UK
   }

   \date{Received; accepted}

\abstract{We have undertaken an HST Space Telescope Imaging Spectrograph survey of 54 late type spiral galaxies to study the scaling relations between black holes and their host spheroids at the low mass end. Our aim is to measure black hole masses or to set upper limits for a sizeable sample of spiral galaxies.
In this paper we present new Space Telescope Imaging Spectrograph (STIS) observations of three spiral galaxies, NGC 4303, NGC 3310 and NGC 4258. The bright optical emission lines H$\alpha$ $\lambda$ $6564 \AA$, [NII] $\lambda$$\lambda$ $6549,6585 \AA$ and [SII] $\lambda$$\lambda$ $ 6718,6732 \AA$  were used to study the kinematics of the ionized gas in the nuclear region of each galaxy with a $\sim 0.07\arcsec$ spatial resolution. Our STIS data for NGC 4258 were analyzed in conjunction with archival ones to compare the gas kinematical estimate of the black hole mass with the accurate value from $H_{2}0$-maser observations.

In NGC 3310, the observed gas kinematics is well matched by a circularly rotating disk model but we are only able to set an upper limit to the BH mass which, taking into account the allowed disk inclinations, varies in the range $5.0 \times 10^{6} - 4.2 \times 10^{7} M_{\odot}$ at the 95\% confidence level. 

In NGC 4303 the kinematical data require the presence of a BH with mass $M_{BH}=(5.0)^{+0.87}_{-2.26}\times 10^{6}M_{\odot}$ (for a disk inclination $i=70$ deg) but the weak agreement between data and disk model does not allow us to consider this measurement completely reliable.
If the allowed inclination values are taken into account, $M_{BH}$ varies in the range $6.0 \times 10^{5} - 1.6 \times 10^{7} M_{\odot}$ at the 95\% confidence level.

In NGC 4258, the observed kinematics require the presence of a black hole with $M_{BH}= (7.9)^{+6.2}_{-3.5} \times 10^{7}M_{\odot}$ ($i=60$ deg) and, taking into account reasonable limits for the inclination, $M_{BH}$ is in the range $2.5\times 10^{7}$ -- $2.6 \times 10^{8}M_{\odot}$ at the 95\% confidence level. This result is in good agreement with the published value $(3.9 \pm 0.1) \times 10^{7} M_{\odot}$, derived from $H_{2}O$-maser observations. As in the case of NGC 4303, the agreement between observed and model kinematics is not strong but this does not affect the recovery of the correct $M_{BH}$ value. 

Our attempt at measuring BH masses in these 3 late type Sbc spiral galaxies has shown that these measurements are very challenging and at the limit of the highest spatial resolution currently available. Nonetheless our estimates are in good agreement with the scaling relations between black holes and their host spheroids suggesting that (i) they are reliable and (ii) black holes in spiral galaxies follows the same scaling relations as those in more massive early-type galaxies. A crucial test for the gas kinematical method, the correct recovery of the known BH mass in NGC 4258, has been successful.}


  \titlerunning{Supermassive black holes in spiral galaxies} 
  \authorrunning{G.~Pastorini \emph{et. al}}                                                                         
  \maketitle


\section{\label{introduzione}Introduction}

It is becoming increasingly clear that the energy output from supermassive
black holes (BH) at galaxy centers plays a pivotal role in the formation
and evolution of galaxies.

It is now widely accepted that the mechanism powering Active Galactic Nuclei (AGN) is accretion of matter onto a BH \citep{salpeter64}, with masses in the $10^{6}-10^{10}M_{\odot}$ range \citep{ferrarese05}.  This paradigm, combined with the cosmological evolution of AGNs, implies that most, if not all, local galaxies should host a BH in their nuclei (e.g. ~\citealt{marconi04} and references therein).

BH are detected in about $40$ galaxies, mostly early-types (E/S0; \citealt{ferrarese05} and references therein). The BH mass ($M_{BH}$) is closely related to the stellar velocity dispersion ($\sigma_{*}$; \citealt{ferrarese00,gebhardt00}) and to the host bulge mass or luminosity ($L_{sph}$, \citealt{kr,MeH}). These correlations reveal a tight link between the growth of central BH and the formation and evolution of the host galaxy. These locally detected BHs were probably grown during AGN activity \citep{marconi04} and during these phases they had a major impact on the host galaxy: the feedback from the accreting BH (i.e.\ from the AGN) can be so strong as to suppress star formation in the host galaxy and is possibly the physical mechanism responsible for setting the close relations between BH and host galaxy structural parameters (e.g.\ \citealt{menci,granato,matteo}.) 

The number of BH detections in spiral galaxies is small (currently $7$, \citealt{ferrarese05}) and it is clearly mandatory to increase this number in order to understand how common BHs are in spirals and if they follow the same correlations as early type galaxies. Even constraining upper limits on $M_{BH}$ could be considered as a significant improvement given the scarcity of measurements and the difficulties in performing the required observations in spiral galaxies.  Moreover, $M_{BH}$-galaxy correlations are studied only in the $3 \times 10^{6}- 3 \times 10^{9} M_{\odot}$ mass range and it is expected that spiral galaxies might host low mass black holes below $10^{7}M_{\odot}$.

The direct methods adopted to measure $M_{BH}$ in the nearby universe use gas or stellar kinematics to gather information on the gravitational potential in the nuclear region of the galaxy.
The stellar kinematical method has the advantage that stars are present in all galactic nuclei and their motion is always gravitational. The drawback is that it requires relatively long observation times in order to obtain high quality observations and that stellar dynamical models are very complex, potentially plagued by indeterminacy \citep{valluri,cretton}.
Conversely, the gas kinematical method is relatively simple, it requires relatively short observation times for the brightest
emission line nuclei, even if not all galaxy nuclei present detectable emission lines. Another important drawback is that non-circular or non-gravitational motions can completely invalidate this method. 
Since the observed correlations are based on BH masses obtained with different methods, it is important to check whether these methods provide consistent and robust results. A comparison between $M_{BH}$ measurements obtained with gas and stellar kinematics so far has been performed only for the Milky Way \citep{genzel,genzel6} and Centaurus A \citep{centauro,silge}, where the two independent estimates are in excellent agreement.
This comparison has been attempted also for IC 1459, NGC 4335 and NGC 3379 \citep{cappel02,verdoes2,shapiro} but in those cases the gas kinematical method does not provide reliable MBH estimates because the observed velocity field is not reproduced by a circularly rotating disk.

All methods used to directly detect BHs and measure their masses require high spatial resolution in order to resolve the BH sphere of influence, i.e.\ the region where the gravitational influence of the BH dominates over that of the host galaxy. The radius of the BH sphere of influence is traditionally defined as $r_{BH}=GM_{BH}/\sigma^{2}$, where $G$ is the gravitational constant and $\sigma$ is the stellar velocity dispersion, and is usually $< 1\arcsec$.  Very high spatial resolution is provided by the \emph{Hubble Space Telescope} (HST) whose angular resolution is approximately $\delta\theta = 0.07\arcsec$ (Full Width at Half Maximum, FWHM, of the Point Spread Function, PSF) at $\lambda \sim 6500\AA$.
Let us assume that a spiral galaxy, located at a distance of 20 Mpc, hosts a BH with $M_{BH}=10^{7}M_{\odot}$ and follows the $M_{BH}-\sigma$ correlation \citep{trem02}; the expected value of $\sigma$ is then $\sim 105$ km/s. The radius of the black hole sphere of influence expected for this object is thus $r_{BH}\sim 0.04\arcsec$ very close to the HST angular resolution, demonstrating how challenging these observations are. 

We have undertaken a spectroscopic survey of 54 spirals using the Space Telescope Imaging Spectrograph (STIS) on HST. Our sample was extracted from a comprehensive ground-based study by D.J. Axon et al. (in preparation) who obtained H$\alpha$ and [NII] rotation curves, at seeing-limited resolution of $1\arcsec$, of 128 Sb, SBb, Sc and SBc spiral galaxies selected from RC3. By restricting ourselves to galaxies with recession velocities $V< 2000$ km s$^{-1}$,  we obtained a volume-limited sample of 54 spirals known to have nuclear gas disks. The sample spans a wide range in bulge mass and concentration. The systemic velocity cutoff was chosen so that we can probe close to the nuclei of these galaxies and detect low-mass BHs. 

This paper is part of a series reporting the results of our HST program. In Paper I \citep{4041} we reported the modeling of the central mass concentration in NGC 4041, the first galaxy to be observed as part of the program. Paper I also contained a detailed description of the modeling techniques that we use to determine BH masses.
In Paper II \citep{huge} we presented the atlas of STIS spectra obtained from the successful observations and we created color maps from the optical STIS images when archival Near-Infrared Camera and Multi-Object Spectrometer (NICMOS) images were available. In Paper III \citep{scarlata} we presented the STIS images acquired in the program and analyzed surface brightness profiles derived from the images. In Paper IV \citep{huge2} we used both color information from these images and the spectra presented here to investigate the age of the central stellar populations.  In Paper V \citep{atki} we reported the BH mass measurements in two Sbc galaxies of the sample, NGC 1300 and NGC 2748. Here (Paper VI) we present the analysis of three additional Sbc galaxies from our sample: NGC 3310, NGC 4303 and NGC 4258, selected from the remaining sources because they combine high S/N spectra with evidence for rotation.

NGC 3310 (Arp217, UGC 5786), located in the vicinity of the Ursa Major cluster, is an Sbc  galaxy ("SAB(r)bc pec" in the RC3 catalogue by \citealt{vacouleur91}) with an inclination of the galactic disk of about $i\sim40$ deg \citep{sanchez}. 
It is a very well known starburst galaxy characterized by a disturbed morphology (see \citealt{elemegreen} for a recent review) which suggests that star formation has been triggered by a collision with a dwarf companion during the last $\sim 10^{8}$ yr \citep{kregel,balick}. Its central starbust region \citep{diaz,elemegreen,chandar}, as well as its young stellar clusters \citep{grijisa,grijisb}, have been studied in detail revealing the presence of very young stars (from $\sim 10^7$ to $10^8$ yr).
The bright inner region is dominated by a two-armed open spiral pattern observed in H$\alpha$ and the inner part of this well-developed pattern connects to a $\sim900$ pc diameter starburst ring, surrounding the blue compact nucleus \citep{kruit,balick}.
The gas kinematics in the galactic disk is disturbed, characterized by non-circular motions and strong streaming along the spiral arms, strengthening the idea of a recent merger event \citep{kregel,mulder}. Concerning the nuclear region, the rotation center of the gas is offset with respect to the stellar continuum isophotes by $\sim 1.5\arcsec \pm 0.3\arcsec$ toward PA 142 deg \citep{kruita}.
The STIS acquisition images of NGC 3310 obtained with our HST project indicate the presence of a strong dust lane crossing the center of the galaxy \citep{scarlata}.
From the Lyon/Meudon Extragalactic Database\footnote{Available at http://leda.univ-lyon1.fr} (LEDA) NGC 3310 has an average heliocentric velocity of 1216 km/s, after correction for Local Group infall onto Virgo. Assuming $H_0= 70$ km s$^{-1}$ Mpc$^{-1}$, this implies a distance of $D\sim 17.4$ Mpc (i.e. $1\arcsec\sim 84$ pc). 

The late-type Sbc galaxy NGC 4303 (M61) is a double-barred galaxy in the Virgo Cluster \citep{bingeli} at distance of 16.1 Mpc ($1\arcsec\sim 78$ pc, see e.g.~\citealt{schinner02}). A large-scale bar of about 2.5 kpc ($\sim 35\arcsec$) lies inside the outer spiral arms \citep{laine,erwin}. In addition to the large-scale bar, this galaxy hosts a second inner bar of about 0.2 kpc ($\sim 2\arcsec$) length that is surrounded by a circumnuclear starburst ring/spiral of $\sim 0.5$ kpc ($\sim 6\arcsec$) diameter \citep{wada,perez,colina97}.
Based on the optical emission line ratios it is classified as a Seyfert 2/LINER borderline AGN \citep{filippenko,arribas}
and, in fact, it constitutes a nice example of the starburst-AGN connection. A young ($\sim 4$ Myr), massive stellar cluster and a low-luminosity active nucleus seem to coexist within the central $3$ pc ($\sim 0.04\arcsec$) of the galaxy with the starburst contribution being the dominant one in the UV spectral region \citep{colina02}. The AGN presence is revealed through the luminous near-infrared point source \citep{colina02} and through hard X-ray emission \citep{bailon}.
The gas kinematics of the galactic disk is affected by the presence of the large bar but, in the nuclear region, the gas seems to show ordered rotation. From integral field observations, \cite{arribas} find that the H$\beta$ velocity field at $\sim 1.5\arcsec$ resolution is consistent with a circularly rotating massive disk of $\sim 300$ pc radius ($\sim 4\arcsec$).
CO observations confirm this behavior and indicate that, at a few arcsec resolution, the gas within the central 0.7 kpc ($\sim 10\arcsec$) is following almost pure circular rotation with the dynamical center located within 1\arcsec\ from the near-IR point source \citep{koda,schinner02}.
At 2\arcsec\ resolution, \cite{schinner02} find the presence of an asymmetric spiral pattern inside a radius of $\sim 400$ pc ($\sim 5\arcsec$) which is consistent either with a self-gravitating $m=1$ mode (hence gas kinematics would be unaffected by the small inner bar) or with a ring of star formation heavily extincted north of the nucleus.

NGC 4258 (M106) is a SABbc nearby galaxy ($D=7.2\pm 0.3$ Mpc, i.e.~$1\arcsec\sim35$ pc \citealt{hernest}) spectroscopically classified as a $1.9$ Seyfert galaxy \citep{ho}.
This galaxy is very well known because it represents the best case for a nuclear supermassive BH after the Milky Way. \citet{myoshi}, \cite{greenhilla,greenhillb} and \cite{hernest} studied the H$_2$O maser line emission of a rotating gas disk near the galaxy center. This disk is nearly edge-on ($i=81\pm \ 1$ deg) and its kinematics is nicely explained by circular rotation around a $(3.9\pm 0.1) \times 10^{7}M_{\odot}$ mass confined within $0.13$ pc of the nucleus, indicating very compelling evidence for a supermassive nuclear black hole. 
The active nucleus has been detected through radio \citep{turner,cecil}, X-ray \citep{fruscione,wilson,maki} and infrared observations \citep{chary,chary2}. The strong polarization of the relatively broad optical emission lines further supports the existence of an obscured active nucleus in NGC 4258 \citep{wilkes}. In particular, the AGN is characterized by a radio jet 
which propagates perpendicularly to the maser disk \citep{cecil}.
On large scales, "anomalous arms" were discovered through H$\alpha$ imaging by \citet{courtes}, which noted their unusual diffuse and amorphous appearance, in strong contrast with the knotty structure  of normal spiral arms due to the presence of  HII-regions (see \citealt{wilson} for a recent review on the anomalous arms). \cite{wilson} interpreted the anomalous arms as jet-cloud interaction between the radio jet and the galactic disk: the jets shock the normal interstellar gas along the first 350 pc ($\sim 10\arcsec$) of their length, causing the hot, X-ray emitting cocoons surrounding the radio jet. To summarize, in the nuclear region of NGC 4258, the radio jet propagates perpendicularly to the maser disk at PA $\sim -5$ deg and the shock edge between the hot X-ray cocoon and the dense gas in the galactic disk is at PA $-33$ deg.

%
The aim of this paper is to present the gas kinematical analysis of these three spiral galaxies in order to measure the mass of their nuclear BHs. The case of NGC 4258 is particularly significant because this galaxy is an important test for the gas kinematical method since its BH mass is accurately known from high spatial resolution maser observations described above.

The structure of the paper is as follows. In Sec.~\ref{osservazioni} we present our HST/STIS and archival HST/NICMOS-ACS observations as well as the data reduction procedures. In Sec.~\ref{fitlinee} we describe the kinematical analysis and we present the kinematical parameters measured from our data. In Sec.~\ref{stelle} we describe the method followed to obtain the stellar contribution to the gravitational potential 
for each galaxy. In Sec.~\ref{modelli} we describe the kinematical modeling and how it is possible to measure $M_{BH}$. In Sec.~\ref{ris.gen} we present the results for all galaxies. These results are discussed and compared with the expectations from the $M_{BH}$-galaxy correlations in Sec.~\ref{discuss} and, in Sec.~\ref{conclusioni}, we summarize our work and present our conclusions.


\section{\label{osservazioni}Observations and data reduction}

\begin{figure*}
\centering
\includegraphics[width=0.31\textwidth]{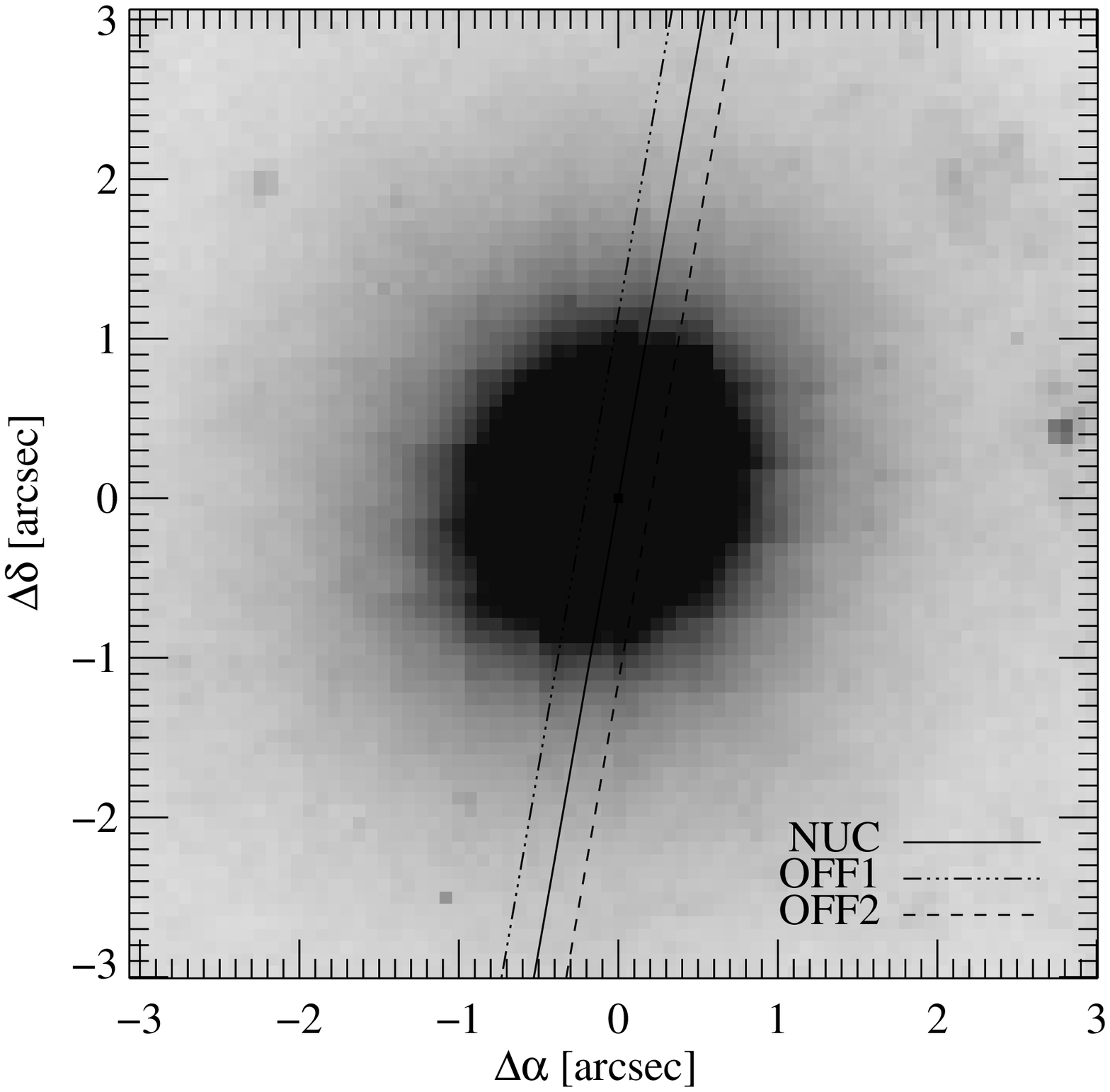}
\includegraphics[width=0.31\textwidth]{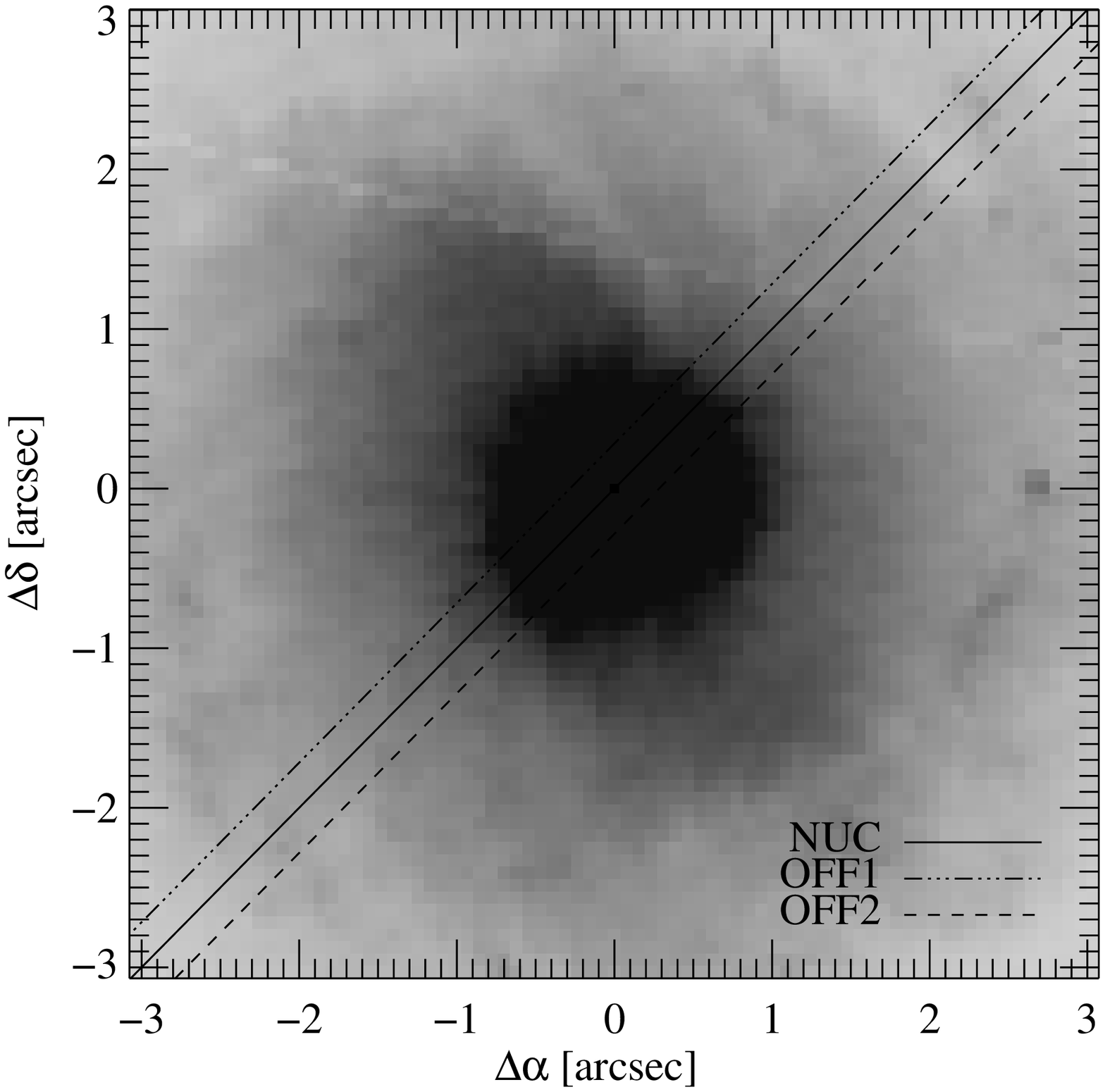}
\includegraphics[width=0.31\textwidth]{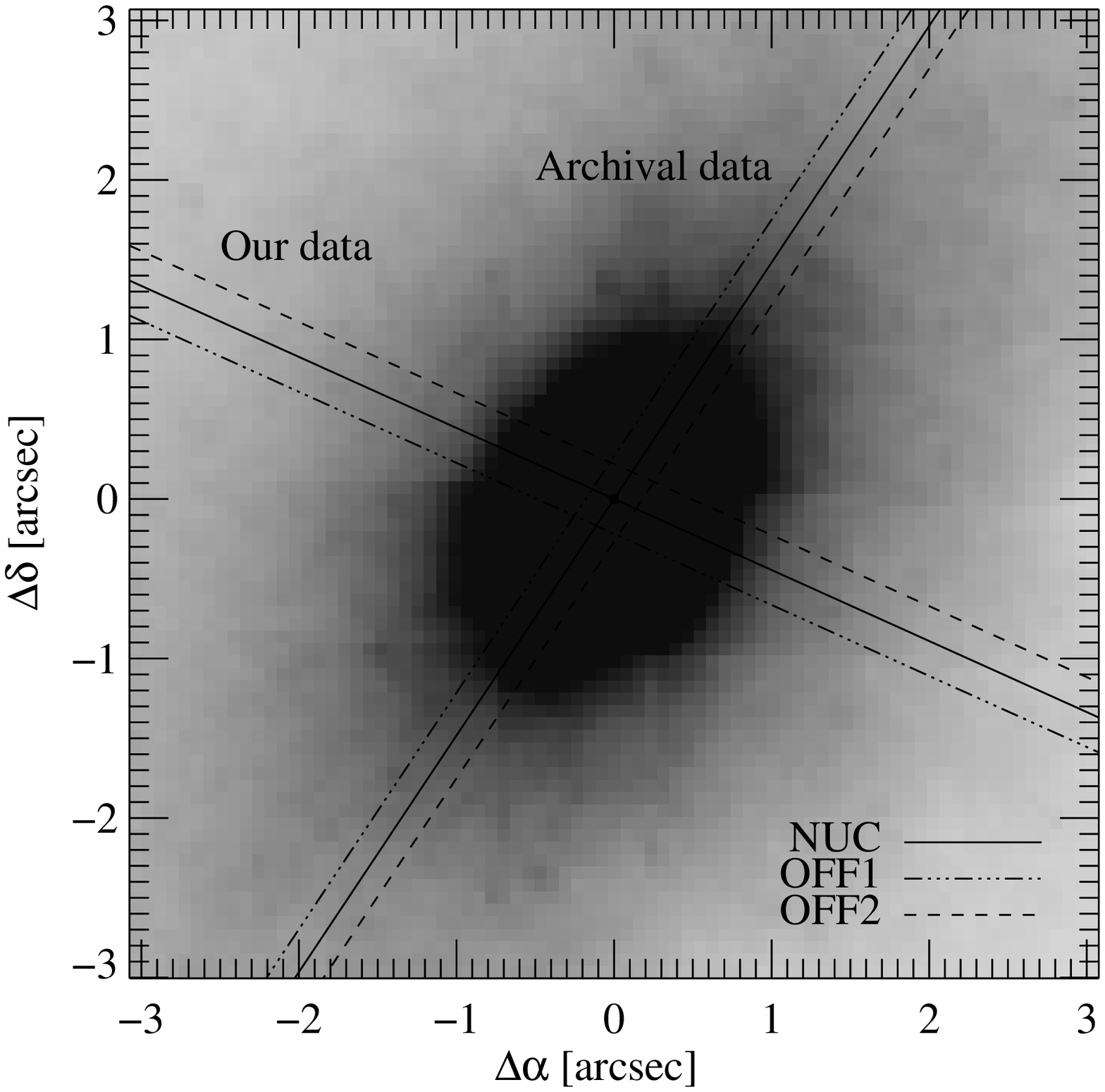}
\caption{\label{fig:posizioni} Left panel: HST-NICMOS F160W image of NGC 3310 with the three different parallel positions of the slits used for spectra acquisition. North is up and East is left. The gray-scale was chosen to have darker points for higher intensity values. The slit positions used in the observations are labeled as NUC, OFF1 and OFF2 (see text for more details). Central and right panels: NGC 4303 and NGC 4258, notations as in left panel.}
\end{figure*}

\begin{figure*}
\centering
\includegraphics[width=0.4\linewidth]{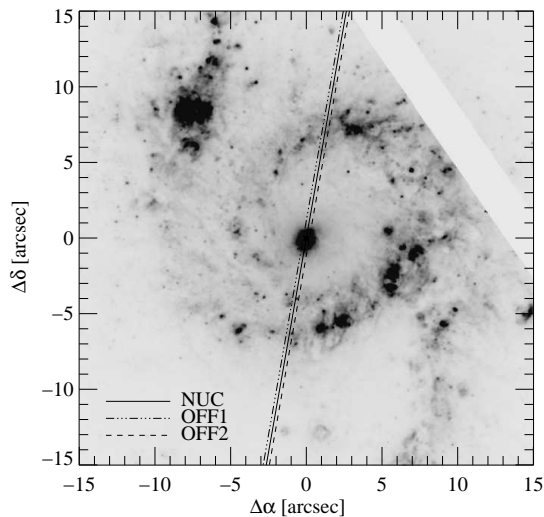}
\caption{\label{riga} HST-ACS narrow band $H\alpha$ image of NGC 3310 with slit positions overlaid. North is up and East is left. }
\end{figure*}

For each galaxy we analyze our own HST-STIS spectra and archival HST-NICMOS images (Fig.~\ref{fig:posizioni}; see also \citealt{huge}). In the case of NGC 3310 we consider also an archival narrow band H$\alpha$ image acquired with HST-ACS (Fig.~\ref{riga}). In the case of NGC 4258 we combine our HST-STIS data with archival ones.
In Tab.~\ref{log} we present a list of all available observations and the following subsections are dedicated to an accurate description of the analysis of spectra and images.

\subsection{\label{stis}STIS Observations}

The three galaxies were observed with STIS onboard the HST.  The adopted observational strategy consisted in obtaining spectra at three parallel positions with the central slit centered on the galaxy nucleus, and the flanking ones offset by $0.2\arcsec$. The archival STIS observations of NGC4258 were obtained with a similar strategy but the offset of the flanking slits was $0.15"$.
These different slit positions are shown in Fig.~\ref{fig:posizioni} and labeled 'NUC', 'OFF1' and 'OFF2'. The log of the observations is presented in Tab.~\ref{log}.

\begin{table*}
\caption{Log of spectroscopic observations}
\label{log}
\centering
\begin{tabular}{c c c c c c c c c c}
\hline\hline
 & ID & PID& Rootname & Date & $\theta$ [deg]$^{a}$ & $t_{exp}$ [s]  & $Offset~[\arcsec]$  & Slit size $[\arcsec]$ & Binning \\
\hline
NGC 3310 & NUC  & 8228& o5h17030,40 & Feb 11, 2000 & 170 & 864 & 0.0 & 0.197 & 2x2  \\
NGC 3310 & OFF1 & 8228& o5h17010,20 & Feb 11, 2000 & 170 & 480 & -0.2& 0.197 & 2x2 \\
NGC 3310 & OFF2 & 8228& o5h17050,60 & Feb 11, 2000 & 170 & 576 & 0.2 & 0.197 & 2x2 \\
\hline
NGC 4303 & NUC  & 8228& o5h735030,40 & March 31, 2000 & 135 & 720 & 0.0 & 0.197 & 2x2  \\
NGC 4303 & OFF1 & 8228& o5h735010,20 & March 31, 2000 & 135 & 480 & -0.2& 0.197 & 2x2 \\
NGC 4303 & OFF2 & 8228& o5h735050,60 & March 31, 2000 & 135 & 576 & 0.2 & 0.197 & 2x2 \\
\hline
NGC 4258 & NUC  &8228 & o5h734030,40 & August 01, 2000 & 246 & 864 & 0.0   & 0.197 & 1x1 \\
NGC 4528 & OFF1 &8228 & o5h734010,20 & August 01, 2000 & 246 & 576 & -0.2  & 0.197 & 2x2 \\
NGC 4258 & OFF2 &8228 & o5h734050,60 & August 01, 2000 & 246 & 576 &  0.2  & 0.197 & 2x2 \\
NGC 4258 & NUC  &8591 & o67104010,20,30 & March 16, 2001  & 146 & 144 & 0.0 & 0.095 & 1x1\\
NGC 4528 & OFF1 &8591 & o67104040 & March 16, 2001  & 146 & 158 & 0.15  & 0.197 & 1x1\\
NGC 4258 & OFF2 &8591 & o67104050 & March 16, 2001  & 146 & 120 & -0.15 & 0.197 & 1x1\\
\hline
\\
\multicolumn{10}{l}{\small{$^{a}$ Position angle of the slit (N to E). }}\\
\end{tabular}
\end{table*}

At each slit position, we obtained 2 spectra with the galaxy nucleus placed in different positions along the slit, shifted by an integer number of pixels, in order to remove cosmic-ray hits and hot pixels.  
Each spectrum was obtained with the G750M grating and the $0.2"$ slit which provided a dispersion of $\delta\lambda=1.108$\AA\  pix$^{-1}$ and a spectral resolution of $\emph{R}=\lambda / (2\delta\lambda) \sim 3000$. 
In the case of the archival NUC position in NGC4258, the slit size was $0.1\arcsec$. All the spectra, except for the NGC4258 archival and our nuclear ones, were obtained with a 2x2 on-chip binning resulting in a pixel spatial scale of
$0.101\arcsec$  pix$^{-1}$. See Tab.~\ref{log} for details on slit size and binning.

The data were first reduced using the STIS pipeline, \emph{calstis} \citep{handbook}, and  the most up-to-date reference files. The raw spectra were first reprocessed through the calstis pipeline and standard pipeline tasks were used to obtain flat-field corrected images. The two exposures taken at a given slit position were then realigned with a shift along the slit direction (by an integer number of pixels), and the pipeline task \emph{ocrreject} was used to reject cosmic rays and hot pixels. Subsequent calibration procedures followed the standard pipeline reduction described in the STIS Instrument Handbook \citep{handbook}; i.e., the spectra were wavelength calibrated and corrected for two-dimensional distortions. The expected accuracy of the wavelength calibration is 0.1-0.3 pixels within a single exposure and 0.2-0.5 pixels among different exposures \citep{handbook}, which converts into$\sim 3-8$ km s$^{-1}$ (relative) and $\sim 5-13$ km s$^{-1}$ (absolute). The relative error on the wavelength calibration is negligible for the data presented here because our analysis is restricted to the small detector region including H$\alpha$,  [NII]  and [SII] ($\Delta\lambda < 200$ \AA). 

\subsection{\label{images} Images}

HST-NICMOS images obtained with Camera 2 \citep{nicmos} and the F160W filter ($\sim H$ band) were retrieved from the HST archive. Their spatial scale is $0.076\arcsec$/pix and they have a field-of-view of $\sim 20 \times 20\arcsec$.
The images, with the slit positions superimposed, are shown in Fig.~\ref{fig:posizioni} and their details in Tab.~\ref{log_ima}. Imaging data reduction is described in detail by \cite{huge}.

\begin{table*}
\caption{Log of imaging observations}
\label{log_ima}
\centering
\begin{tabular}{c c c c c c c c c}
\hline\hline
          &  PID  & Rootname   & Date                & Instrument & Camera& Filter& $\theta(^{\circ})^{a}$ & $t_{exp}(s)$\\
\hline
NGC 4303 & 7330   & n3z132u010 & April 03, 1998      & HST-NICMOS & NIC2  & F160W &-103.42                 & 320  \\
\hline
NGC 3310 & 7268   & n49j06tkq  & November 12, 1997   & HST-NICMOS & NIC2  & F160W & 80.53                  & 128  \\
NGC 3310 & 9892   &  j8og20010 & October 21, 2003    & HST-ACS    & WFC1  & F658N & 0.014                  & 950   \\
\hline
NGC 4258 & 7230   & n46801040  & November 21, 1997   & HST-NICMOS & NIC2  & F160W & 85.54                  & 56  \\
\hline
\multicolumn{9}{l}{\small{$^{a}$ Position angle of image $y$ axes (deg e of n).}}\\
\end{tabular}
\end{table*}

For NGC 3310 we have also retrieved a narrow-band filter (F658N) image obtained with ACS (camera WFC1) \citep{acs}. The image was recalibrated using the ACS pipeline \citep{acs}~ with the most up-to-date calibration files. The final spatial scale of the calibrated image is $0.05"$/pix and its total field of view is $\sim 300\arcsec \times 300\arcsec$. The image is displayed in Fig.~\ref{riga} and its details in Tab.~\ref{log_ima}.


\section{\label{RI}Results}

In this section we present the results from the spectroscopic and imaging observations and the preliminary steps required to perform the kinematical modeling which can result in a BH mass estimate. In particular in Sec.~\ref{fitlinee} we first describe the procedure adopted to estimate the kinematical moments along the slits (surface brightness, velocity, velocity dispersion and asymmetry parameters) and we then present the results for each galaxy. In Sec.~\ref{stelle} we briefly describe the procedure used to derive the contribution of the stellar mass to the gravitational potential and we then present the results for each galaxy.

\subsection{\label{fitlinee}Kinematical analysis}

The emission lines analyzed to extract the kinematical parameters are H$\alpha$ $\lambda 6564$\AA, [NII] $\lambda\lambda 6548,6583$\AA\ and [SII] $\lambda\lambda$ $6716,6731\AA$.
Line-of-sight velocities, velocity dispersions, and surface brightnesses along each slit were obtained by fitting the emission lines in each row of the continuum-subtracted two-dimensional spectra.
In all cases the emission lines were modeled with a Gauss-Hermite series \citep{marel:hermexp,cappellaro2004}:

\begin{equation}
\emph{L(v)}=\frac{e^{-\frac{1}{2}y^2}}{\sigma\sqrt{2\pi}}\left[1+\sum_{m=3}^{4}h_{m}H_{m}(y)\right] 
\end{equation}

where $v$ is the velocity, $y=(v-v_0)/\sigma$ and $H_{m}$  with $m=3,4$ are the Hermite  polynomials of order 3 and 4, respectively \citep{cappellaro2004}.
The best fit parameters (line flux, average velocity, velocity dispersion, $h_3$ and $h_4$) were determined
by minimizing a penalized $\chi^2$ as described in \citealt{cappellaro2004}. The Gauss-Hermite expansion is usually adopted in the analysis of stellar kinematics from absorption lines; \cite{centauro} provide a complete description of the application of this method to the analysis of emission line spectra.

During the fit the only constraints we imposed were to link the kinematical parameters of lines belonging to the same doublet. Lines from the [NII] or [SII] doublet were constrained to have the same velocity, velocity dispersion and $h_3$, $h_4$. In the case of [NII] we also imposed a fixed flux ratio of $1:3$ between the faintest and the brightest line of the doublet. The velocity, velocity dispersion, $h_3$ and $h_4$ for H$\alpha$ and for the [NII] and [SII] doublets were then left free to vary during the fit.  Finally, for each slit position we considered the flux weighted average of all kinematical quantities.

\begin{figure*} 
\centering
\epsfig{file=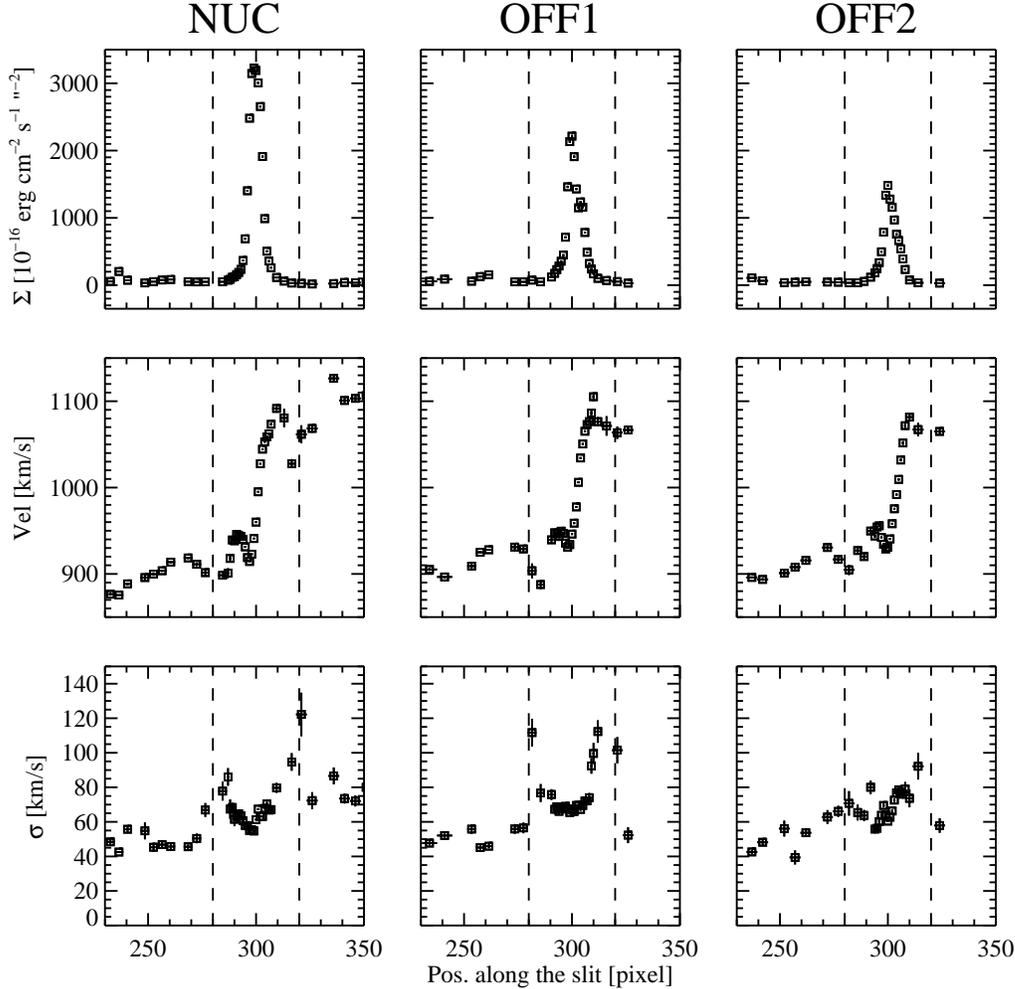, width=0.9\linewidth}
\caption{\label{andamenti_3310}Observed kinematical moments for NGC 3310 obtained with the fitting procedure described in the text. The first row shows the average emission line surface brightness (units of $10^{-16} erg s^{-1}cm^{-2}arcsec^{-2}$). The second row shows the velocity along the line of sight (units of km/s) and the third one the velocity dispersion of the line (units of km/s). From left to the right, the three columns refer, respectively, to the NUC, OFF1 and OFF2 slit positions indicated in Fig.~\ref{fig:posizioni}. The vertical errors-bars, when visible, represent the formal errors of the fit, while the horizontal bars are the sizes of the bins used to extracted spectrum (in general one detector pixel). The vertical deshed lines delimit the nuclear region, which we have identified on the basis of the surface brightness distributions in the upper panels and which will be analyzed in the kinematical model.}
\end{figure*}

\begin{figure*} 
\centerline{\epsfig{file=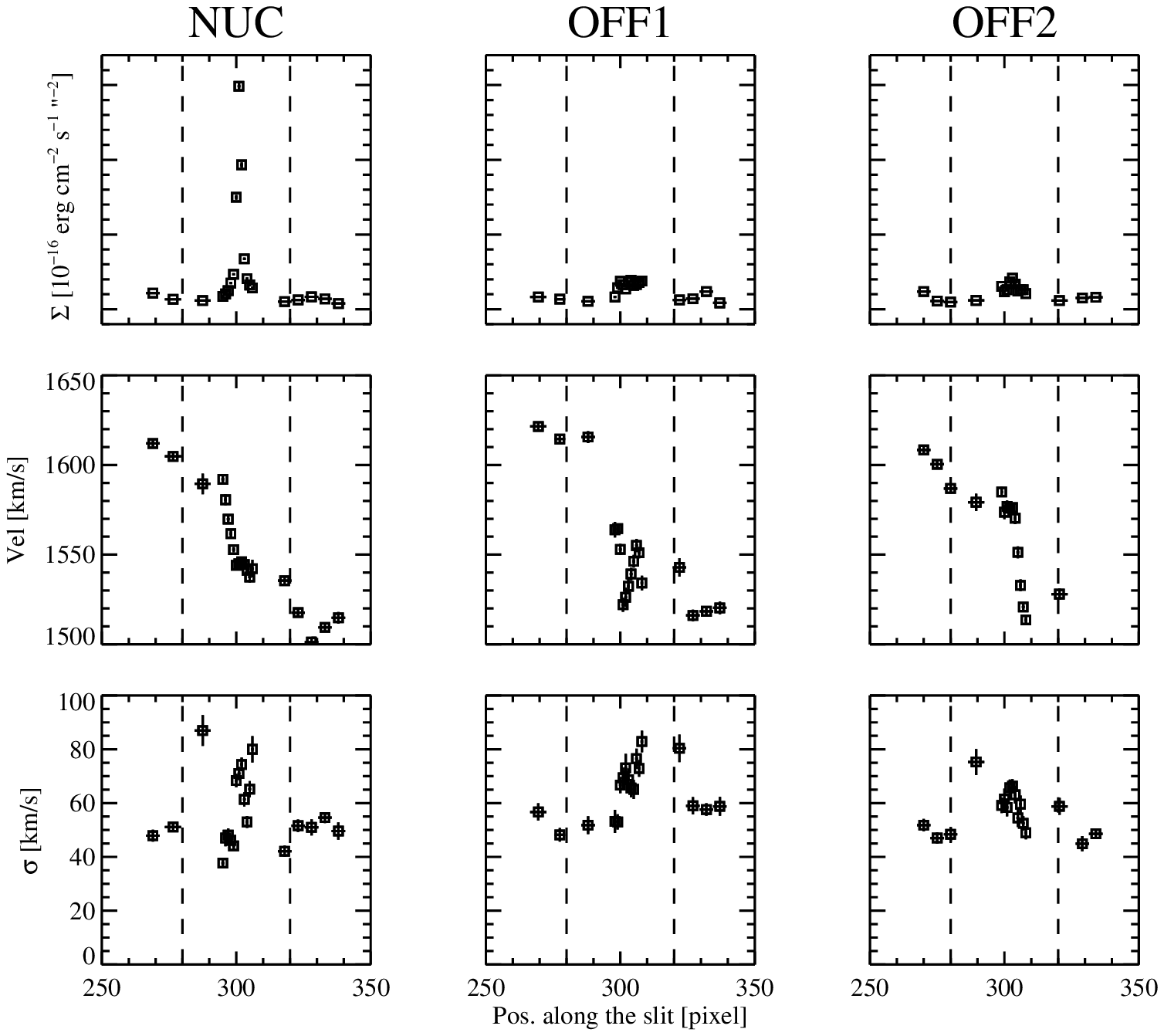, width=0.9\linewidth}}
\caption{\label{andamenti_4303}Observed kinematical moments for NGC 4303. Notation as in Fig.~\ref{andamenti_3310}.}
\end{figure*}

 Fig.~\ref{andamenti_3310} and Fig.~\ref{andamenti_4303} present the resulting kinematical quantities measured along all slit positions for NGC 3310 and NGC 4303.  The kinematical analysis of NGC 4258 followed a slightly different procedure and is presented at the end of this section.

The use of the Gauss-Hermite expansion allowed us to reproduce asymmetries in the emission line profiles in a more robust way than, e.g., the use of two Gaussian functions (see the discussion in \citealt{cappellaro2004}).
In the case of NGC 3310 and NGC 4303 the coefficients of the Hermite polynomials $h_3$, $h_4$ are small ($< 0.1-0.2$) so that the fitted emission line profiles are similar to pure Gaussian functions
and, indeed, the rotation curves obtained by imposing $h_3=h_4=0$ are not significantly different, only less smooth.

Both NGC 3310 and NGC 4303 are characterized by very peaked surface brightness distributions which suggest the existence of two different regions: a nuclear one, identified by the peak in surface brightness, and an extended one characterized by a surface brightness which is much smoother and more diffuse. A similar situation was found in NGC 4041 \citep{4041} and is consistent with the findings by \cite{knapen} at ground-based spatial resolution. 
In Fig.~\ref{andamenti_4303} and Fig.~\ref{andamenti_3310} the 'nuclear' region, identified on the basis of the surface brightness distribution (see Sec.~\ref{ris.gen}), is delimited by vertical dashed lines.

The velocity curves are different for the two galaxies. In NGC 3310, the nuclear rotation curve has the typical \emph{S} feature expected from a rotating disk (e.g.~\citealt{centauro}) with an amplitude of $\sim 200$ km/s. On the contrary, the velocity curve of NGC 4303 seems less regular and the \emph{S} shape appears more distorted. The amplitude is also much smaller, $\sim 50$ km/s.  Both galaxies show a large scale velocity gradient, such as that expected from the rotation of the galactic disk, but the rotation curves are disturbed.
The velocity dispersions of the gas do not show the behavior expected from a rotating disk (smooth plateau and peak at the position of the BH) but are more or less irregular suggesting that turbulent or non circular motions might affect the kinematics. 

In the case of NGC 4258 we adopted a modified procedure with respect to the one used for NGC 3310 and NGC 4303. In this galaxy the H$\alpha$ and [NII] lines are severely blended and, being a Seyfert 1.9, one has to take into account the possible presence of a broad line.

\begin{figure*}
\centering
\epsfig{file=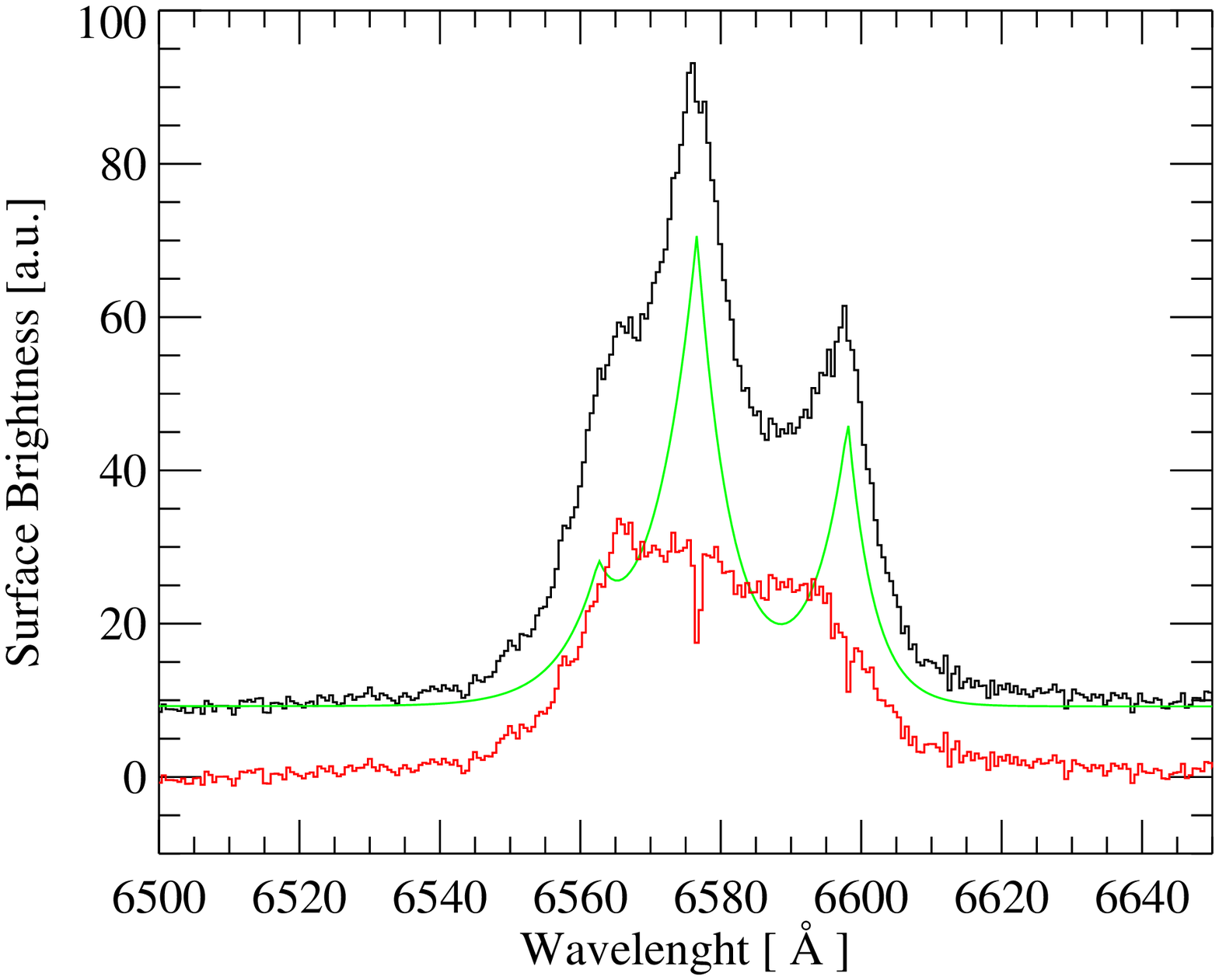, width=0.45\linewidth}
\epsfig{file=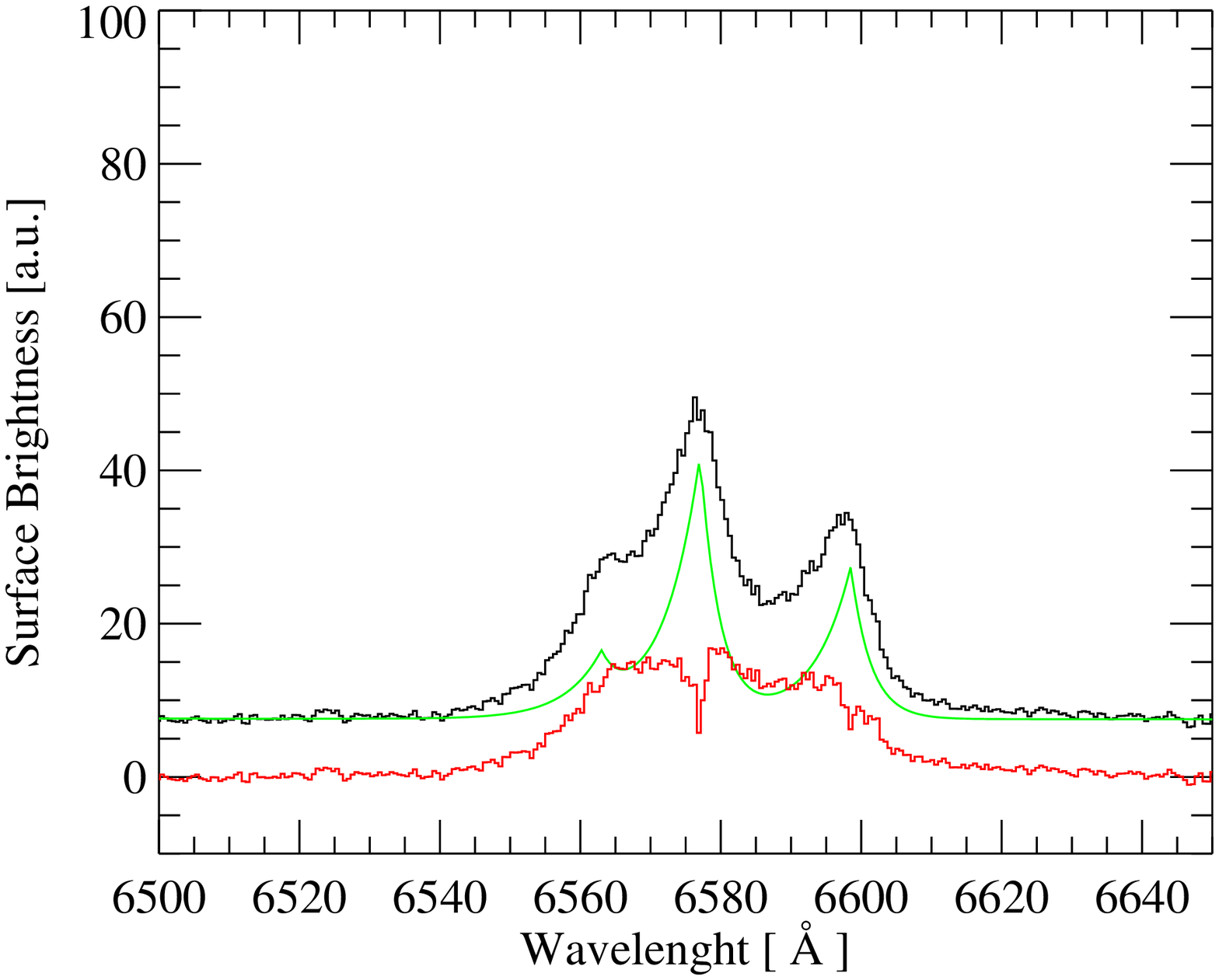, width=0.45\linewidth}
\caption{\label{broad} The black line shows the observed NGC 4258 nuclear spectrum, the green one the narrow lines used for subtraction; the residuals are drawn in red. These residuals represent the Broad H$\alpha$ Line. The left and the right panels refer, respectively, to archival and our own data.}
\end{figure*}

\begin{figure*}
\centering
\epsfig{file=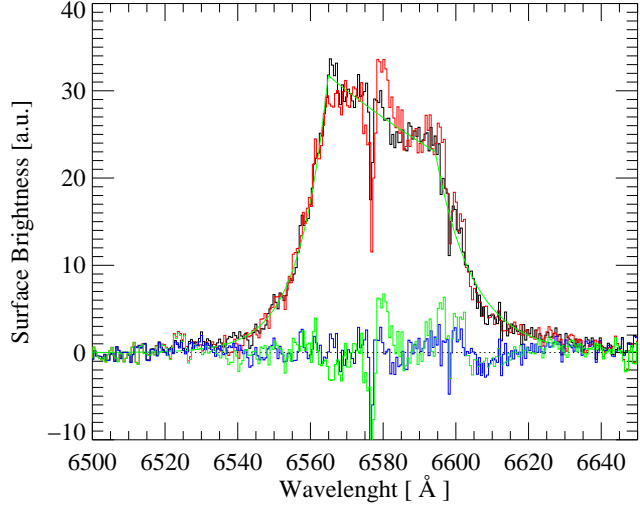, width=0.48\linewidth}
\caption{\label{paragone} The black and red lines show, respectively, the broad H$\alpha$ component obtained by subtracting the narrow lines from archival and our own nuclear spectra of NGC 4258. The green and blue lines show, respectively, the fitted broken-power-law models and their residuals.}
\end{figure*}

We first assume that H$\alpha$, [NII] and [SII] share the same kinematics. Although there might be differences due to the fact that H$\alpha$ can be emitted also by gas at different stages of excitation, our assumption will in any case result in average kinematical quantities, weighted with the line surface brightness.
We collapse the central 4 or 6 rows in order to obtain a high S/N nuclear spectrum and we fit the [SII] doublet with the same constraints adopted for NGC 3310 and NGC 4303. The kinematical parameters found from the [SII] doublet are then used to reconstruct the H$\alpha$ and [NII] narrow lines which are then subtracted from the spectrum. The flux of the H$\alpha$ and [NII] narrow lines is adjusted so that the residual spectrum does not show obvious structure due to under/over subtracted narrow lines. The residual spectrum is that showing the broad line and any other components which have a different kinematics to that of the [SII] doublet (see Fig.~\ref{broad}). It is interesting to note that the broad H$\alpha$ line (FWHM$\sim 1600$ km/s) does not have a Gaussian profile but a profile resembling that of double humped broad line galaxies (e. g. \citealt{heraclus,strateva} and references therein.)
\citet{nagao} modelled the BLR of their quasar templates with a double power law function. Here, we find that a good model profile for the BLR is a three piece broken powerlaw as shown in Fig.~\ref{paragone}.

\begin{figure*}
\centering
\epsfig{file=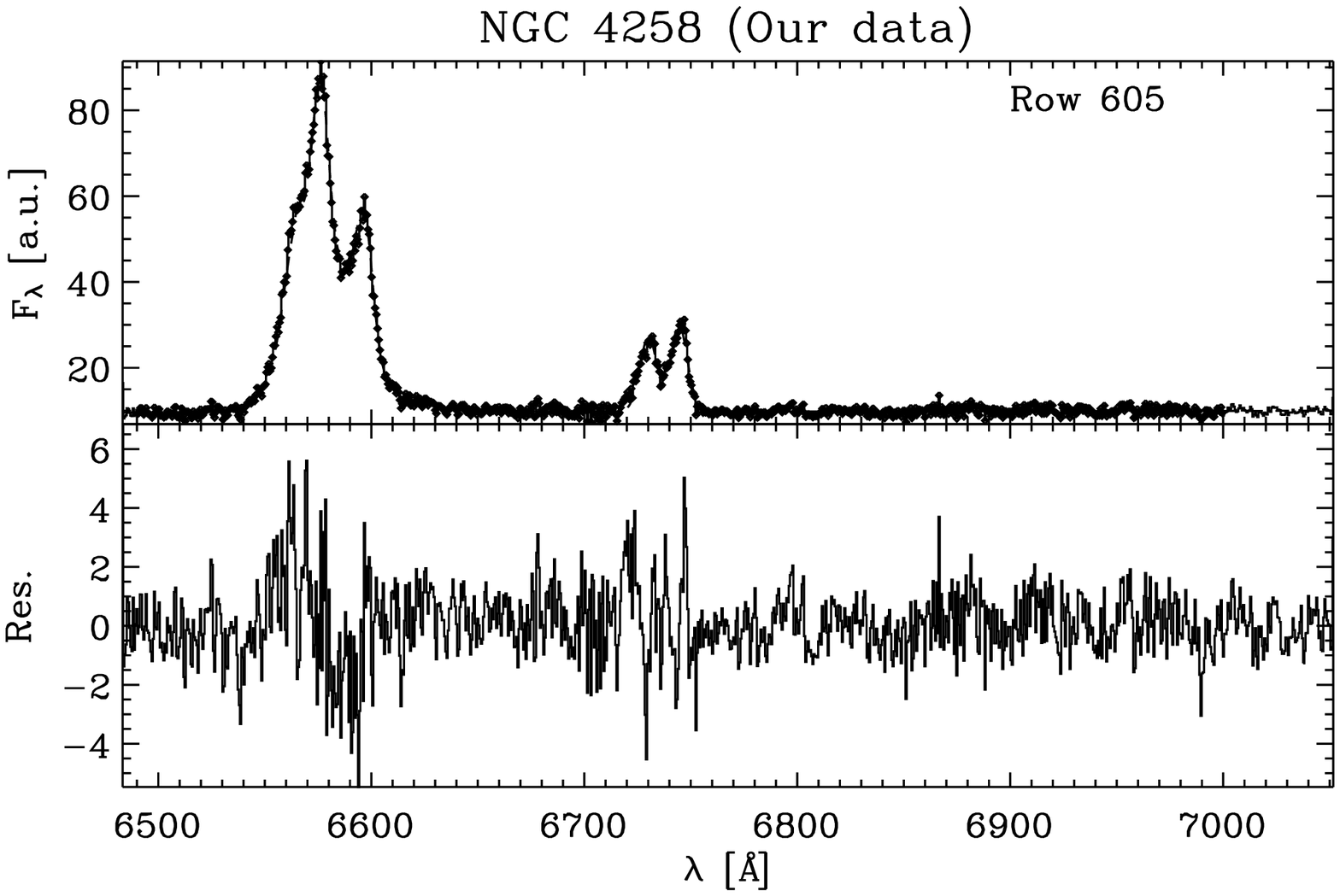, width=0.45\linewidth}
\epsfig{file=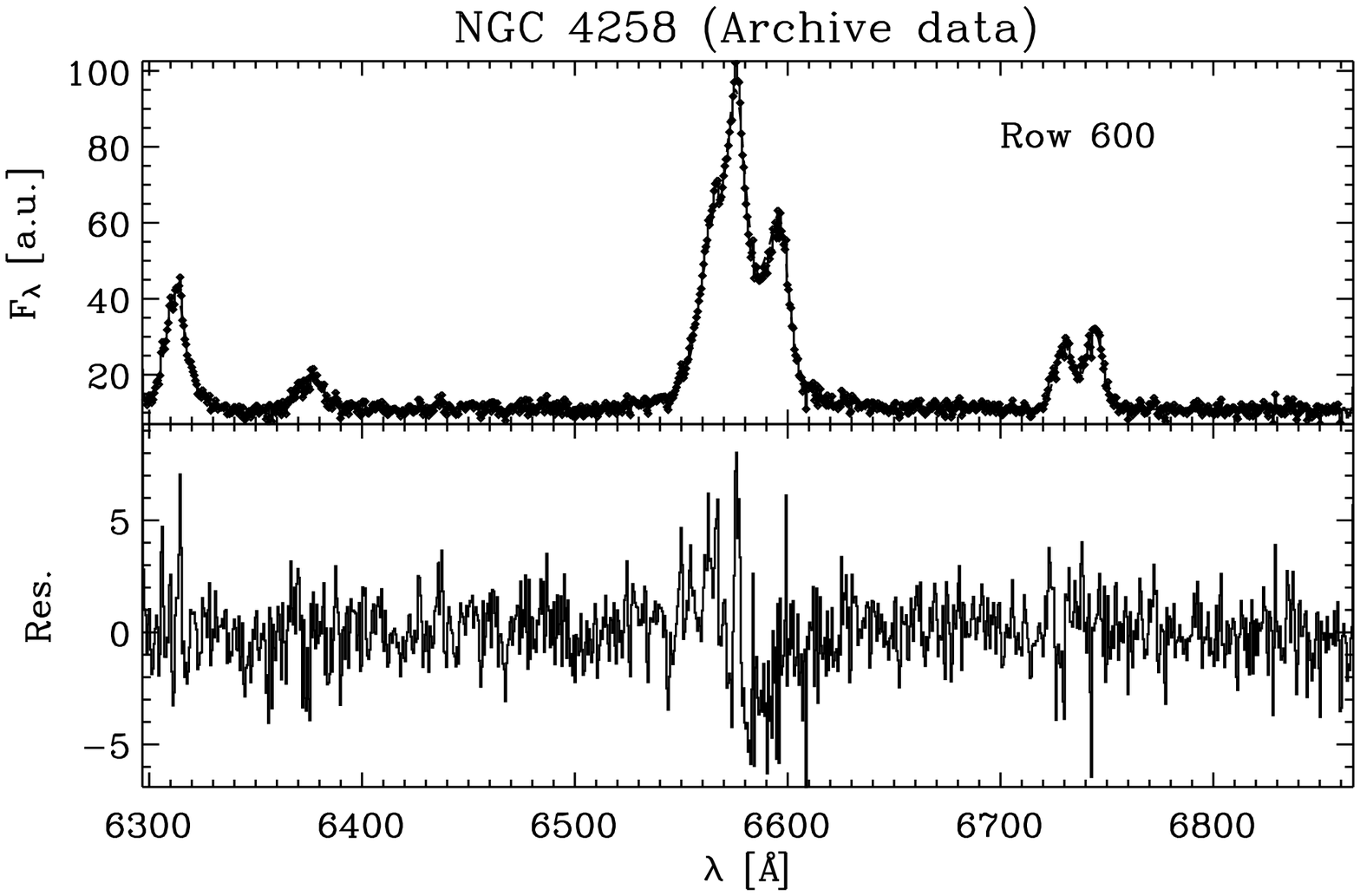, width=0.45\linewidth}
\caption{\label{es:4258} Example of line fitting of the H$\alpha$, [NII] and [SII] lines in the nuclear region of NGC 4258. The left and right panels refer, respectively, to our and archival data. The lower window of each panel shows the residuals of the fit.}
\end{figure*}

Note that the BLR profiles derived from our and archival spectra are consistent, within the uncertainties, in shape and normalization. If the broad line flux did not vary significantly from one observation to another, the above similarity implies that the NUC slits were well centered on the nucleus position.
We then refitted all emission lines in nuclear spectra with the components (modeled with the Gauss-Hermite expansion and constrained to share the same kinematics) and a broad H$\alpha$ modeled as described above. Except for the flux, all parameters of the broad H$\alpha$ line derived from these fits were then kept fixed in the subsequent row by row fits.
$h_3$ and $h_4$ were set to 0 in the low S/N spectra of NUC and in all spectra of OFF1,2.
Fig.~\ref{es:4258} shows an example of the fit for ours and the archival data.

\begin{figure*}
\centering
\epsfig{file=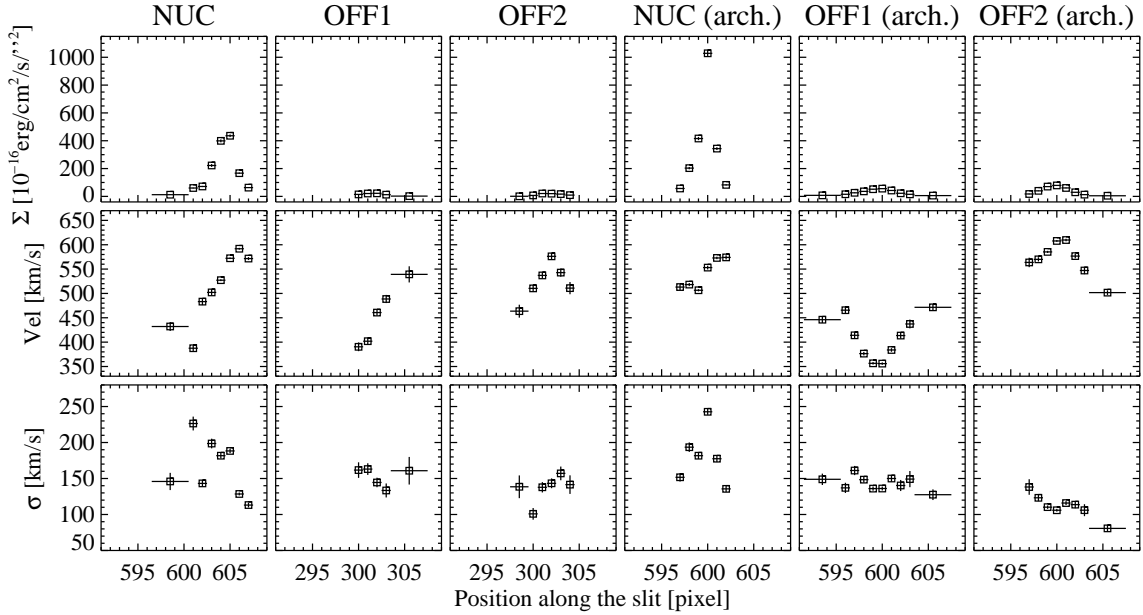,angle=90, width=0.85\linewidth}
\caption{\label{andamenti_4258_ours}Observed kinematical moments for NGC 4258: the first three columns refers to our data, the other ones represent the archival data. Archival and our own data were acquired with the slit placed at two different position angles as shown in Fig.\ref{fig:posizioni}.}
\end{figure*}

The kinematics extracted with this fitting procedure are shown in Fig.~\ref{andamenti_4258_ours}. 
In contrast with NGC 3310 and NGC 4303, the rotation curves are limited only to within $0.5\arcsec$ of the central source. As for the two other spiral galaxies, we detect the bright nuclear emission but we lack the fainter extended component. 
With the limited spatial coverage, the rotation curves appear more regular than in the cases of NGC 3310 and NGC 4303.
The velocity amplitudes of the rotation curves in our data are about $\sim 200$ km/s while these drop to $\sim 100$ km/s for the archival data. This is not surprising since our spectra and the archival ones are taken at different position angles.
The velocity dispersions, apart from some deviant points, seem to exhibit a nuclear peak at about $\sim 200-250 $ km/s
and much smaller constant values at the offnuclear positions $\sim 150$ km/s with a behavior similar to that of a rotating disk.

A possible cause of concern for the accuracy of our kinematical measurement is given by the fact that the point spread function of HST is undersampled by the pixels of the STIS CCD.
When spectral images with unresolved components are rectified, interpolation along the undersampled spatial axis produces artifacts in the rectified image. An example of the artifacts in the peak row of a G750M image is shown in Figure 11.9 (Section 11.3.5) in the STIS Instrument Handbook \citep{handbook}. The spectrum in each row of the image undulates across the detector as flux is alternately retained, gained, or lost by the interpolation. Adjacent rows have undulations that are out of phase, since the interpolation inappropriately pushes flux out of one row and into the next. Undulations were obvious in the continuum spectrum of NGC4303. They presented no problems for the analysis of the emission lines, however. The lines were so narrow that the underlying continuum was well represented by a short linear segment. NGC 4258 was a somewhat more problematic case, since it has a spatially unresolved spectral feature (a broad $H_{\alpha}$ component) which is broad relative to the structure of the undulations. This undulating feature must be subtracted from the rows of the spectrum so that the parameters of the narrow lines can be measured. The undulation structure cannot be directly observed in this case, since it is mixed in with the spatial and spectral structure of the $H_{\alpha}$ + [NII] blend. We assessed its impact by artificially dividing our nuclear spectra with several observed undulation patterns and remeasured kinematical parameters. The impact on the velocity measurements is in general small. 
Moreover, the size of the residuals in the original fits sets limits to the magnitude 
of the neglected undulations.  Our tests indicate that undulations of this 
magnitude would not significantly affect the velocity measurements.

\subsection{\label{stelle}Light Profile}

In order to account for the contribution of the stars to the gravitational
potential in the nuclear region, the stellar luminosity density has to be
derived from the observed surface brightness distribution. The inversion
procedure to derive this density distribution from the observed surface brightness is not unique
if the gravitational potential does not have a spherical symmetry, such as can be revealed by
circular isophotes. Assuming that the gravitational potential is an oblate
spheroid, the inversion depends on the knowledge of the potential axial ratio,
$q$, and the inclination of its principal plane with respect to the line of
sight, $i$.  As these two quantities are related by the observed isophote
ellipticity, we are left with the freedom of assuming different galaxy
inclinations to the line of sight.  
Following \cite{van98}, we 
assumed an oblate spheroid density distribution parameterized as:
\begin{equation}\label{eqrho1}
\rho(m) = \rho_0\left(\frac{m}{r_b}\right)^{-\alpha} \left[1+\left(\frac{m}{r_b}\right)^2\right]^{-\beta}
\end{equation}
$m$ is given by $m^2 = x^2+y^2+z^2/q^2$ ; $xyz$ is a reference system with
the $xy$ plane corresponding to the principal plane of the potential and $q$ is the intrinsic axial ratio.  When $q=1$ the model is spherical and $m$ corresponds to the radius.
The above stellar density is integrated along the line of sight to provide the surface brightness distribution on the plane of the sky. This is then convolved with the instrumental Point Spread Function (PSF) (derived from TinyTim, \citealt{tinytim}) of the telescope+instrument system and averaged over the detector pixel size to obtain the observed surface brightness distribution. At this point, the derived model light profile can be directly compared with observed ones (for a detailed description of the inversion and fit, see \citealt{4041}).

\begin{figure*} 
\centerline{\epsfig{file=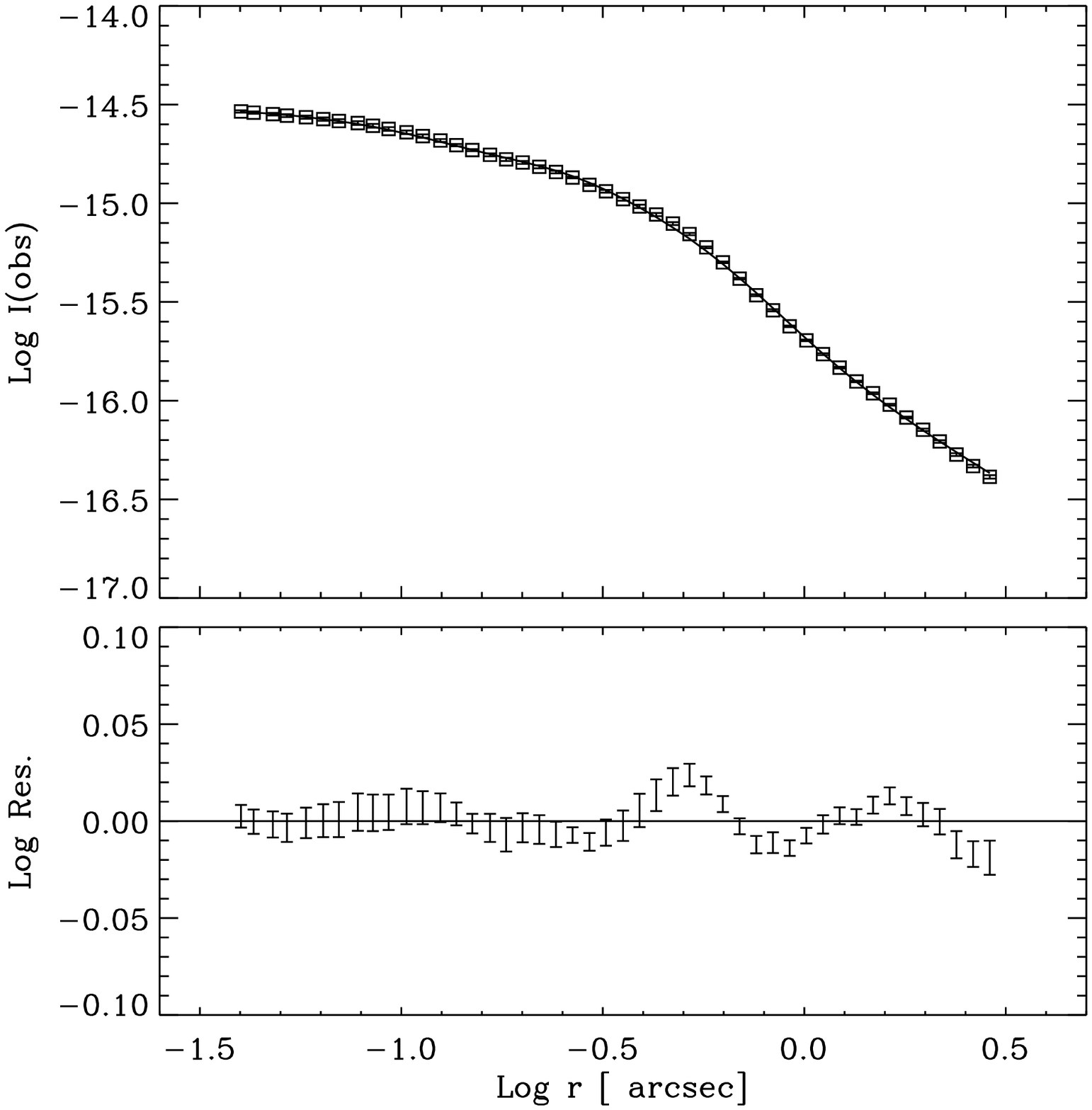, width=0.35\linewidth}\epsfig{file=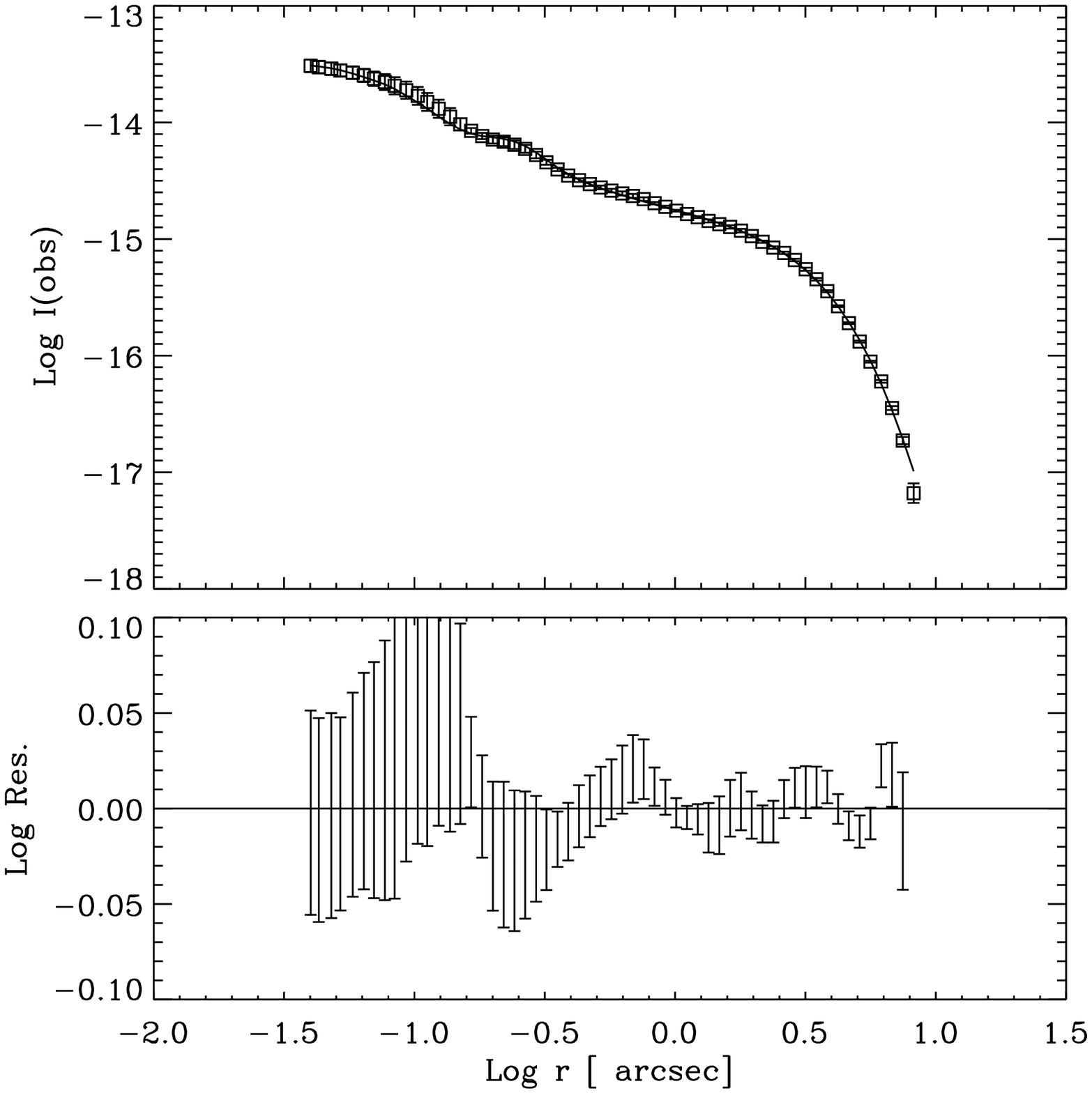, width=0.35\linewidth}\epsfig{file=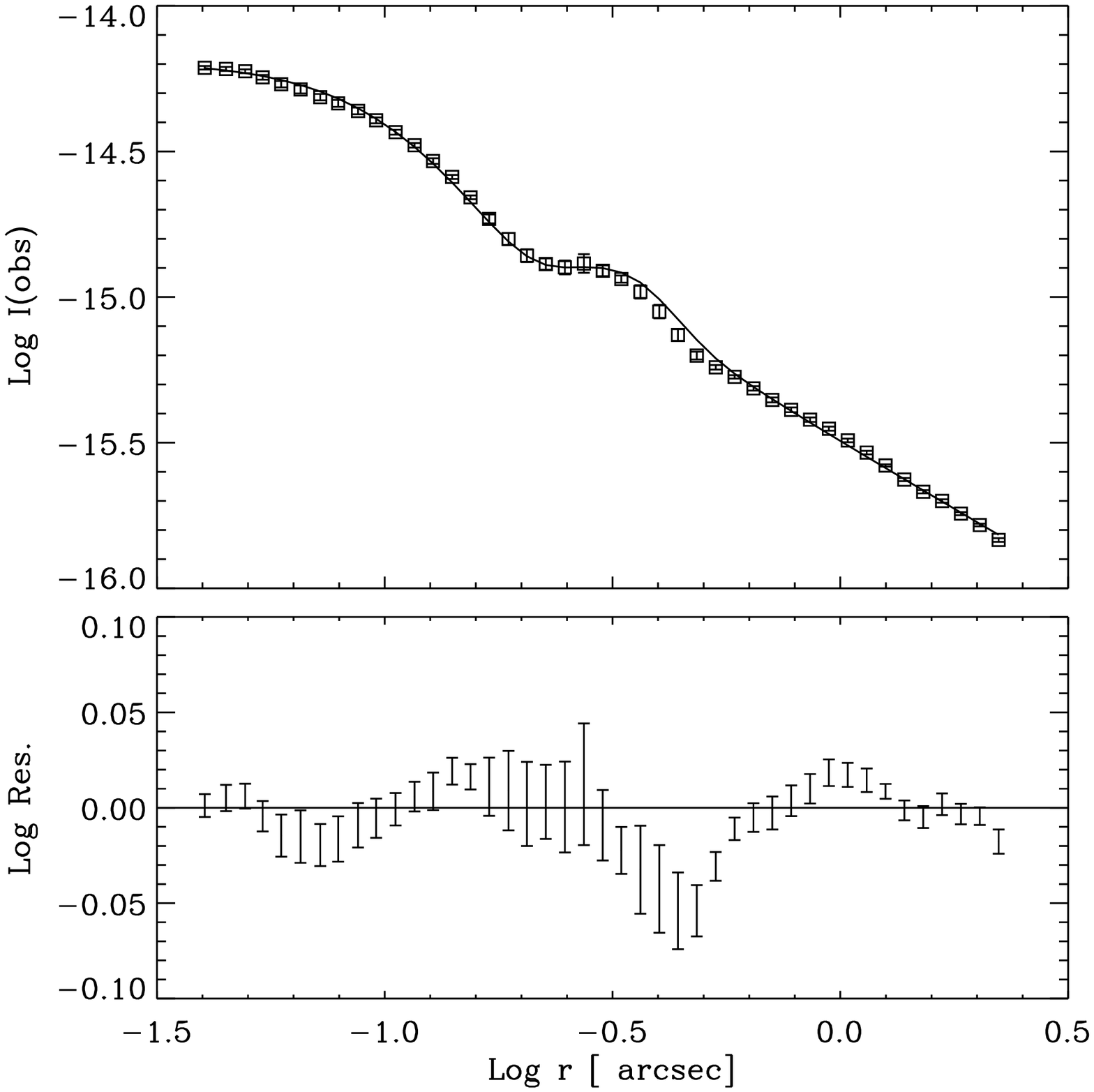, width=0.35\linewidth}}
\caption{\label{profili} Surface brightness profiles ($erg\,cm^{-2}\,s^{-1}\,arcsec^{-2}$) of the continuum emission from the F160W NICMOS images. The left, central and right panels refers to NGC 3310, NGC 4303 and NGC 4258, respectively.
Dots with error bars (when visible) denote the observed data while the solid lines indicate the model fits (see text). The lower plots show the residuals after subtraction of the model fits.}
\end{figure*}

We considered the galaxy light profiles obtained by using the IRAF/STSDAS program \emph{ellipse} to fit elliptical isophotes to the F160W continuum NICMOS images of NGC 3310, NGC 4303 and NGC 4258.
The ellipticity of the isophotes in the nuclear region of all our galaxies (excluding the contribution from the central unresolved source, when present)
is very close to zero ($\epsilon=0.15,0.05, 0.0$ for NGC 4303, NGC 3310 and NGC 4258, respectively) allowing us to assume spherical symmetry (i.e. $q=1$).
In the case of NGC 3310, the isophote fit was limited to the region included by the circumnuclear ring visible in Fig.~\ref{riga}, in order not to be affected by the bright spiral arms present on large scale.
The observed surface brightness profiles are shown in Fig.~\ref{profili}.

\begin{table*}
\caption{Best-fit parameters for the stellar light density model. Mass densities ($\rho_{0}$, $\rho_{0}^{*}$) are derived from luminosity density  under the assumption that $\Gamma=M_{\odot}/L_{\odot,H}$. $\chi^{2}_{red}$ is $\chi^{2}/dof $ where \emph{dof} represents the degrees of freedom of the fits (47 for NGC 4303, 37 for NGC 3310 and NGC 4258). 
}
\label{tab:risultati_stelle}
\centering
\begin{tabular}{c c c c c c c c c c c c c}
\hline\hline
          &  $q^{a}$ &  $\log\rho_{0}^{b}$   & $r_{b}^{c}$ &  $\alpha$ & $\beta$ & $\gamma$ & $\log\rho_{0}^{*a} $ & $r_{b}^{*b} $ & $\alpha^{*}$ & $\beta^{*}$ & $F_{PSF}^d$ & $\chi_{red}^{2}$ \\
\hline
NGC 4303 & $1.0^{\star}$  & 2.28  & 19.90 & 2.35  & 11.90 & $2.00^{\star}$ & -0.76  & 8.51  & 0.21 &  7.87 & 9.62  & 13.09\\
NGC 3310 & $1.0^{\star}$  & 3.43  & 0.66  & -0.71 & 3.67  & $2.00^{\star}$ & 3.40   & 0.23  & 1.79 &  0.22 & 4.40  & 52.66\\
NGC 4258 & $1.0^{\star}$  & 1.59  & 4.55  & 1.86  & 0.14  & $0.000^{\star}$& ---    & ---   & ---- &  ---- & ----- & 48.83\\
\hline
\multicolumn{13}{l}{\small{$^{a}$ Observed axial ratio (defined as $q^{2}=1-e^{2}$ where $e$ is the ellipticity).}} \\
\multicolumn{13}{l}{\small{$^{b}$ In units of $M_{\odot}$ pc$^{-3}$.}}\\
\multicolumn{13}{l}{\small{$^{c}$ In units of arcsec.}}\\
\multicolumn{13}{l}{\small{$^{d}$ Point source flux in units of $10^{-17}$ erg cm$^{-2}$ s$^{-1}$ \AA$^{-1}$.}}\\
\multicolumn{13}{l}{\small{$^{\star}$ Parameter was held fixed, as explained in detail in ~\ref{stelle}.}}\\
\end{tabular}
\end{table*}

We followed the above procedure 
to fit the observed surface brightness profiles within $\sim 3\arcsec$ of the continuum peak. This is adequate to our goals since, as shown in the next section, the kinematical model fitting is performed only in the inner $\sim 1\arcsec$. The best-fit parameters for all galaxies are presented in Tab.~\ref{tab:risultati_stelle} where mass densities are computed for $\Gamma=M/L=M_{\odot}/L_{\odot,H},$ (its real value will be determined during the kinematical fitting procedure in Sec.~\ref{modelli}).  
The comparison between models and observed data is shown in Fig.~\ref{profili}.

A nuclear point source was added for NGC 4303 and NGC 4258 in order to account for the AGN emission.  Indeed 
its presence is clearly indicated  by the little bump at $\log r \sim -0.5$, which represents the first Airy ring of the HST PSF.
In the following analysis, we will assume that the central point sources are unresolved emission from the AGN, and therefore they do not contribute to the total stellar light. An alternative possibility is that these central point sources are unresolved star clusters common in late-type spirals \citep{boeker}. However NGC 4303 and NGC 4258 are known AGNs (see Sec.\ref{introduzione}). It has been shown by \citet{quillen} that unresolved infrared sources are found in the great majority of HST NICMOS images of Seyfert galaxies and that their luminosities strongly correlate with both the hard X-ray and $[O III]$ line luminosity. This result strongly suggests a dominant AGN contribution to the IR emission. We thus consider unlikely that the point sources are due to unresolved star clusters.

In NGC 4258, the observed brightness profile is well reproduced by the spherical stellar light distribution of Eq. ~\ref{eqrho1} and by a central point source. However in the cases of NGC 3310 and NGC 4303 a good match of the observed profiles required an additional component described by Eq. ~\ref{eqrho1} but characterized by  $\rho_{0}^{*},r_{b}^{*},\alpha^{*},\beta^{*}$.

\begin{figure*} 
\centerline{\epsfig{file=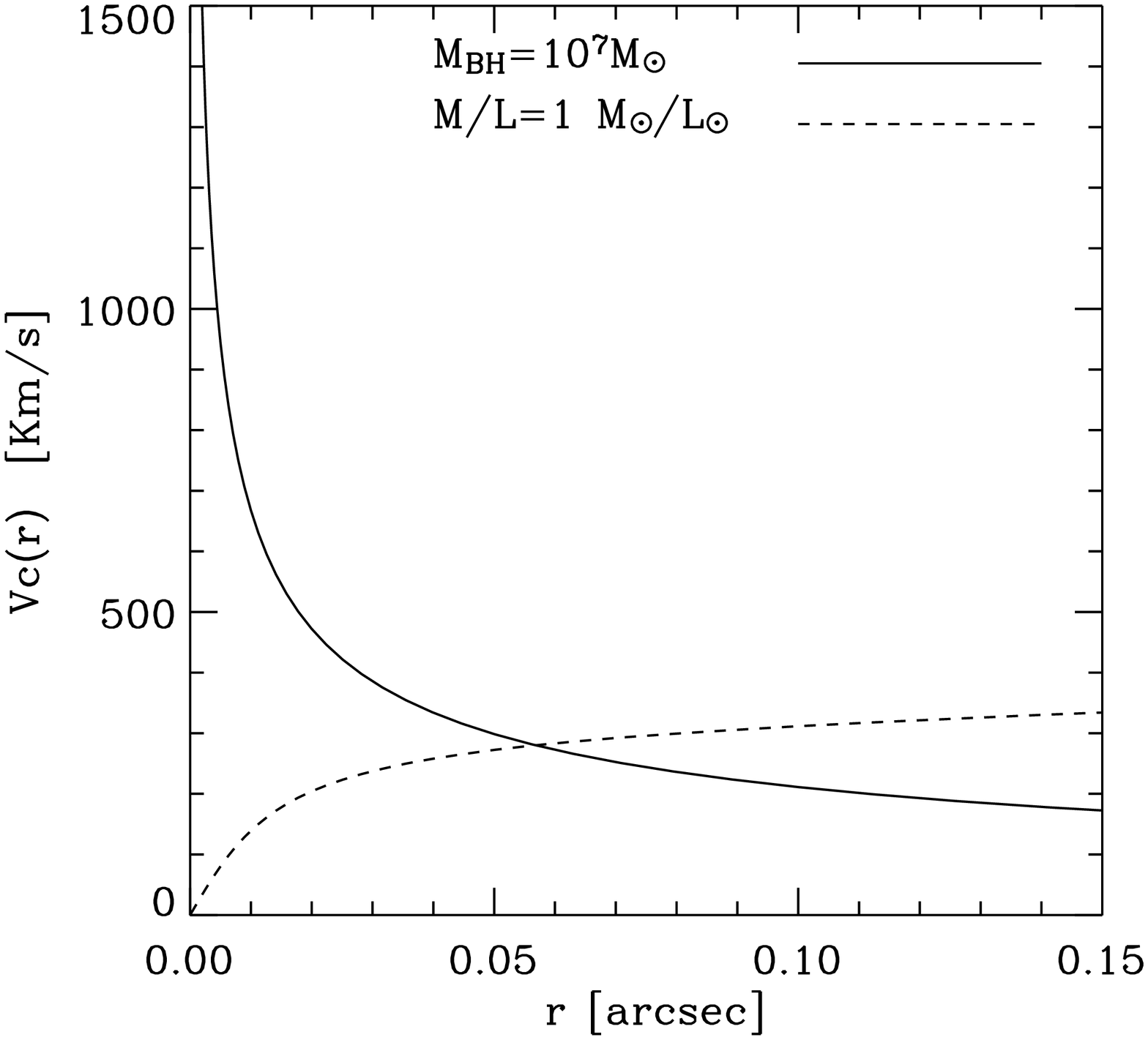, width=0.35\linewidth}\epsfig{file=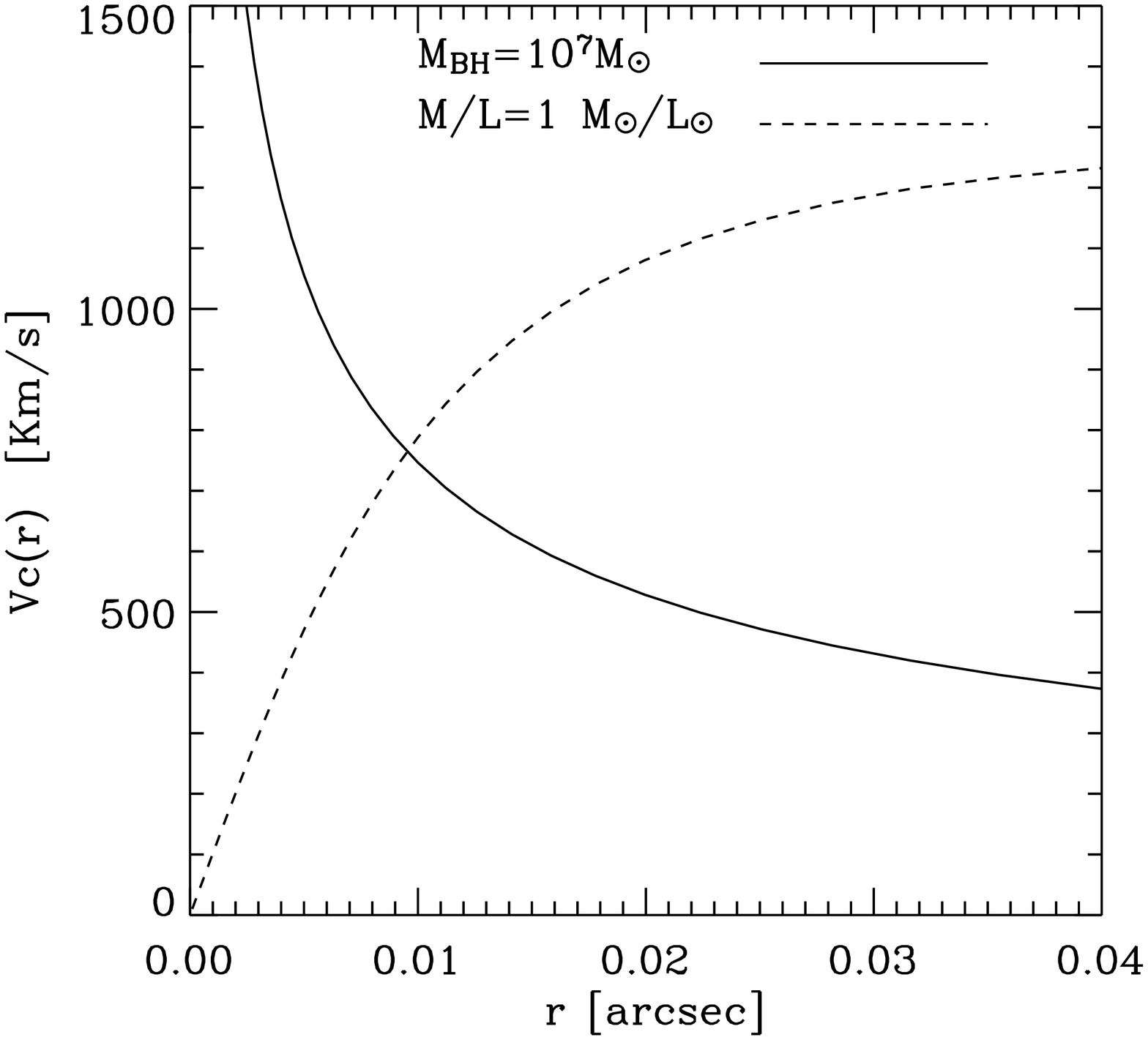, width=0.35\linewidth} \epsfig{file=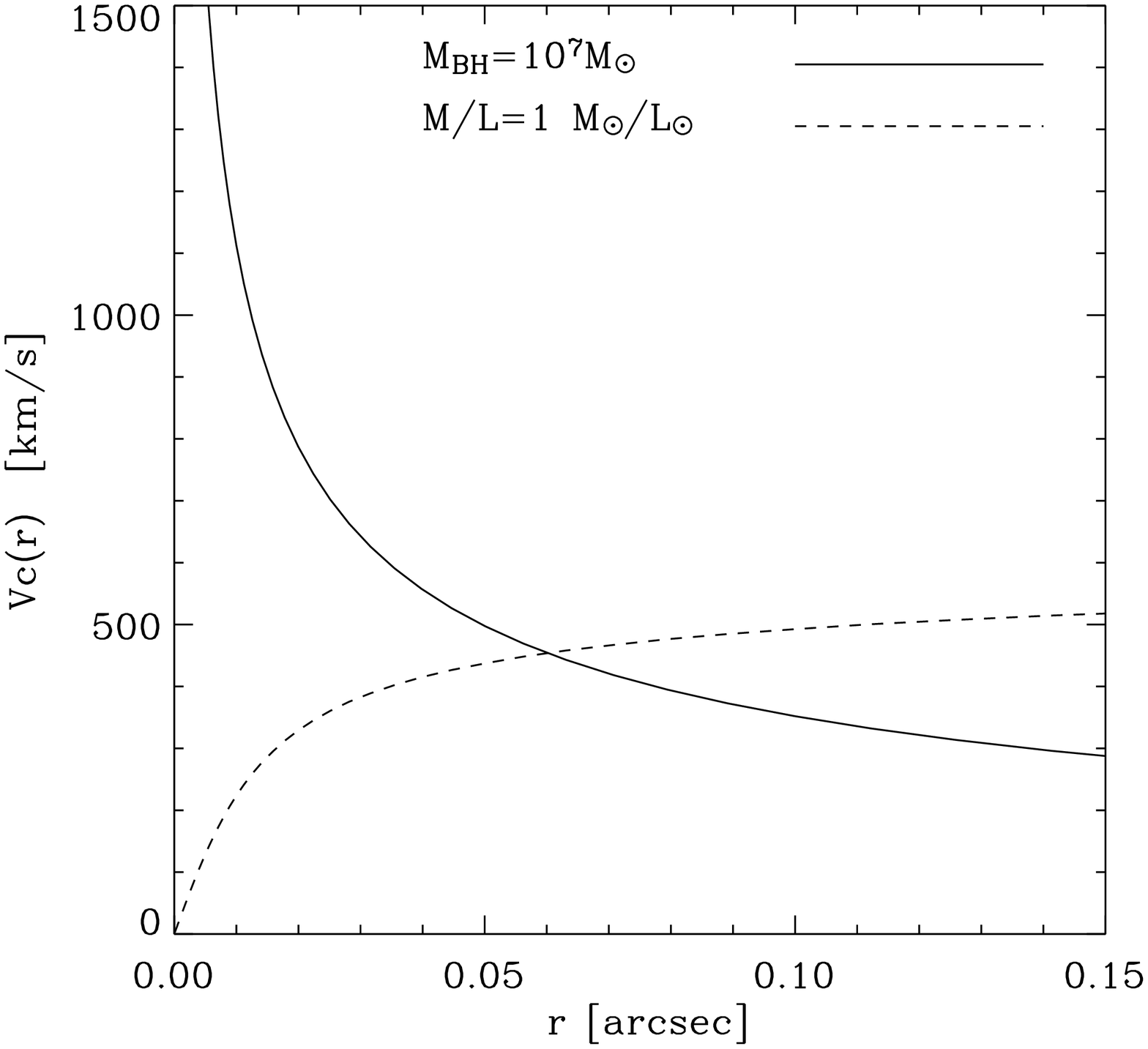, width=0.35\linewidth}}
\caption{\label{vcircolare} Circular velocities expected from the stellar gravitational potential on the principal planes (dashed lines) with the assumption of unit $M/L$.
 The solid lines represents the circular velocities expected from the gravitational potential of a BH with
 $M_{BH}=10^7M_{\odot}$. The left, central and right panels refer to NGC 3310, 4303 and 4258, respectively.}
\end{figure*}

Given $\rho(r)$, one can easily derive $M(r)$ the amount of stellar mass contained within the radius $r$.
Fig.~\ref{vcircolare} shows, for all galaxies, the circular velocity due to the stars as a function of the distance from the nucleus computed assuming $M/L=M_{\odot}/L_{L\odot,H}$ (solid line). The dashed line represents the circular velocity expected from a BH with mass $M_{BH}=10^7M_{\odot}$.

\section{\label{modelli}Kinematical Models}
 
The procedure used to model the kinematical data was first described by \citet{macchetto97} and then refined by several authors \citep{van98,barth01,4041}. This procedure, described in detail in \citet{4041}, assumes that the ionized gas is circularly rotating in a thin disk placed in the principal plane of gravitational galaxy potential. We neglect any hydrodynamical effects like gas pressure. The gravitational potential is made by two components: the stellar one (whose mass distribution was determined in Sec.~\ref{stelle}) and a dark mass concentration (the putative BH), unresolved with the spatial resolution of our observations. In computing the rotation curves we take into account the finite spatial resolution of observations, the intrinsic surface brightness distribution (which is the weight to average kinematical quantities) and we integrate over the slit and pixel area. The free parameters characterizing the best fitting model are found by standard $\chi^{2}$ minimization. 

In Sec.~\ref{modelli_ISBD} we determine the Intrinsic Surface Brightness Distribution (ISBD hereafter) which plays a very important role in the modeling (see \citealt{centauro} for a detailed discussion).
In Sec.~\ref{ris.gen} we describe the fitting of the rotation curves. \cite{centauro} have shown that the velocity dispersion depends on the adopted ISBD, therefore we decided to restrict the model fitting only to velocity.
Moreover the observed velocity dispersions of our galaxies are small (see Fig.~\ref{andamenti_3310}-\ref{andamenti_4303} and  Fig.~\ref{andamenti_4258_ours}), not-monotonic and likely affected by turbulent motions; we will discuss this issue further in Sec.~\ref{discuss}.
Summarizing, the free parameters during the model kinematical fitting are:
\begin{itemize}
\item
the black hole mass $M_{BH}$;
\item
the stellar mass to light ratio $M/L$;
\item
the coordinates  of the BH position $(X,Y)$ on the plane of the sky where the nuclear slit is centered at (0,0);
\item
the galaxy systemic velocity $V_{sys}$;
\item
the position angle of the disk line of nodes $\theta$; 
\end{itemize}

The inclination $i$ of the gaseous disk cannot be treated as a free parameter. Since the amplitude of the rotation curve scales as $v(r)^{2}\sim M(r)sin^{2}(i)$, $i$ is tightly coupled with $M_{BH}$. The residual, weak dependence of the projected velocity on $\cos i$ (Eq.\ B2 and B4 in Appendix B of \citealt{marconi04}) can be used to derive $i$ only when high S/N, integral field kinematical data are available (e.g.\ \citealt{barth01}) but this is not the case of the data presented here. For this reasons we decided to work, for each galaxy, with an initial fixed value of the inclination and we then investigated the effects of varying $i$ on the final best fit parameters.

Usually the PSF adopted in the modelling is the one at $\lambda$ 6700 $\AA$ directly obtained with the TinyTIM modeling software \citep{tinytim}. However, \citet{dressel} pointed out that, by default, TinyTIM computes the PSF with the presence of a Lyot stop which is not present when the G750M grating is used. The Lyot stop is mounted in front of the optical elements (e.g.\ G750L) close to the pupil plane and it supplements the focal plane mask used in coronographic mode by suppressing residual scattered and diffracted starlight from the collimated beam. It has a broadening effect on the PSF: with the Lyot-stop, in fact, the detector sees a smaller diameter primary mirror, has a broader Airy disk and a slightly more distant first Airy ring. A comparison of the radial profile of light for model imaging the PSF at 6600 $\AA$  with and without the Lyot stop, is shown in Fig.~1 of \citet{dressel}. Therefore we followed the procedure suggested by \citet{dressel} to obtain from TinyTIM the G750M PSF without the Lyot-stop.

\subsection{\label{modelli_ISBD}The Intrinsic Surface Brightness Distribution}

The observed kinematical quantities are averages over apertures defined by the
slit width and the detector pixel size along the slit. The ISBD is thus a fundamental ingredient
in the modeling of the kinematical quantities because it is the weight of the
averaging process (see, e.g., the discussion in \citealt{centauro}).
 
\cite{barth01} derived the ISBD by deconvolving a continuum subtracted emission
line image and were successful in reproducing the microstructure of the
rotation curves thus reducing the overall $\chi^{2}$ of their best fitting
model. However, apart from the microstructure of the rotation curves which has
little impact on the final $M_{BH}$ estimate, \cite{4041} showed that it is
crucial to adopt a good ISBD in the central region where the velocity gradients
are the largest. This has a strong impact on the quality of the fit and
possibly on the final $M_{BH}$ estimate. When the observed line surface
brightness distribution along the slit is strongly peaked in the nuclear region
and has a profile which is a little different from that of an unresolved source,
one cannot simply use a deconvolved emission line image but should try to
reproduce the observed line surface brightness distribution with a
parameterized intrinsic one and take into account the instrumental effects.
\\Recently \citet{centauro} showed that the adopted ISBD, provided that it
reproduces the observed one within the errors, does not affect the final BH
mass estimate.  In their paper, they showed that systematic errors on $M_{BH}$
due to the adopted ISBD were of the order of $0.05 - 0.08$ in $\log M_{BH}$.
Notwithstanding the ISBD has an important effect on the quality of
the velocity fit although not on the $M_{BH}$ value. 

We  parameterized the ISBD with a combination of analytical functions.  We used a combination of
exponential, Gaussian and constant components (labeled respectively
\emph{expo}, \emph{gauss} and \emph{const} in tables of results) and their
expression are, respectively:

\begin{equation}
\left\{
\begin{array}{ll}
I(r)=0 &   r \leq r_{h}\\
I(r)=I_{0}e^{-r/r_{0}} & r > r_{h}
\end{array}
\right.
\end{equation}

\begin{equation}
I(r)=\frac{I_{0}}{2\pi\sigma^{2}}e^{\frac{-(r-r_{h})^{2}}{r_{0}^{2}}}
\end{equation}

\begin{equation}
\left\{
\begin{array}{ll}
I(r)=0 &   r \leq r_{h}\\
I(r)=I_{0} & r > r_{h}
\end{array}
\right.
\end{equation}

$I_{0}$ is the amplitude of the component and $r_{0}$ is its scale radius.
The presence of a nuclear hole in the ISBD, modeled with a non-zero $r_{h}$, could be caused by varying ionization conditions of the gas (e.g.\ close to the nucleus the gas is too ionized to emit the line of interest) or by absence of the gas itself. 
In the previous equations other important parameters are 'hidden': these are  $q$ and $\theta$ which are used to provide ISBDs with elliptical isophotes. $q$ is the ratio between the minor and major axis of the isophotes and $\theta$ is the PA of the major axis. In practice:

\begin{equation}
\label{r}
r = \left[X^{2}+Y^{2}\right]^{1/2}
\end{equation} 

where the coordinates $(X,Y)$ are defined by:

\begin{equation}
\left\{
\begin{array}{ll}
\label{x''}
X=(x-x_{0})\cos\theta  + (y-y_{0})\sin\theta \\
Y=\left(-(x-x_{0})\cos\theta + (y-y_{0})\sin\theta\right)/q
\end{array}
\right.
\end{equation}

$(x_0,y_0)$ represent the position of rotation center (the putative black hole) in a reference frame on the plane of the sky with $x$ and $y$ aligned along the North and East directions respectively.

Summarizing the free parameters for each ISBD component are:
\begin{itemize}
\item
the amplitude $I_{0}$;
\item
the scale radius $r_{0}$; 
\item
the hole radius $r_{h}$; 
\item
the axial ratio $q$;   	  
\item
the position angle $\theta$;
\item
the coordinates of the symmetry center $(x_{0},y_{0})$;    
\end{itemize}

\section{\label{ris.gen}Kinematical results}

In this section we first determine the intrinsic surface brightness distribution for each galaxy and then we fit the observed velocity.

\subsection{\label{ris.3310}Analysis of NGC 3310}

As explained in the Introduction, the galactic disk of NGC 3310 is characterized by non-circular motions with strong streaming along the spiral arms (e.g.~\citealt{kregel} and references therein). Fig.~\ref{riga} clearly shows that the spiral arms depart from a $\sim 6\arcsec$ ring, the starburst ring, and suggests that, beyond that radius, motions might be significantly non-circular. Within the ring (which might indicate the corotation radius) the morphology of the emission line region is characterized by a peaked nuclear blob, about $1.2\arcsec$ in size, which appears unrelated with the starburst ring. In order to avoid as much as possible obvious contamination by non circular motions we have decided to restrict our analysis to the nuclear blob, i.e.\ to the central 1.2\arcsec region. This region, which includes the high surface brightness peak shown in Fig.~\ref{andamenti_3310} is delimited by a dashed line.

\begin{figure*} 
\centering
\epsfig{file=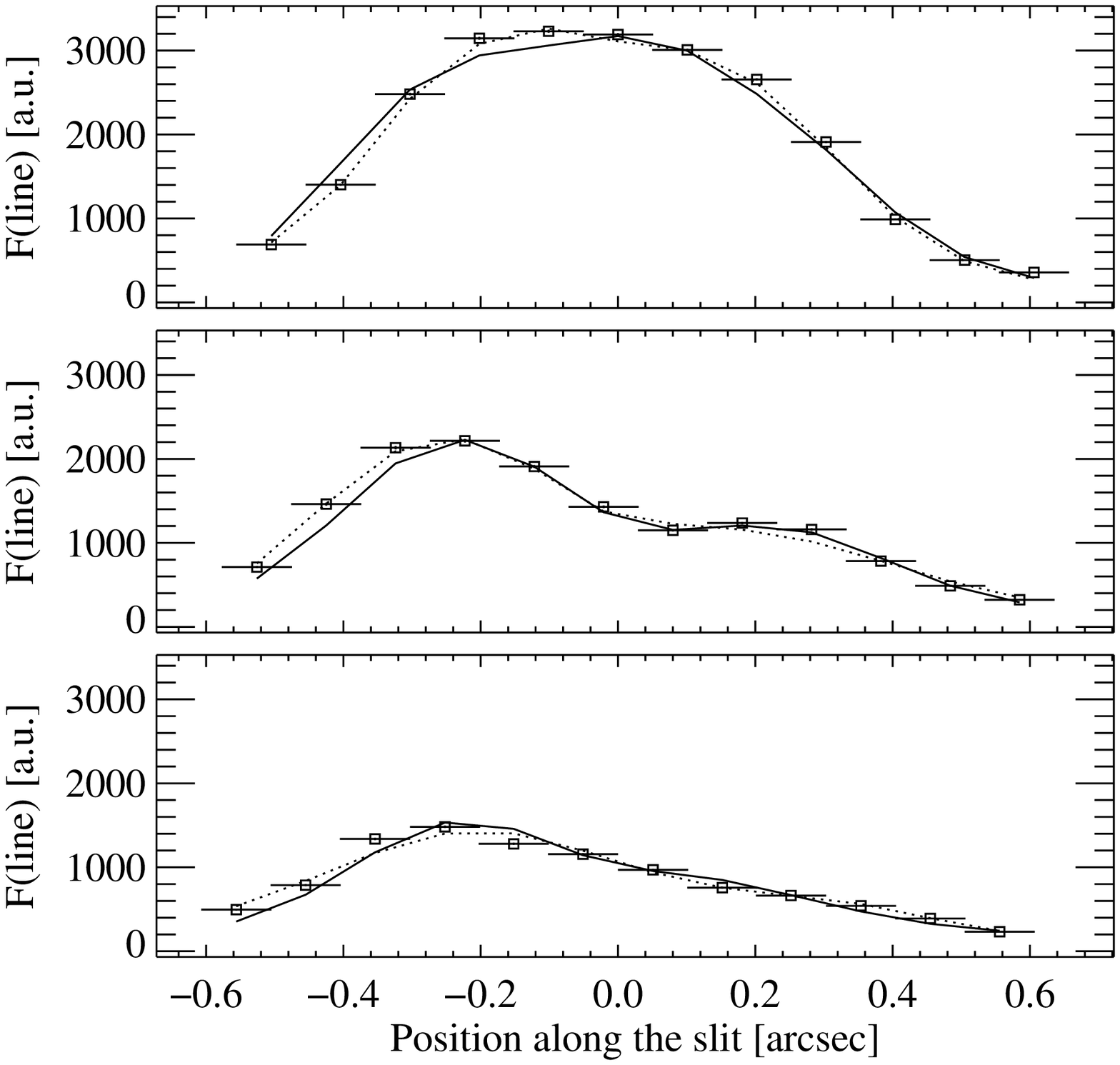, width=0.3\linewidth}
\epsfig{file=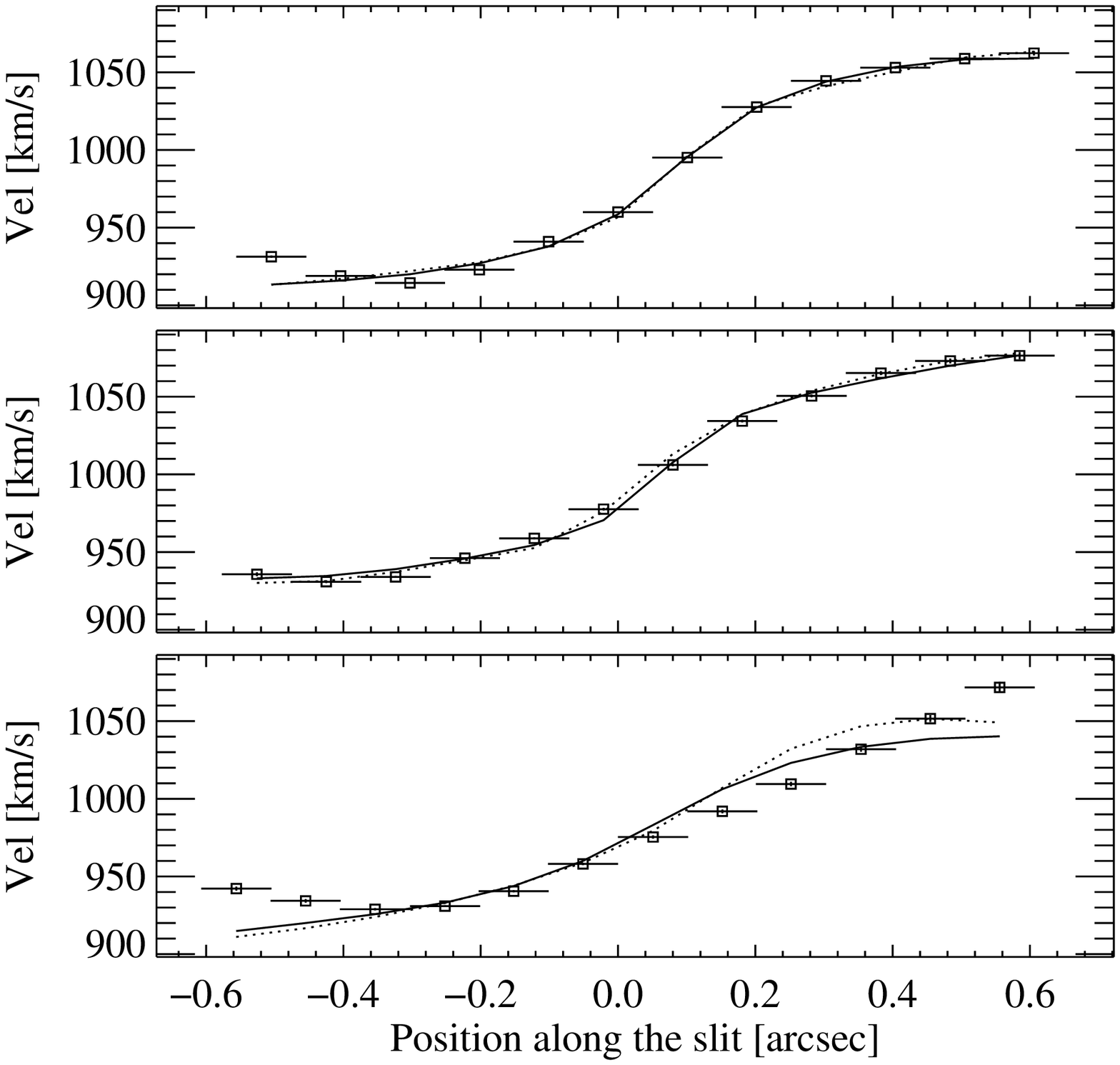, width=0.3\linewidth}
\epsfig{file=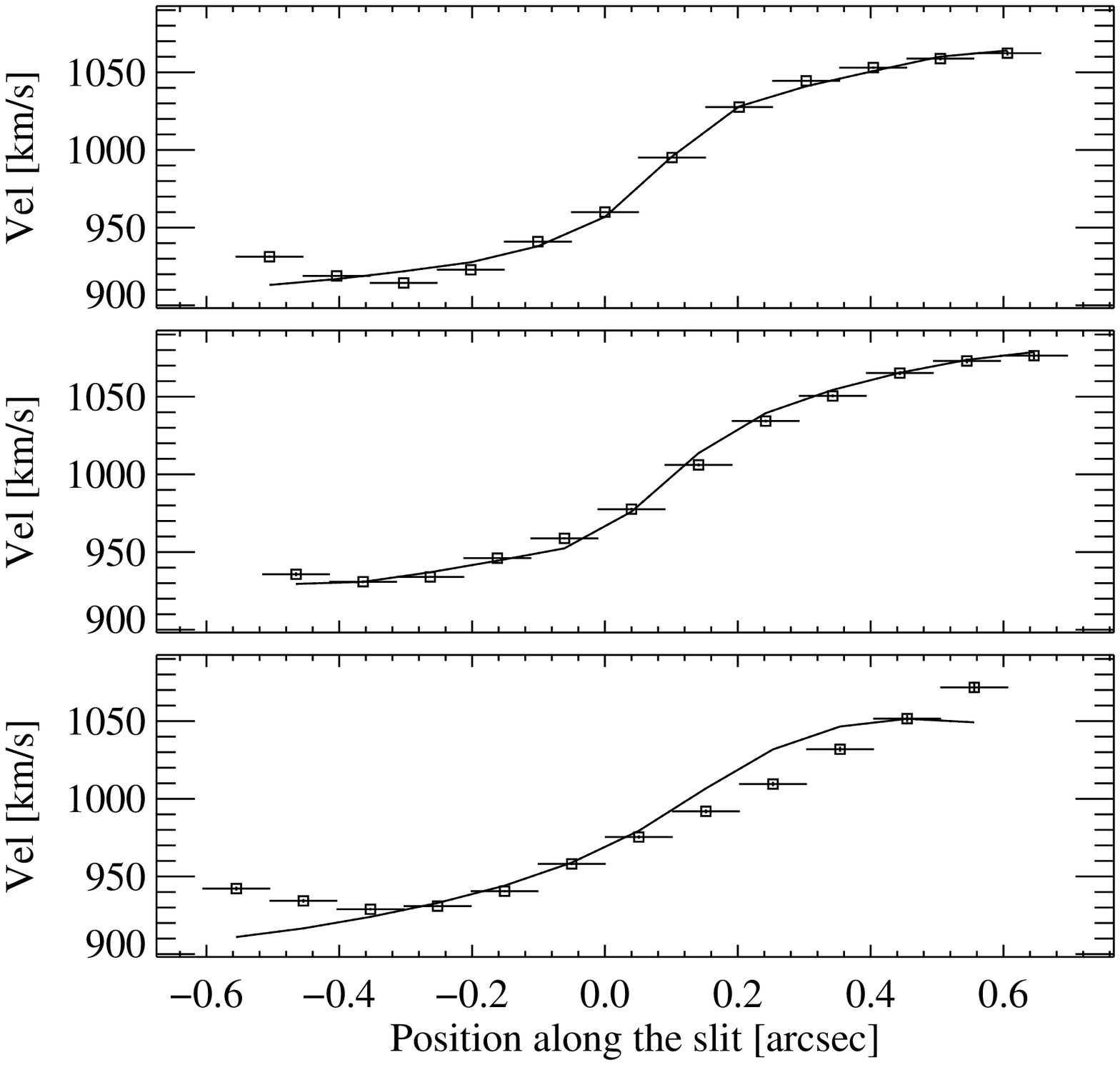, width=0.3\linewidth}
\caption{\label{fig.fluxmod3310}NGC 3310. \emph{Left panel}: best-fit models of the emission line surface brightness distribution compared with the observed data (dots with error bars) at the three different slit positions. The solid and dotted lines refer to ISBD A and B, respectively. From top to bottom the panels refer to NUC, OFF1 and OFF2 respectively. \emph{Central panel}: best-fit models and observed data for the rotation curve obtained with $A$ (solid line) and $B$ (dotted line) flux distributions, assuming $i=40$ deg.
\emph{Right panel}:\label{fig:60} best fit model of velocity obtained using ISBD $A$ with the 'best' inclination ($i=70$ deg) at the three different positions of the slit.}
\end{figure*}

To determine the best ISBD we made several attempts with different compositions of the analytical functions described above. The best choices are the following: the one called $A$ (composed of two Gaussian components, an exponential and a constant one) and the one called $B$ (composed of two exponential components, two Gaussian ones and a constant component). Their resulting best fitting parameters are shown in Tab.~\ref{table.flux3310} and, in the left panel of Fig.~\ref{fig.fluxmod3310}, we compare the model with the observed flux distribution.

\begin{table*}
\caption{NGC 3310: best fitting parameters of the two ISBDs adopted in the kinematical model fitting. 
}
\label{table.flux3310}
\centering
\begin{tabular}{cccccccccc}
\hline\hline
$Id$ &  $function$ & $i$ & $I_{0i}^{a}$ & $r_{0i}(\arcsec)$ & $r_{hi}(\arcsec)$ & $x_{0i}(\arcsec)$ & $y_{0i}(\arcsec)$ & $\theta_{i}(^{\circ})$ & $q_{i}$  \\
\hline
A  & gauss   &  1  & 3003  & 0.37 &  0.0   & -0.19   &  0.09 & -18.9 & 55.14   \\
   & gauss   &  2  & 2717  & 0.40 &  0.06  &  0.24   &  0.01 &  79.2 & 43.08    \\
   & expo    &  3  & 7029  & 0.09 &  0.02  & -0.04   & -0.02 &  4.6  & 71.27     \\
   & const   &  4  & 186   & --   & --     & --      &   --  &   --  &   --       \\
\cline{1-10}
B & gauss  &  1 & 19591      &  0.29  & 0.05 & -0.23 & -0.02 &  -0.5 & 88.12  \\
  & gauss  &  2 & 2631       &  0.68  & 0.04 & -0.02 &  0.00 & -31.2 & 72.94   \\
  & expo   &  3 & 8265       &  0.10  & 0.06 &  0.23 &  0.09 & -30.5 & 27.98    \\
  & expo   &  4 & 2796       &  0.76  & 0.18 & -3.43 &  1.72 & -12.4 & 87.70     \\
  & const  &  5 & 151 & --     & --   & --    & --    & --    & --         \\
\hline
\multicolumn{10}{l}{\small{$^{a}$ Units of erg s$^{-1}$ cm$^{-2}$ arcsec$^{-2}$ \AA$^{-1}$}}\\
\end{tabular}
\end{table*}

As explained in Sec.~\ref{modelli} we fitted the velocities with fixed disk inclination and, as starting point
for $i$, we adopted the inclination of the large scale galaxy disk ($i=40$ deg). With both ISBD A and B, it is possible to reproduce the observed rotation curves (right panels of Fig.~\ref{fig.fluxmod3310}).  Tab.~\ref{table.vel3310} shows the corresponding best fit parameters. The reduced $\chi^{2}$ of the fits are very large at variance with the apparently good agreement between data and models. The reason of this apparent contradiction is to be found in the small velocity errors, the formal ones of the line profile fitting. These errors are extremely small, of the order of  $1$ km/s, and corresponding to $1/50$ pixel. Although the $S/N$ ratio of our spectra is large, such small errors values are not realistic and will be corrected as described below.

\begin{table*}
\caption{NGC3310: best fitting parameters of the kinematical model obtained with different ISBDs and $i=40^{\circ}$.}
\label{table.vel3310}
\centering
\begin{tabular}{c c c c c c c c}
\hline\hline
$Flux^{a}$ &  $x_{0}(\arcsec)$ & $y_{0}(\arcsec)$ &  $\log M_{BH}^{b}$ & $\log M/L^{c}$ & $\theta(^{\circ})$ & $V_{sys}$(km/s) & $\chi^{2}_{red}$ \\
\hline
A  &   -0.04  &  0.10  &  4.58 & -0.29 & -19.8 &  990.37 & 101  \\
B  &   -0.08  &  0.07  &  4.74 & -0.26 &  -4.3 &  988.72 & 135  \\
\cline{1-8}
\hline
\multicolumn{8}{l}{\small{$^{a}$ Adopted ISBD.}}\\
\multicolumn{8}{l}{\small{$^{b}$ Units of $M_{\odot}$}}\\
\multicolumn{8}{l}{\small{$^{c}$ Units of $\log{M_{\odot}/L_{\odot,H}}$}}\\
\end{tabular}
\end{table*}

\begin{figure*} 
\centerline{\epsfig{file=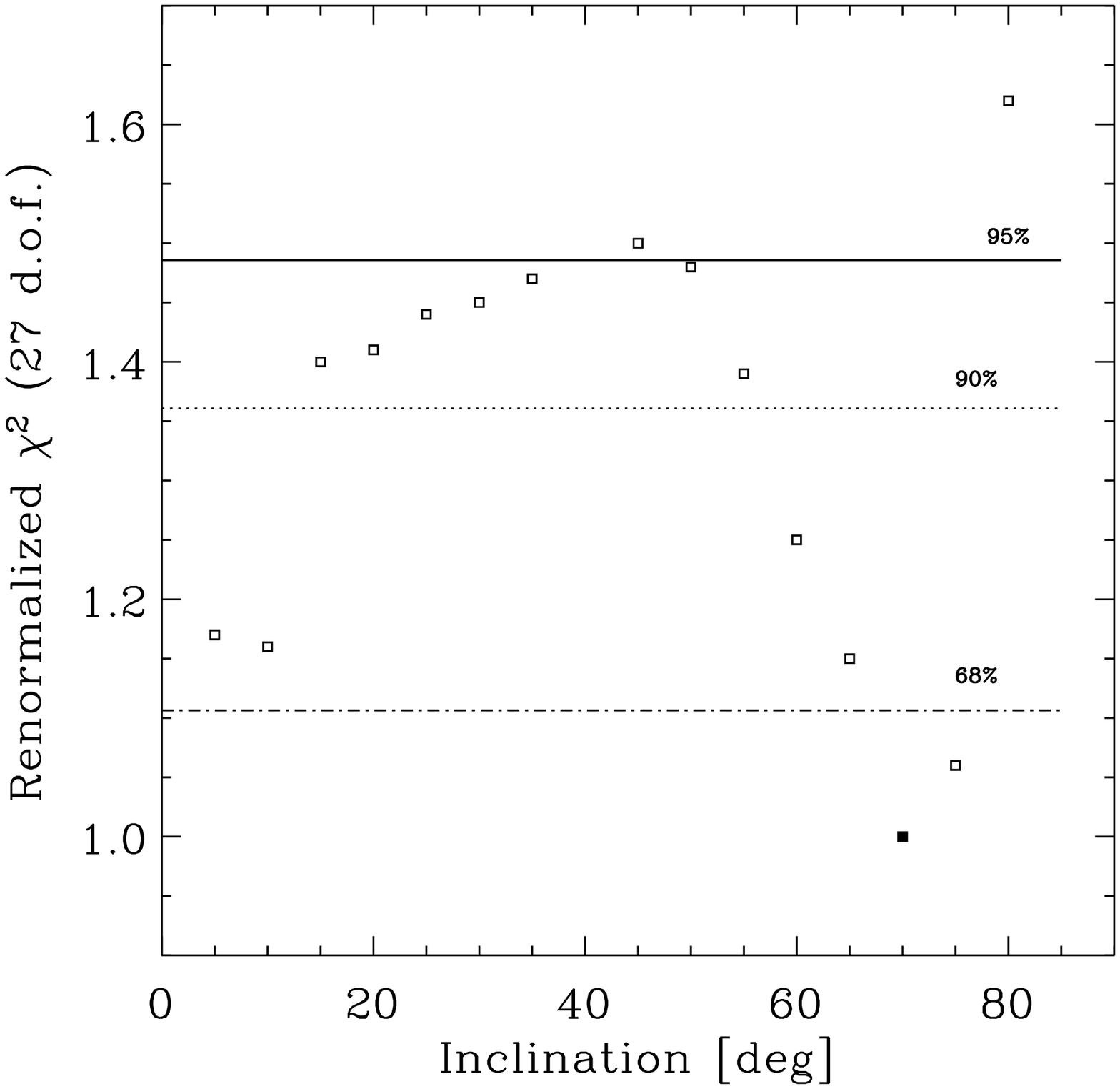, width=0.4\linewidth}\epsfig{file=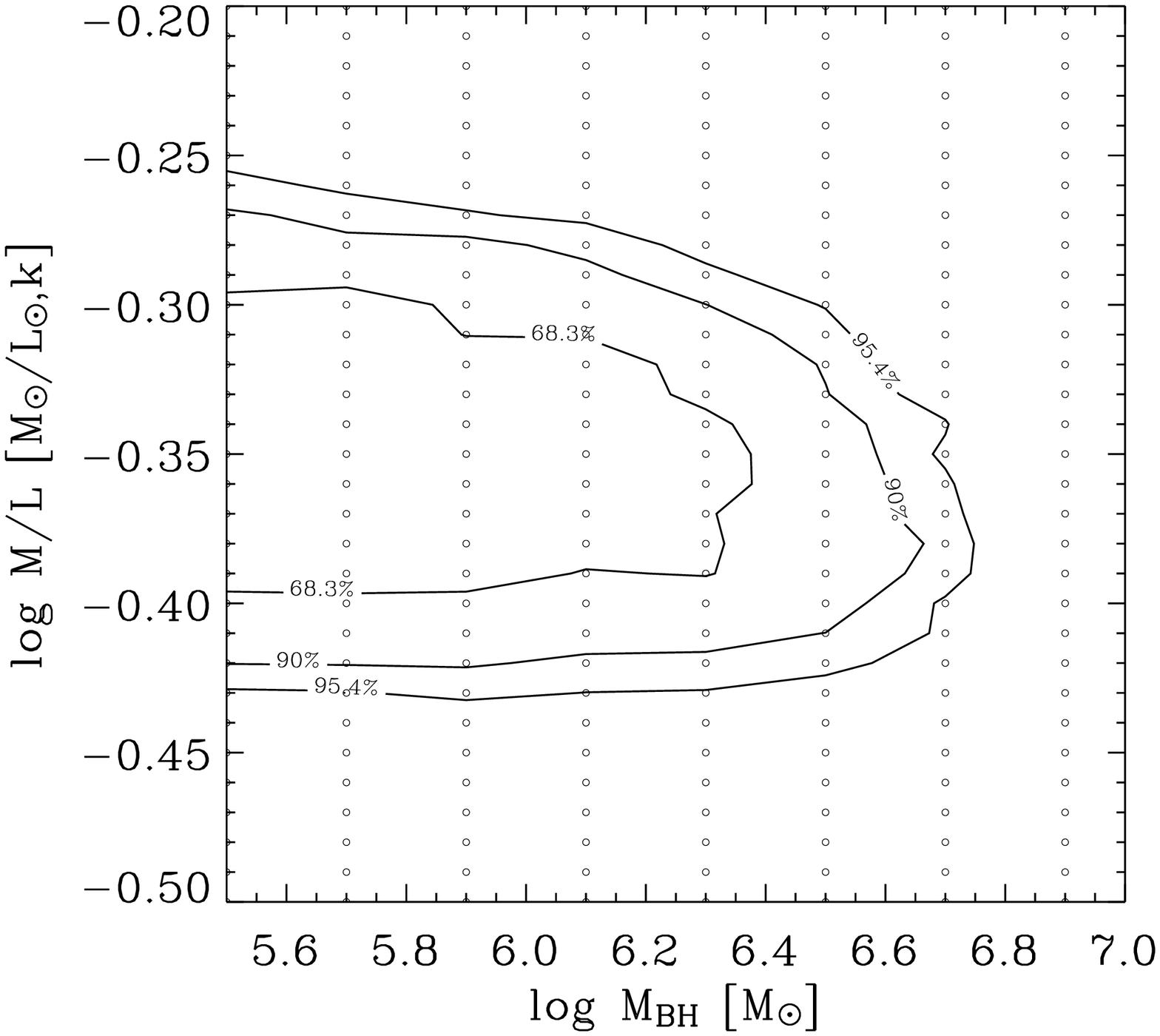, width=0.4\linewidth}}
\caption{\label{fig.incl_chi_3310} NGC 3310. \emph{Left}: dependence of $\chi^{2}$ on the adopted disk inclination. Horizontal lines indicate the 95 \%, 90\% and 68\%  confidence levels. The filled square represents the point corresponding to a $\chi^{2}_{rescaled}=1$ for $i=70$ deg. The $\chi^2$ value at $i=85$ deg is out of the plot range. \emph{Right}: $\chi^{2}$ contours for the joint variation of $M_{BH}$ and $M/L$ (with $i=70$ deg). Contour levels are for $\chi^{2}=\chi^{2}_{\emph{min}}+2.3,4.61,6.17$ corresponding to 68.3\%, 90\% and 95.4\% confidence levels. The dots indicate the $M_{BH}$ and $M/L$ values for which the $\chi^{2}$ minimization was actually computed. }
\end{figure*}

The above kinematical fit was computed with the inclination $i$ fixed at the value of the large scale disk inclination ($i=40$ deg \citealt{sanchez}). Recall that the choice of fixing the inclination is justified by its tight coupling with mass [$v(r)^2 \sim M(r) sin^{2}(i)$, see Sec.~\ref{modelli}].  We now verify the dependence of our model against the choice of this parameter. In particular we wish to check whether the quality of the fit might improve at different inclination.
We thus fit the rotation curves with different fixed values of the inclination, from $i=5$ deg to $i=85$ deg, using the $A$ flux distribution. Tab.~\ref{table.3310_inc} shows the best fit parameters obtained with the different inclinations. In order to establish the statistical significance of $\chi^2$ variations among different models we renormalized $\chi^2$ by adding in quadrature an extra velocity error to the data, $\Delta\nu_{0}$. The value of $\Delta\nu_{0}$ is found by imposing that the model with the lowest $\chi^2$ value has $\chi^2/dof=1$, where $dof$ is the number of degrees of freedom (see also \citealt{marconi04}).
With \emph{dof}=30, the best fit models which are acceptable with a significance of 95\% 
have $\chi^{2}_{rescaled}\le 1.48$. The dependence of $\chi^{2}$ on the adopted disk inclination is reported in the left panel of Fig.~\ref{fig.incl_chi_3310}: the filled square represents the inclination value $i=70$ deg corresponding to the lowest $\chi^{2}$ value.

\begin{table*}
\caption{\label{table.3310_inc} NGC3310: effect of \emph{i} variation on the best fit parameters and $\chi_{red}^{2}$. $\chi^2$ values have been rescaled by an extra error on velocity $\Delta\nu_{0}$. All models adopt ISBD $A$.}
\centering
\begin{tabular}{cccccc}
\hline\hline
 $i(^{\circ})$ & $\log M_{BH}^{a}$  &  $\log M/L^{b}$ & $V_{\emph{sys}}$ & $\theta(^{\circ})$ & $(\chi^{2}_{rescaled})^{c} $  \\
\hline\hline
\multicolumn{6}{c}{Fit of velocity $(\Delta\nu_{0}=8.59 km s^{-1})^{d}$}\\
5  & -6.18 & 1.85  & 1023.18 & -62.58 & 1.17   \\
10 &  1.46 & 1.26  & 1024.01 & -62.62 & 1.16  \\
15 &  3.77 & 0.51  &  992.36 & -28.51 & 1.40  \\
20 &  3.56 & 0.26  &  991.82 & -26.87 & 1.41  \\
25 &  4.13 & 0.08  &  991.85 & -26.79 & 1.44  \\
30 &  4.46 & -0.06 &  991.86 & -26.33 & 1.45  \\
35 &  4.60 & -0.16 &  992.74 & -28.92 & 1.47  \\
45 &  4.89 & -0.39 &  988.71 & -13.01 & 1.50  \\
50 &  5.10 & -0.44 &  986.44 & -4.33  & 1.48  \\
55 &  4.85 & -0.46 &  985.06 & 1.08   & 1.39  \\
60 &  4.62 & -0.45 &  983.31 & 5.87   & 1.25  \\
65 &  4.50 & -0.45 &  982.19 & 6.99   & 1.15  \\
70 &  4.23 & -0.41 &  981.66 & 8.56   & 1.00  \\
75 &  5.20 & -0.35 &  980.36 & 8.04   & 1.06  \\
80 &  5.51 & -0.23 &  977.99 & 6.36   & 1.62  \\
85 &  5.61 &  0.03 &  975.09 & 3.37   & 3.97  \\
\hline
\multicolumn{6}{l}{\small{$^{a}$ Units of $M_{\odot}$.}}\\
\multicolumn{6}{l}{\small{$^{b}$ Units of $M_{\odot}/L_{\odot,H}$.}}\\
\multicolumn{6}{l}{\small{$^{c}$ $\chi^{2}$ rescaled with errors computed as $\Delta\nu_{i}^{'2}$=$\Delta\nu_{i}^{2}+\Delta\nu_{0}^{2}$.}}\\
\multicolumn{6}{l}{\small{$^{d}$ Systematic error on velocity adopted to renormalize $\chi^{2}$}}\\
\multicolumn{6}{l}{\small{~~($\ll 49$ km/s, velocity corresponding to the pixel size).}}\\
\end{tabular}
\end{table*}

With the $\chi^2$ renormalization described above, all inclinations with $i<80$ deg are acceptable at the 95\%\ level and, apparently, our analysis produces an upper bound to the inclination.  However, the mass-to-light ratio increases sharply with decreasing inclination and, indeed, also $M/L$ scales with $(\sin i)^{-2}$ as $M_{BH}$.
We then considered the evolutionary synthesis models for stellar population derived by \cite{vazquez} in order to derive a lower bound on $i$ based on astrophysically acceptable $M/L$ values. The typical stellar populations have $H-K\sim 0.25$ and the value of $M/L$ monotonically increases with galaxy age reaching $\log M/L_K< 0.6$ i.e.~$\log M/L_H< 0.25$ \citep{vazquez}. Note that, for the sake of simplicity, we assume that the F160W band can be approximated with the standard $H$ band.
This physical upper limit on $M/L$ can be translated into a lower limit on $i$ which corresponds to an inclination of $\sim 20$ deg.
To summarize, the constraints on inclination are that $20$ deg $ < i < 80 $ deg.
The model with the lowest $\chi^2$ has $i=70$ deg and the comparison between observed and model rotation curves is shown in Fig.~\ref{fig:60}.
The inclination value of $i\sim40$ deg, found by \citet{sanchez}, refers to the large scale disk while the inner nuclear disk (which is clearly separated from the large scale structure, Fig.~\ref{riga}), is seen at a higher inclination.  
The best fit black hole mass for the $70$ deg-model is $\log M_{BH}[M_{\odot}]=4.2$ and, in general, all inclinations require similarly small black hole masses ($\log M_{BH}[M_{\odot}]<5.6$). Such values are unrealistically small and, indeed, we are only able to set an upper limit to $M_{BH}$. 

In order to estimate statistical errors associated with the best fit parameters, we present in the right panel of Fig.~\ref{fig.incl_chi_3310} the contours for the joint variation of $M_{BH}$ and $M/L$. The grid has been computed with ISBD $A$ and $i=70$ deg, fixing the 2 parameters of interest ($M_{BH}$ and $M/L$) and varying the others to minimize $\chi^2$. Before computing confidence levels we have renormalized $\chi^2$ as described above so that the model with the lowest $\chi^2$ value has $\chi^2$ reduced equal to 1. Confidence levels for 2 parameters of interest are then found following \cite{avni}: $\chi^2 = \chi^2_{min} +$ 2.3, 4.61, 6.17 for confidence levels 68.3\%, 90\% and 95.4\%, respectively. 
Fig.~\ref{fig.incl_chi_3310} shows that it is not possible to measure the BH mass but only place an upper limit to it.
By choosing a confidence level of $68.3\%$, the stellar mass to light ratio for this galaxy results $M/L=0.47^{+0.04}_{-0.07}M_{\odot}/L_{\odot,H}$ which is the typical H-band value for spiral bulges ($0.6\pm 0.2$ in K, which correspond to $0.3\pm0.1$ for $H-K=0.2$; \citealt{moriondo}).
For the black hole mass upper limit we chose a confidence level of $95\%$ and the corresponding estimate results $M_{BH} < 5.62 \times 10^{6} M_{\odot}$.

We then verified that at different inclinations the statistical errors on $\log M/L$ were roughly the same and that the upper limits on $M_{BH}$ were scaling with $(\sin i)^{-2}$ as expected. By taking into account the allowed $i$ variation ($20$ deg $ < i < 80 $ deg) the upper limit on the black hole mass varies in the range $5.0\times 10^6$ -- $4.2\times 10^7 M_{\odot}$.

\begin{figure*}
\centering
\includegraphics[width=0.5\linewidth]{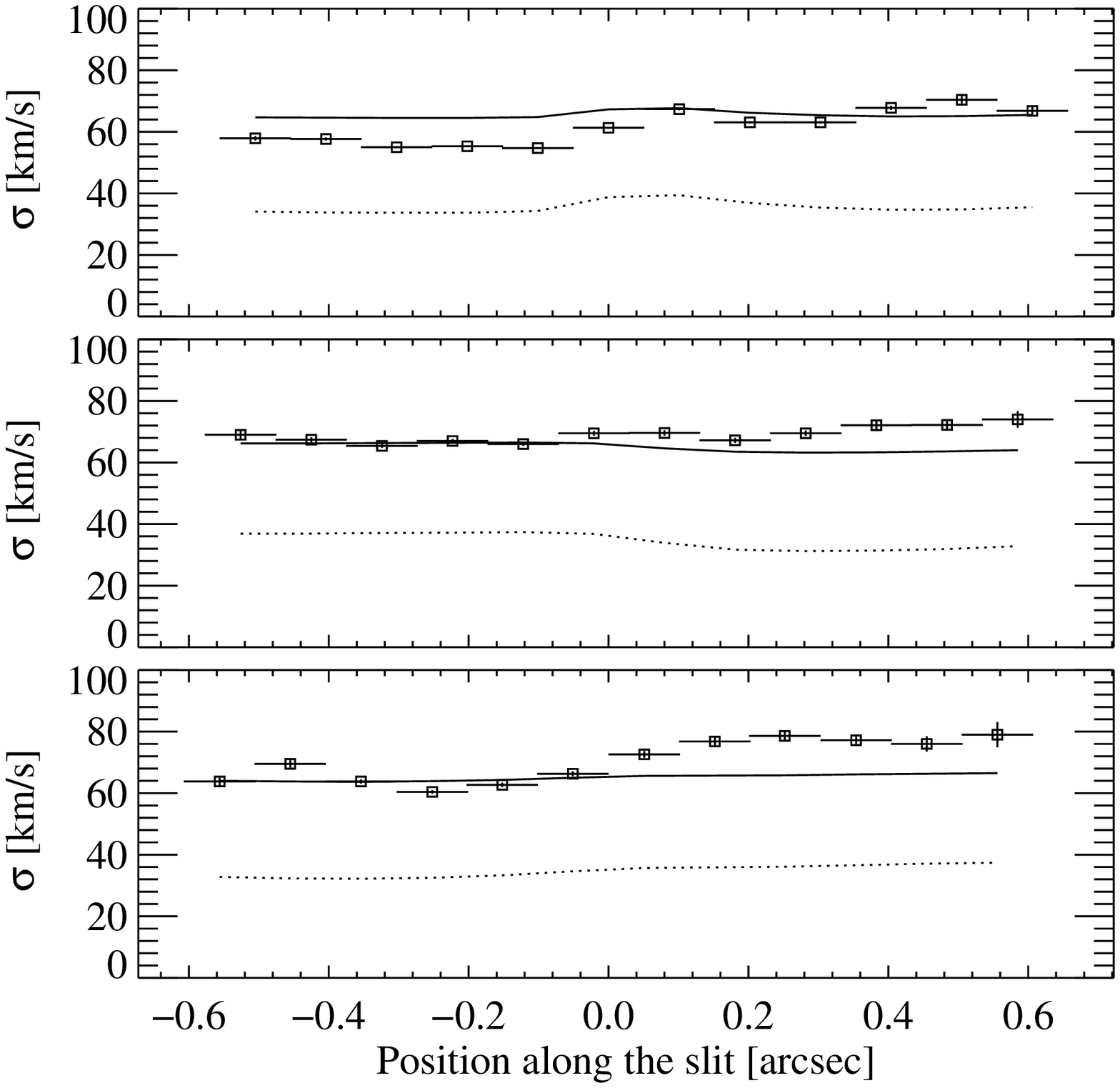}
\caption{\label{3310sigma}NGC 3310. Expected velocity dispersions for the best fitting rotating disk model with $i=70$ deg compared with observations (dots with error bars) assuming a constant intrinsic dispersion of $55$ km/s (solid line) compared with the dispersion expected from pure rotation (dotted line).}
\end{figure*}

In Fig.~\ref{3310sigma} we plot the expected velocity dispersion for the best fitting model at $i=70$ deg which includes intrumental broadening and
the contribution of unresolved rotation (dotted line). In order to match observations we added an intrinsic constant dispersion of $55$ km/s (solid line).
Overall data points do not show a deviation of more than $20$ km/s from the model values, after the intrinsic dispersion was taken into account. 

\subsection{\label{ris.4303}Analysis of NGC 4303}

The kinematics of NGC 4303 can be traced up to $4\arcsec$ from the galaxy nucleus identified by the location of the near-IR point source \citep{colina02}. Although at $1-2\arcsec$ spatial resolution the motion within the central region ($\sim 10\arcsec$ of the nucleus) appears ordered and consistent with circular rotation \citep{schinner02,arribas} the presence of the inner bar with a size comparable with the spatial resolution ($\sim 2\arcsec$ size, \citealt{erwin}) suggests caution in the kinematical analysis.
Indeed, the curve of the extended component shows irregular behavior at about 3\arcsec\ from the nucleus (pix $\sim 320-340$ in Fig.~\ref{andamenti_4303}). Moreover, the H$\alpha$+[NII] surface brightness distribution along the slit clearly indicates the presence of a well defined nuclear component as in the case of the other two galaxies in this paper and NGC 4041 \citep{4041}. This then suggests to limit our analysis to the nuclear component, i.e.\ to the central $\sim 1\arcsec$ from the nucleus.

\begin{figure*} 
\centering
\epsfig{file=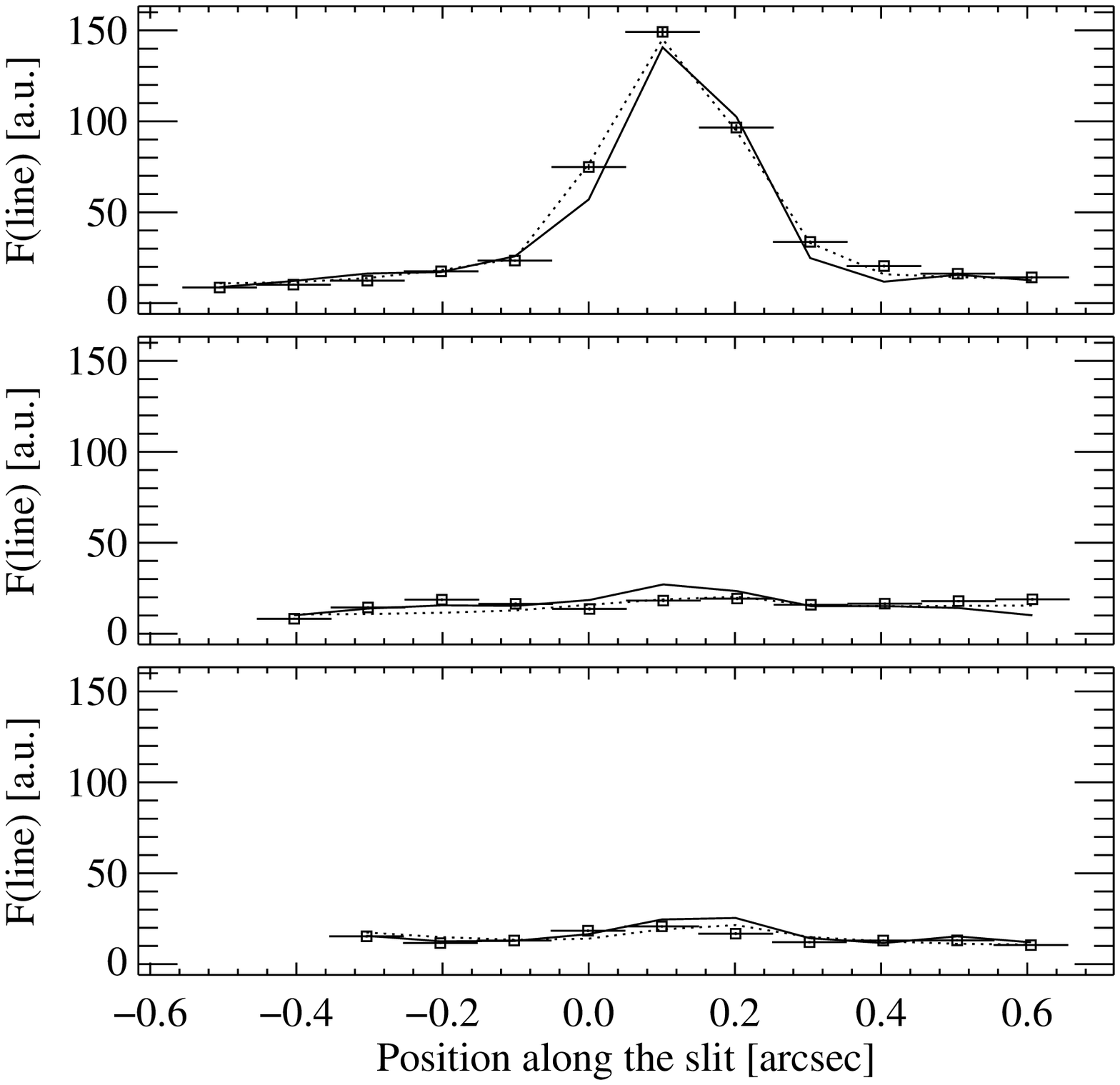, width=0.3\linewidth}
\epsfig{file=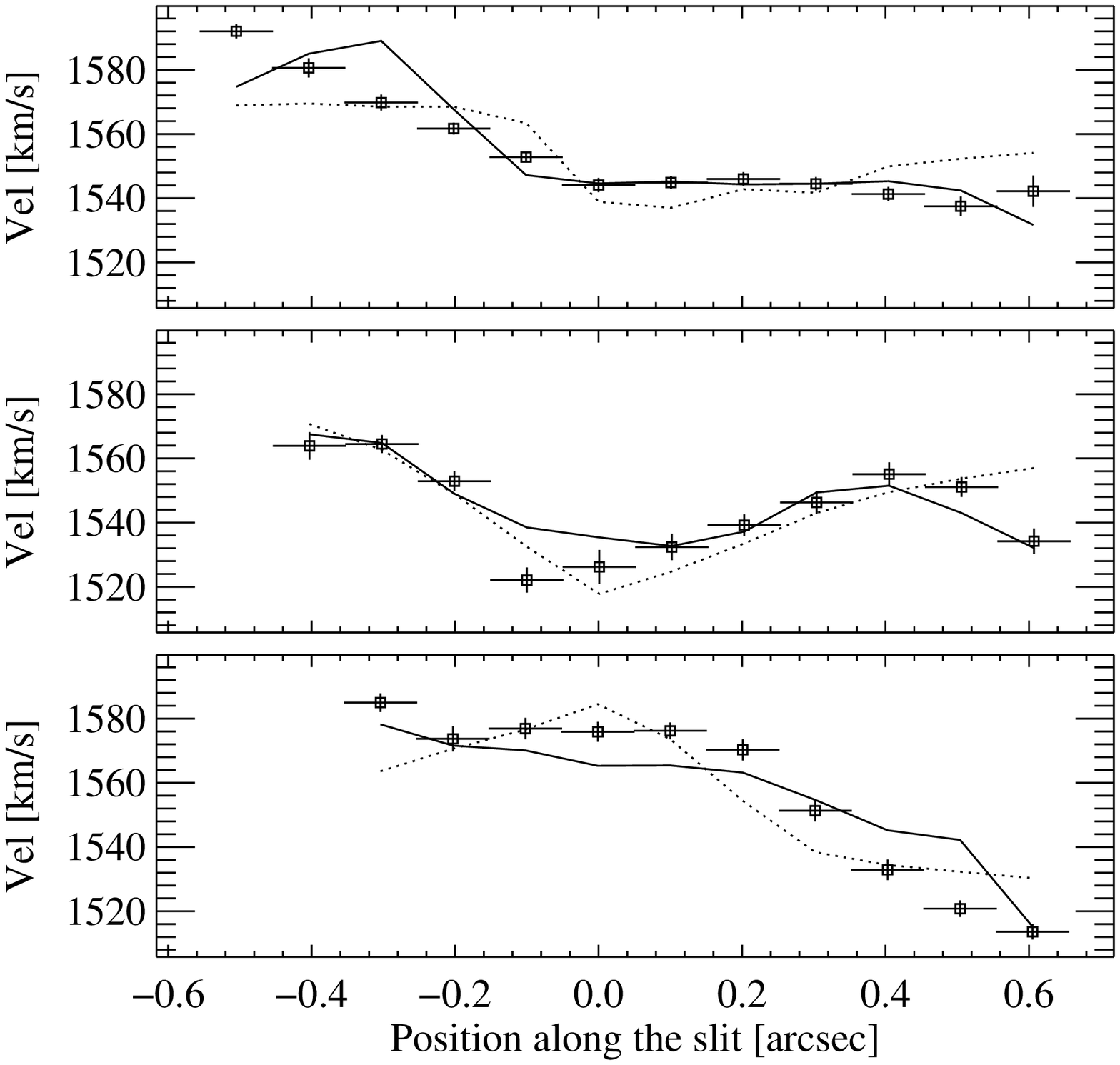, width=0.3\linewidth}
\epsfig{file=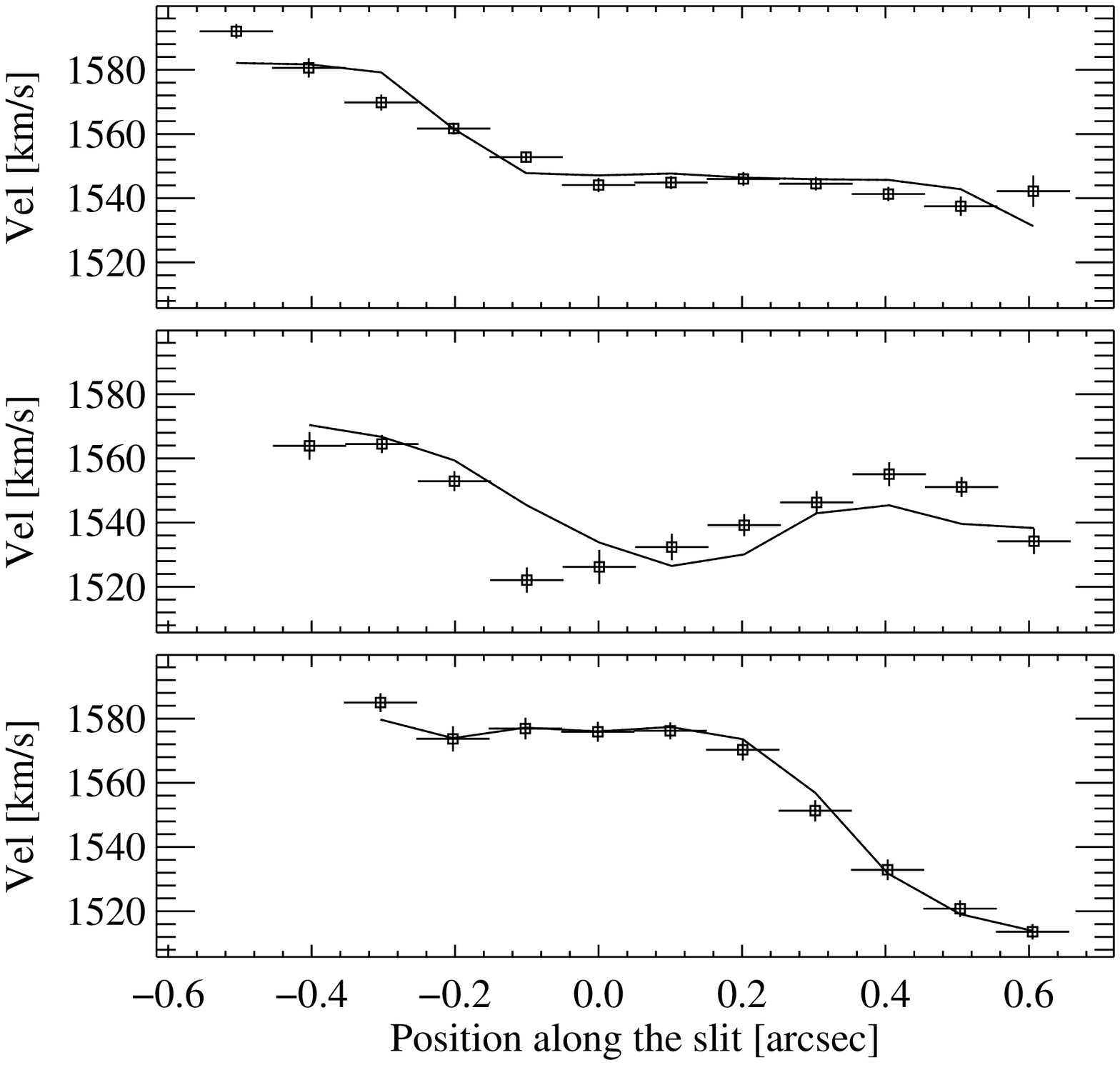, width=0.3\linewidth} 
\caption{\label{fig.4303_fluxmod}\label{fig.4303_solovel}NGC 4303.\emph{Left panel}: best-fit models of the emission line surface brightness distribution compared with the observed data (dots with error bars) at the three different slit positions. The solid and dotted lines refer to ISBD D and E, respectively. From top to bottom the panels refer to NUC, OFF1 and OFF2 respectively.\emph{Center panel}: best-fit models (solid and dotted lines) and observed data (dots with errorbars) for the rotation curves at the three slit positions. The solid and dotted lines denote models obtained with the $D$ and $E$ ISBD, respectively. The models were computed assuming $i=20$ deg. \emph{Right panel}:  as in the left panel, but the solid line is the model computed for the best inclination $i=70$ deg and ISBD $D$.}
\end{figure*}

We follow the same procedure adopted for NGC 3310. We considered different ISBDs but we found that only two of them are able to provide a reasonable fit of the observed data: the one called $D$, composed of three exponential components and a constant one, and the one called $E$, composed of three Gaussian components and a constant one.
Fig.~\ref{fig.4303_fluxmod} shows the comparison between observed and model surface brightness distributions while the best fit ISBD parameters are placed in Tab.~\ref{table.flux}.

\begin{table*}
\caption{NGC 4303: best fitting parameters of the two ISBDs adopted in the kinematical model fitting.}
\label{table.flux}
\centering
\begin{tabular}{c c c c c c c c c c}
\hline\hline
$Id$ &  $function$ & $i$ & $I_{0i}^{a}$ & $r_{0i}(\arcsec)$ & $r_{hi}(\arcsec)$ & $x_{0i}(\arcsec)$ & $y_{0i}(\arcsec)$ & $\theta_{i}(^{\circ})$ & $q_{i}$  \\
\hline
D &  expo  &  1 & 850 & 0.10  &  0.38 & -0.12 & 0.04 & 0.0  & 0.04  \\
  &  expo  &  2 & 857 & 0.03  & 0.07 & 0.05  & -0.05 &-0.04 & 0.04 \\
  &  expo  &  3 & 801 & 0.04  & 0.00 & -0.1 &  0.01 & 0.01 & 0.03  \\
  &  const &  4 & 11 & --  & -- & -- & -- & -- & --  \\
\cline{1-10}
E & gauss  &  1 & 78  & 0.02  &  2.21 & -1.97 & -0.98 & -0.18 & 0.25  \\
  & gauss  &  2 & 728 & 0.03  & 0.02 & -0.12 & 0.17 & 1.89 & 0.04   \\
  & gauss  &  3 & 716 & 0.04  & 0.02 & -0.06 & 0.05 & 1.58 & 0.12  \\
  & const  &  4 & 6  & --  & -- & -- & -- & -- & --  \\
\cline{1-10}
\hline
\multicolumn{10}{l}{\small{$^{a}$ Expressed in erg s$^{-1}$ cm$^{-2}$ arcsec$^{-2}$ $\AA^{-1}$}}\\
\end{tabular}
\end{table*}

As for the case of NGC 3310, we started the velocity fitting with a fixed inclination value, $i=20$ deg corresponding to the large scale value one, and we used both ISBDs described above. The nuclear rotation curves of NGC 4303 have a less regular appearance than NGC 3310 and the agreement with the models might thus be worse. A blind $\chi^2$ minimization can result in non-physical parameters values. In particular the position of the kinematical center of rotation can move away by $>0.5\arcsec$ from the continuum peak which clearly identifies the galaxy nucleus (e.g.\ \citealt{schinner02,colina02}). To avoid this problem we have constrained the $x,y$ coordinates of the kinematical center to be located within $\pm 0.2\arcsec$ of the continuum peak position. 
The best fits of the rotation curves with ISBDs $D$ and $E$ are shown in the left panel of Fig.~\ref{fig.4303_solovel} and the best fitting kinematical parameters are shown in Tab.~\ref{table.vel}.

\begin{table*}
\caption{NGC 4303: best fitting parameters of the kinematical model obtained with different ISBDs and $i=20^{\circ}$.}
\label{table.vel}
\centering
\begin{tabular}{c c c c c c c c}
\hline\hline
$Flux^{a}$ &  $x_{0}(\arcsec)$ & $y_{0}(\arcsec)$ &  $\log M_{BH}^{b}$ & $\log M/L^{c}$ & $\theta(^{\circ})$ & $V_{sys}$(km/s) & $\chi^{2}_{red}$ \\
\hline
D  & 0.2  &  -0.09    & 7.53 &  2.14 & 95.41 &  1560.64 & 13.28  \\
E  & -0.01  &  0.00   & 8.29 & -0.89 & 57.12 &  1556.06 & 20.24   \\
\cline{1-8}
\hline
\multicolumn{8}{l}{\small{$^{a}$ Adopted flux distribution.}}\\
\multicolumn{8}{l}{\small{$^{b}$ Units of $M_{\odot}$.}}\\
\multicolumn{8}{l}{\small{$^{c}$ Units of ${M_{\odot}/L_{\odot,H}}$.}}\\
\end{tabular}
\end{table*}

ISBD $D$ provides a much better fit of the observed velocities and the disagreement in the best fit parameter values among the two models is readily explained by the difference in reduced $\chi^2$ values. Model $D$ has two unsettling characteristics: an unphysically high $M/L$ (which might be due to the low disk inclination) and a PA of the line of nodes which is very close to $90$ deg, a geometry in which only a small fraction of the rotational velocity is projected along the line of sight.

\begin{table*}
\caption{NGC 4303: effect of \emph{i} variation on the best fit parameters and $\chi_{red}^{2}$. $\chi^2$ values have been rescaled by an extra error on velocity $\Delta\nu_{0}$. All models adopt ISBD $D$.}
\label{table.inclinazione}
\centering
\begin{tabular}{c c c c c c}
\hline\hline
$i(^{\circ})$ &  $\log M_{BH}^{a}$ & $\log M/L^{b}$ & $V_{sys}$ (km/s) & $\theta(^{\circ})$ & $(\chi^{2}_{rescaled})^{c}$ \\
\hline
\multicolumn{6}{c}{Fit of velocity $(\Delta\nu_{0}=7.14 Km s^{-1})^{d}$ }\\
5  & 8.86  & 3.90    & 1561.65 &86.35   &   2.08   \\
10 & 8.12  & 3.05    & 1562.14 &96.65   &   1.89  \\
15 & 7.80  & 2.55    & 1561.87 &94.63   &   1.80  \\
25 & 7.36  & 1.54    & 1559.96 &93.41   &   1.60  \\
30 & 7.21  & 0.96    & 1558.90 &94.31   &   1.55  \\
35 & 7.15  & 0.24    & 1558.31 &90.53   &   1.53 \\
40 & 7.04  & -0.15   & 1558.78 &92.93   &   1.41  \\
45 & 6.91  & -0.42   & 1559.90 &96.99   &   1.26 \\
50 & 6.79  & -0.64   & 1560.92 &106.21  &   1.07 \\
55 & 6.73  & -0.77   & 1561.94 &113.80  &   1.04  \\
60 & 6.68  & -0.90   & 1562.90 &119.19  &   1.04 \\
65 & 6.66  & -0.94   & 1564.42 &125.02  &   1.03 \\
70 & 6.69  & -0.96   & 1565.98 &129.78  &   1.00 \\
75 & 6.68  & -0.99   & 1566.69 &131.53  &   1.19  \\
80 & 6.69  & -0.99   & 1566.79 &131.66  &   1.19  \\
85 & 6.18  & -0.87   & 1570.87 &157.37  &   1.33  \\
\hline                                        
\multicolumn{6}{l}{\small{$^{a}$ Units of $M_{\odot}$.}}\\
\multicolumn{6}{l}{\small{$^{b}$ Units of $M_{\odot}/L_{\odot,H}$.}}\\
\multicolumn{6}{l}{\small{$^{c}$ Rescaled $\chi^{2}$ with errors computed as $\Delta\nu_{i}^{'2}$=$\Delta\nu_{i}^{2}+\Delta\nu_{0}^{2}$.}}\\ 
\multicolumn{6}{l}{\small{$^{d}$ Systematic error adopted to renormalize $\chi^{2}$}}\\
\multicolumn{6}{l}{\small{~~($\ll 49$ km/s, velocity corresponding to the pixel size).}}\\
\end{tabular}
\end{table*}

\begin{figure*} 
\centerline{\epsfig{file=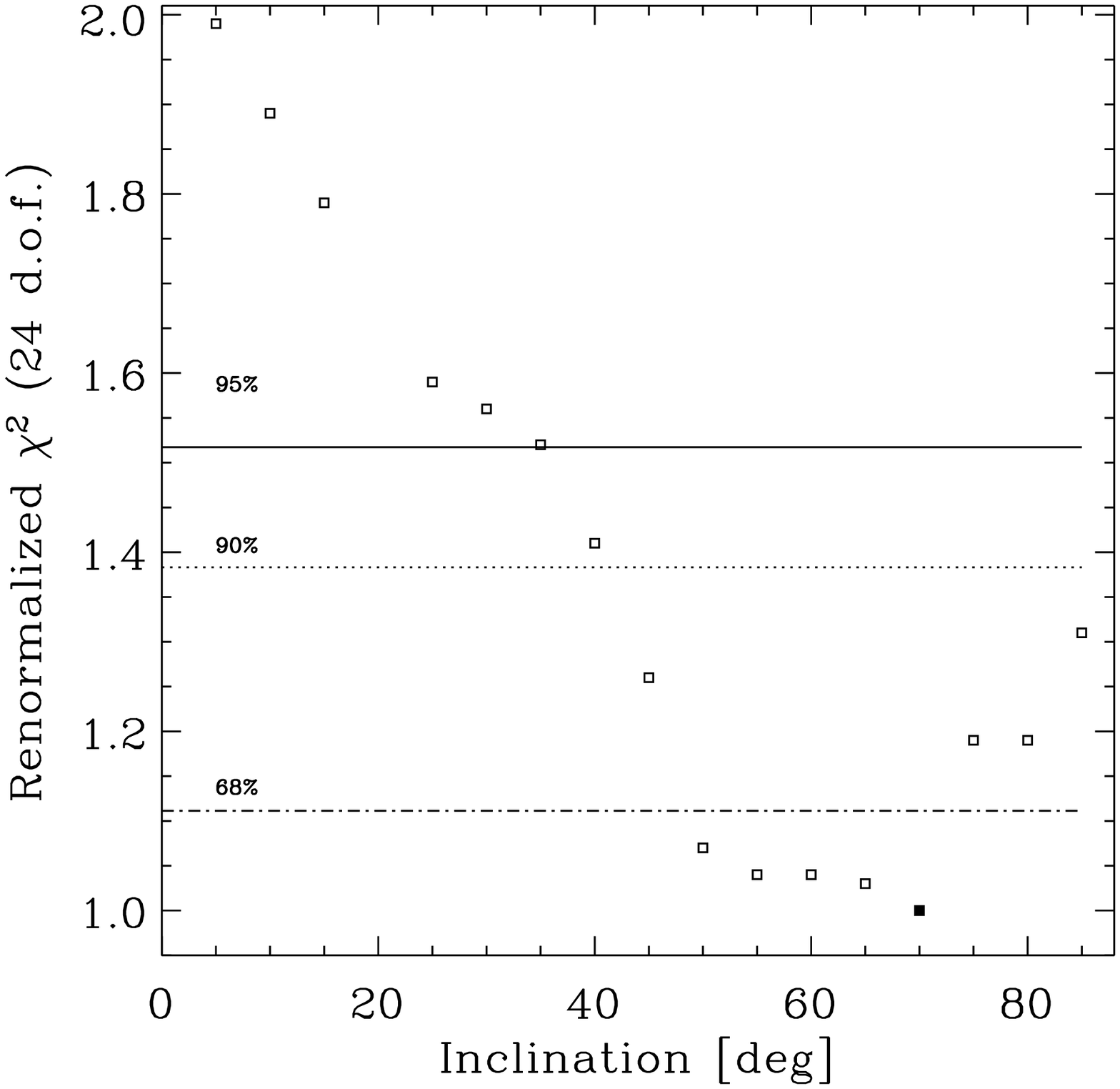, width=0.4\linewidth}\epsfig{file=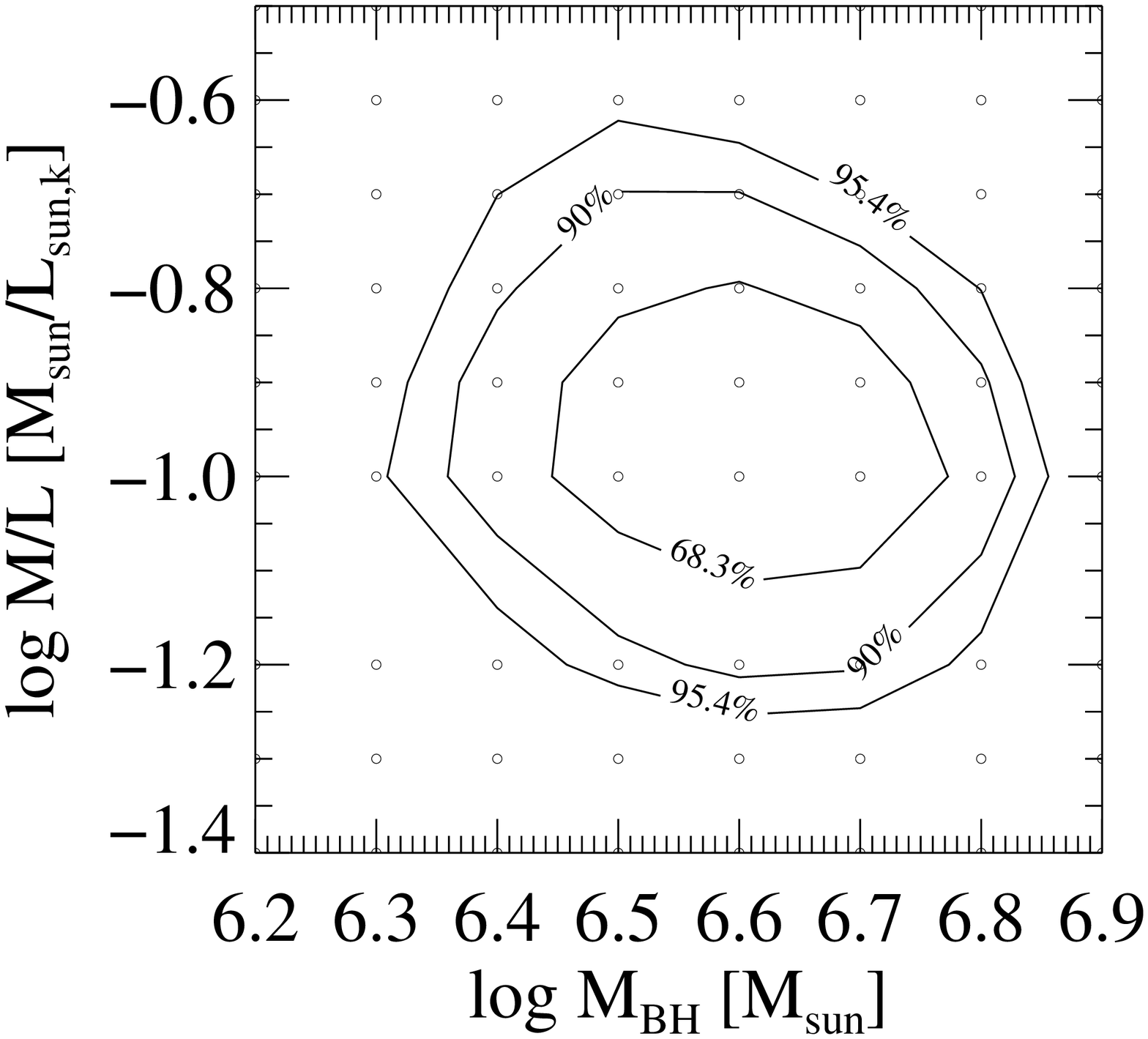, width=0.4\linewidth}}
\caption{\label{fig.incl_chi_4303} NGC 4303. \emph{Left}: dependence of $\chi^{2}$ on the adopted disk inclination. Horizontal lines indicate the 95 \%, 90\% and 68\%  confidence levels. The filled square represents the point corresponding to a $\chi^{2}_{rescaled}=1$ for $i=70$ deg. \emph{Right}: $\chi^{2}$ contours for the joint variation of $M_{BH}$ and $M/L$ (with $i=70$ deg). Contour levels are for $\chi^{2}=\chi^{2}_{\emph{min}}+2.3,4.61,6.17$ corresponding to 68.3\%, 90\% and 95.4\% confidence levels. The dots indicate the $M_{BH}$ and $M/L$ values for which the $\chi^{2}$ minimization was actually computed. }
\end{figure*}

Again, we have investigated whether different inclinations can provide a better fit to the observed rotation curves. Tab.~\ref{table.inclinazione} shows the resulting best fit values obtained for different inclinations
from $i=5$ deg to $i=85$ deg, using the $D$ ISBD. As previously, $\chi^2$ have been rescaled so that the model with the lowest $\chi^2$ has a value of 1 for the reduced $\chi^2$.  The $\chi^{2}$ dependence on the adopted disk inclination is shown in the left panel of Fig.~\ref{fig.incl_chi_4303}, where the model with the lowest $\chi^2$ is marked with a filled square.  In the case of NGC 4303, 24 degrees of freedom imply that the 95\%\ confidence level is found for $\chi^{2}_{rescaled} = 1.52$. Above that value all models are to be discarded with a 95\%\ significance implying $i > 40$ deg.
The model with the lowest $\chi^2$ has $i=70$ deg, $M_{BH}=4.90 \times 10^{6}M_{\odot}$ and is plotted in the right panel of Fig.~\ref{fig.4303_solovel}.
The first modeling attempt, made at $i=20$ deg, should be discarded and the non-physical value of $M/L$ is not problematic. Moreover, at high inclination values the PA of the line of nodes is significantly different from $90$ deg and is similar to the one found by \citet{arribas} and \citet{koda} for the arcsec scale gas rotation ($130-180=-50$ vs $130$ and $-44.5$, respectively). As in the case of NGC 3310, the inclination of the nuclear disk is smaller than that of the large scale galactic disk ($i=45$ deg).

Apart from the consistent kinematical parameters found, the agreement between model and observed rotation curve is good except for OFF1 where discrepancies might indicate the presence of non-circular motions.

In order to estimate statistical errors associated with the best fit parameters, we present in the right panel of Fig.~\ref{fig.incl_chi_4303} the contours for the joint variation of $M_{BH}$ and $M/L$. The grid has been computed with ISBD $D$ and $i=70$ deg, fixing 2 interesting parameters ($M_{BH}$ and $M/L$) and varying the others to minimize $\chi^2$ (see Sec.~\ref{ris.3310}).  By choosing a $68.3\%$ confidence level, $M_{BH}=(5.0)^{+0.87}_{-2.26}\times 10^{6}M_{\odot}$ and $M/L = (0.1)^{+0.06}_{-0.04} M_{\odot}/L_{\odot,H}$, indicative of a very young stellar population in the central region \citep{colina02}.

\begin{figure*} 
\centering
\epsfig{file=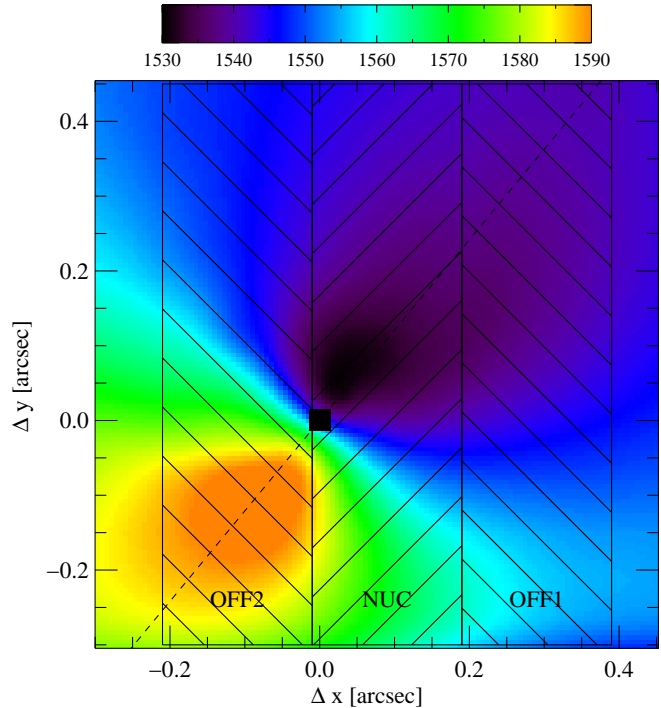, width=0.5\linewidth}
\caption{\label{campo} Map of the velocity field in the best fitting model of NGC 4303.
The velocity field is weighted with the intrinsic line flux distribution and is shown prior to averaging over slit and pixel sizes. The dashed areas denote the three parallel slits. The filled square represents the position of the black hole and the dashed line is the line of nodes.}
\end{figure*}

Fig.~\ref{campo} provides an alternative visualization of the best fitting model. The velocity field weighted with the intrinsic line flux distribution is shown prior to the averaging over slit and pixel sizes. The velocity asymmetries with respect to the lines of nodes (dashed line) are due to the intrinsic line flux distribution, which is not symmetric with respect to the BH position.

\begin{figure*}
\centering
\includegraphics[width=0.4\textwidth]{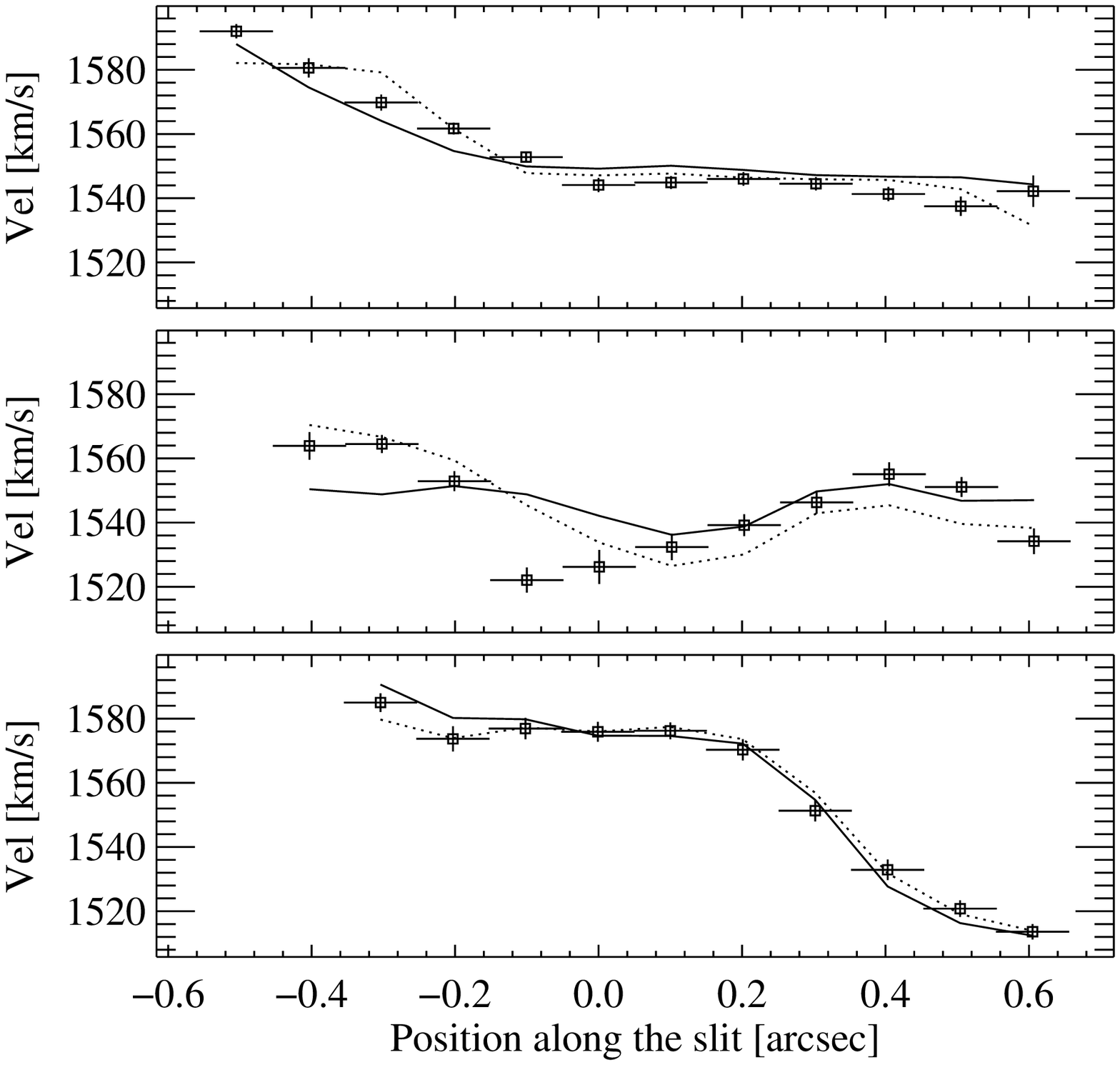}
\includegraphics[width=0.4\textwidth]{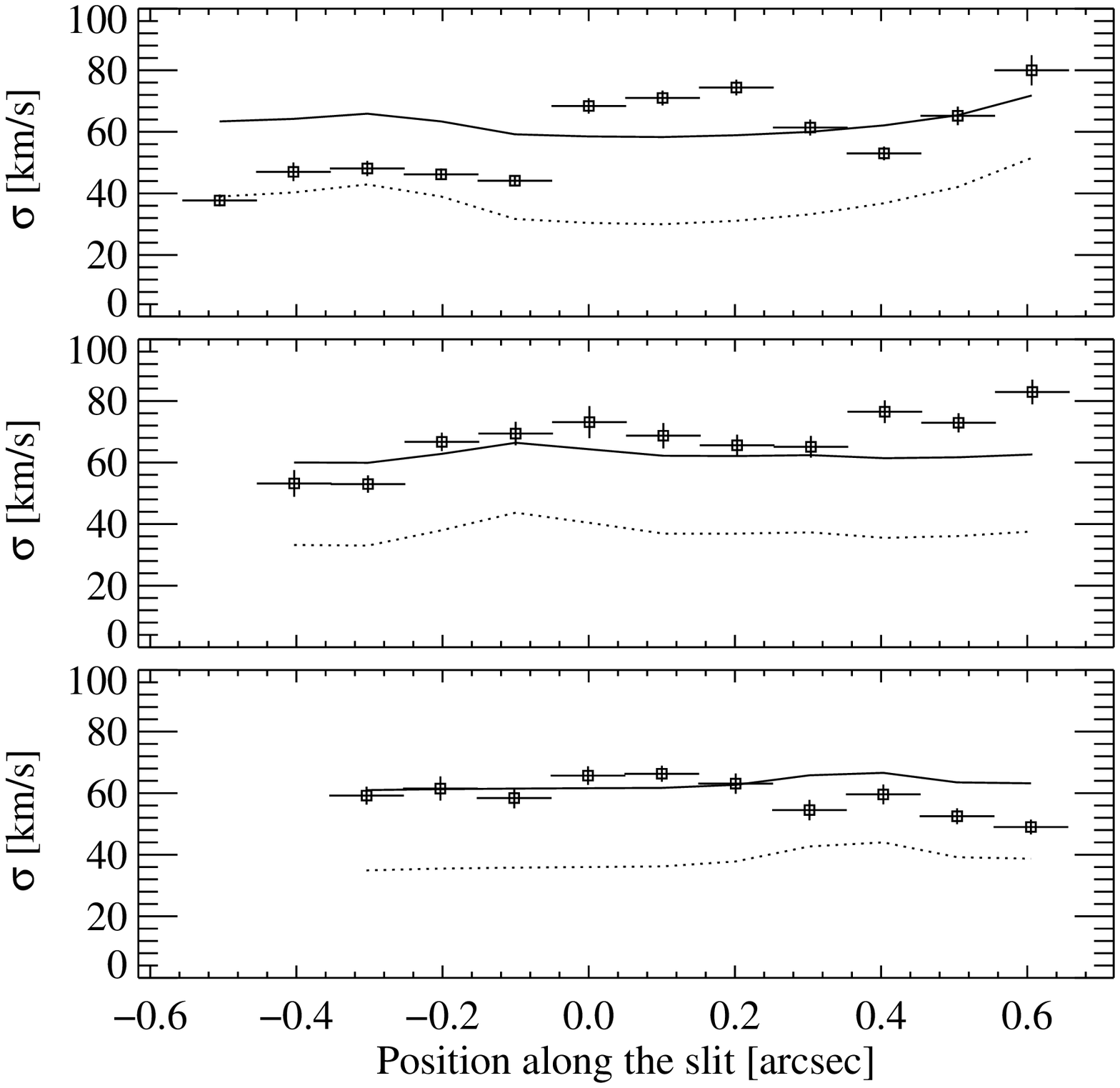}
\caption{\label{nobh} NGC 4303. \emph{Left panel}: best-fit model without a BH at $i=70$ deg and ISBD $D$.
The \emph{dotted line} represents the best fit model with a BH, obtained with the same inclination and ISBD. Notation as in the Fig.~\ref{fig.4303_solovel}. \emph{Right panel}:\label{4303sigma} expected velocity dispersions for the best fitting rotating disk model with $i=70$ deg compared with observations (dots with error bars), assuming a constant intrinsic dispersion of $50$ km/s (solid lines). The dotted line is the velocity dispersion expected only by unresolved rotation.}
\end{figure*}

To check for the robustness of the BH detection we tried to fit the rotation curves without a BH but only with the gravitational potential due to the stars. We worked at $i=70$ deg and with $D$ ISBD. The result of this fit is shown in the left panel of Fig.~\ref{nobh}. At first sight the no--BH fit does not appear significantly worse than the one with a BH but it is nonetheless statistically unacceptable. After rescaling velocity errors with $\delta v = 6.9 km/s$, as described above, the $68.3\%, 90\%, 95.4\%$ confidence levels
are $\chi^{2} = 27.8$, 30.1 and 31.7, respectively while the model without the BH has  $\chi^{2}=33.8$. 
However, given the non perfect agreement between our model and the data, we do not consider our $M_{BH}$ detection completely reliable. 

As in the case of NGC 3310, we verified that at different inclinations the statistical errors on $\log M/L$ and $\log M_{BH}$ were roughly the same. 
The allowed range of inclinations ($i> 40$ deg) is then translated into a possible range for $M_{BH}$ which can vary between  
$6.0 \times 10^{5}$ and $1.6 \times 10^{7}M_{\odot}$ where the extreme have been computed by taking into account also the 95\% confidence errors on $\log M_{BH}$ (-0.4, +0.15).

In Fig.~\ref{4303sigma} we plot the expected velocity dispersion for the best fitting model at $i=70$ deg which includes intrumental broadening and
the contribution of unresolved rotation (dotted line). In order to match observations we added an intrinsic constant dispersion of $50$ km/s (solid line).
Overall data points do not show a deviation of more than $20$ km/s from the model values, after the intrinsic dispersion was taken into account. 

\subsection{\label{ris.4258}Analysis of NGC 4258}

An inspection of Fig.~\ref{andamenti_4258_ours} shows that the available kinematical parameters for this galaxy are limited to the nuclear component i.e.\ within $0.5\arcsec$ of the continuum peak. We followed the same procedure outlined for NGC 3310 and NGC 4303 with the difference that, in this case, we fitted simultaneously ours and archival data. Effectively we fit 6 slit positions at the same time.

The starting assumption is that the nuclear slits are centered on the same point, as confirmed from the fact that the broad H$\alpha$ components have very similar fluxes in the two nuclear spectra (see Fig.~\ref{paragone}).

\begin{figure*} 
\centerline{\epsfig{file=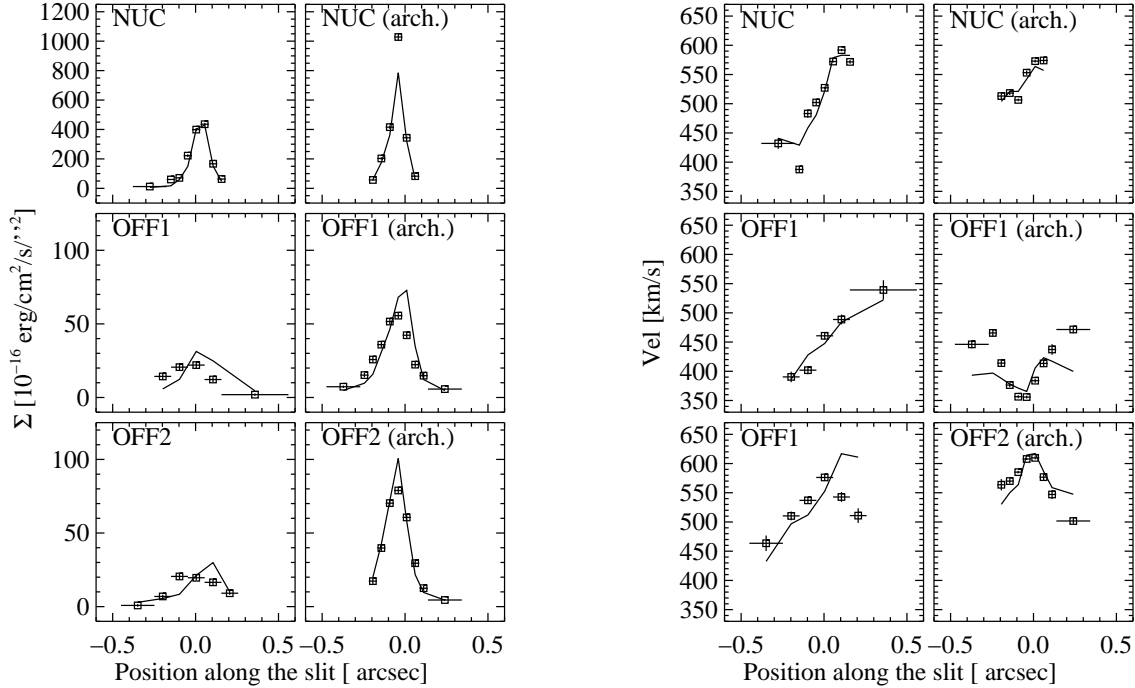, angle=90, width=0.85\linewidth}}
\caption{\label{fig:4258_modf}NGC 4258. \emph{Left panel:} best-fit model (solid line) of the emission line surface brightness distribution compared with the observed data (dots with error bars) at the different slit positions.  From top to bottom the panels refer to NUC, OFF1 and OFF2 respectively. The different columns denote our and archival data. \emph{Right panel:} best-fit model (solid line) and observed data (dots with error bars) for the rotation curves. The model is computed for $i = 80 $ deg. The meaning of the columns is the same as in the left panel.}
\end{figure*}

 We first tried to fit the observed surface brightness distribution with different combinations of analytical functions, but we could find only one ISBD that provided a good agreement with the data. Left panel of Fig.~\ref{fig:4258_modf} and Tab.~\ref{4258_tabf} show the best fit model and the parameter values, respectively. It is clear from the figure that the surface brightness drops by a factor $\sim 10$ from the nuclear to the offset positions. Note that the observed surface brightness in the NUC slit of the archival spectra
is a factor 2 larger than in our data. This is simply due to the fact that our data use the $0.2\arcsec$ slit instead of the $0.1\arcsec$ one.

\begin{table*}
\caption{NGC 4258: best fitting parameters of the ISBD adopted in the kinematical model fitting simultaneously our and archive data.}
\label{4258_tabf}
\centering
\begin{tabular}{c c c c c c c c c c}
\hline\hline
$Id$ &  $function$ & $i$ & $I_{0i}^{a}$ & $r_{0i}(\arcsec)$ & $r_{hi}(\arcsec)$ & $x_{0i}(\arcsec)$ & $y_{0i}(\arcsec)$ & $\theta_{i}(^{\circ})$ & $q_{i}$  \\
\hline
A  & expo   &  1  & 14139  & 0.01 &  0.00   &  0.04   & -0.01& -0.1  & 0.49  \\
   & expo   &  2  & 1173   & 0.03 &  0.00   &  0.07   & -0.04& -0.4  & 0.30   \\
   & const  &  3  & 2      & --   &  --     &   --    &  --  & --    & --      \\
\hline
\multicolumn{10}{l}{\small{$^{a}$ Units of erg s$^{-1}$ cm$^{-2}$ arcsec$^{-2}$ \AA$^{-1}$}}\\
\end{tabular}
\end{table*}

To fit the observed rotation curves, we started assuming that disk inclination is $i=80^{\circ}$ as in the much smaller maser disk \citep{hernest}.  The best-fit model is shown in the right panel of Fig.~\ref{fig:4258_modf} and the best fit parameter values are displayed in Tab.~\ref{4258_tabv}.

\begin{table*}
\caption{NGC 4258: best fitting parameters from the velocity fit for $i=80$ deg fitting simultaneously our and archive data.}
\label{4258_tabv}
\centering
\begin{tabular}{c c c c c c c c}
\hline\hline
$Flux^{a}$ &  $x_{0}(\arcsec)$ & $y_{0}(\arcsec)$ &  $\log M_{BH}^{b}$ & $\log M/L^{c}$ & $\theta(^{\circ})$ & $V_{sys}$(km/s) & $\chi^{2}_{red}$ \\
\hline
A  &   0.02  &  -0.01  &  8.02 & 0.70 & 235.44 & 466.45  & 34.96 \\
\cline{1-8}
\hline
\multicolumn{8}{l}{\small{$^{a}$ Adopted flux distribution}}\\
\multicolumn{8}{l}{\small{$^{b}$ Units of $M_{\odot}$}}\\
\multicolumn{8}{l}{\small{$^{c}$ Units of ${M_{\odot}/L_{\odot,H}}$}}\\
\end{tabular}
\end{table*}

As in the cases of NGC 3310 and NGC 4303, we investigate the influence of the disk inclination on quality of the fit. Tab.~\ref{4258_sivs} shows the best-fit parameters obtained at different values of the inclination. 

\begin{table*}
\caption{\label{4258_sivs}NGC4258: effect of \emph{i} variation on the best fit parameters and $\chi_{red}^{2}$. $\chi^2$ values have been rescaled by an extra error on velocity $\Delta\nu_{0}$.
}
\centering
\begin{tabular}{c c c c c c}
\hline\hline
$i(^{\circ})$ &  $\log M_{BH}^{a}$ & $\log M/L^{b}$ & $V_{sys}$ (km/s) & $\theta(^{\circ})$ & $(\chi^{2}_{rescaled})^{c}$ \\
\hline
\multicolumn{6}{c}{Fit of velocity ($\Delta\nu_{0}=33.74 Km s^{-1}$)$^{d}$ }\\
5  &  9.69 &-0.74  & 465.60  & 239.22  &  1.11 \\
10 &  9.12 &-1.21  & 466.31  & 239.84  &  1.09\\
15 &  8.78 &-1.15  & 466.78  & 239.54  &  1.09  \\
20 &  8.56 &-2.10  & 461.66  & 238.51  &  1.10  \\
25 &  8.37 &-1.80  & 465.01  & 238.95  &  1.08  \\
30 &  8.26 &-1.92  & 464.67  & 238.98  &  1.07 \\
35 &  8.15 &-1.74  & 466.36  & 238.68  &  1.06 \\
40 &  8.06 &-1.83  & 465.47  & 238.35  &  1.05  \\
45 &  8.03 &-2.24  & 463.81  & 237.64  &  1.04 \\
50 &  7.99 &-2.22  & 465.32  & 237.07  &  1.03 \\
55 &  7.97 &-1.96  & 465.13  & 236.85  &  1.02  \\
60 &  7.96 &-2.30  & 467.98  & 237.04  &  1.00 \\
65 &  8.00 &-1.70  & 468.68  & 236.14  &  1.00 \\
70 &  8.01 &-1.64  & 472.64  & 236.69  &  1.00 \\
75 &  8.09 &-1.76  & 472.40  & 236.45  &  1.02  \\
85 &  8.47 &-2.07  & 497.09  & 244.23  &  1.12  \\
\hline                                        
\multicolumn{6}{l}{\small{$^{a}$ Units of $M_{\odot}$.}}\\
\multicolumn{6}{l}{\small{$^{b}$ Units of ${M_{\odot}/L_{\odot,H}}$.}}\\
\multicolumn{6}{l}{\small{$^{c}$ Rescaled $\chi^{2}$ with errors computed as $\Delta\nu_{i}^{'2}$=$\Delta\nu_{i}^{2}+\Delta\nu_{0}^{2}$.}}\\ 
\multicolumn{6}{l}{\small{$^{d}$ Systematic error adopted to renormalize $\chi^{2}$.}}\\
\end{tabular}
\end{table*}

\begin{figure*} 
\centerline{\epsfig{file=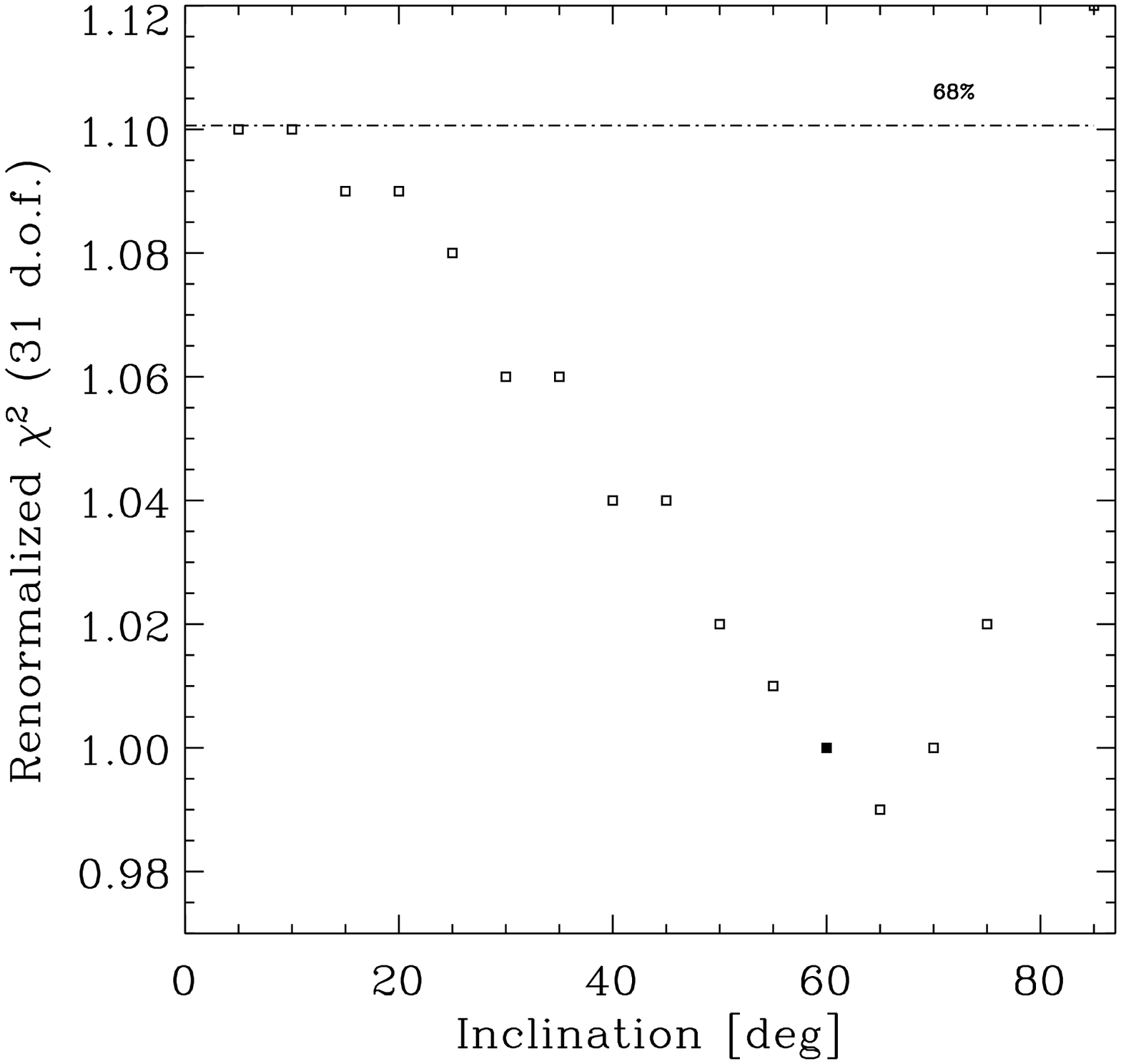, width=0.4\linewidth}\epsfig{file=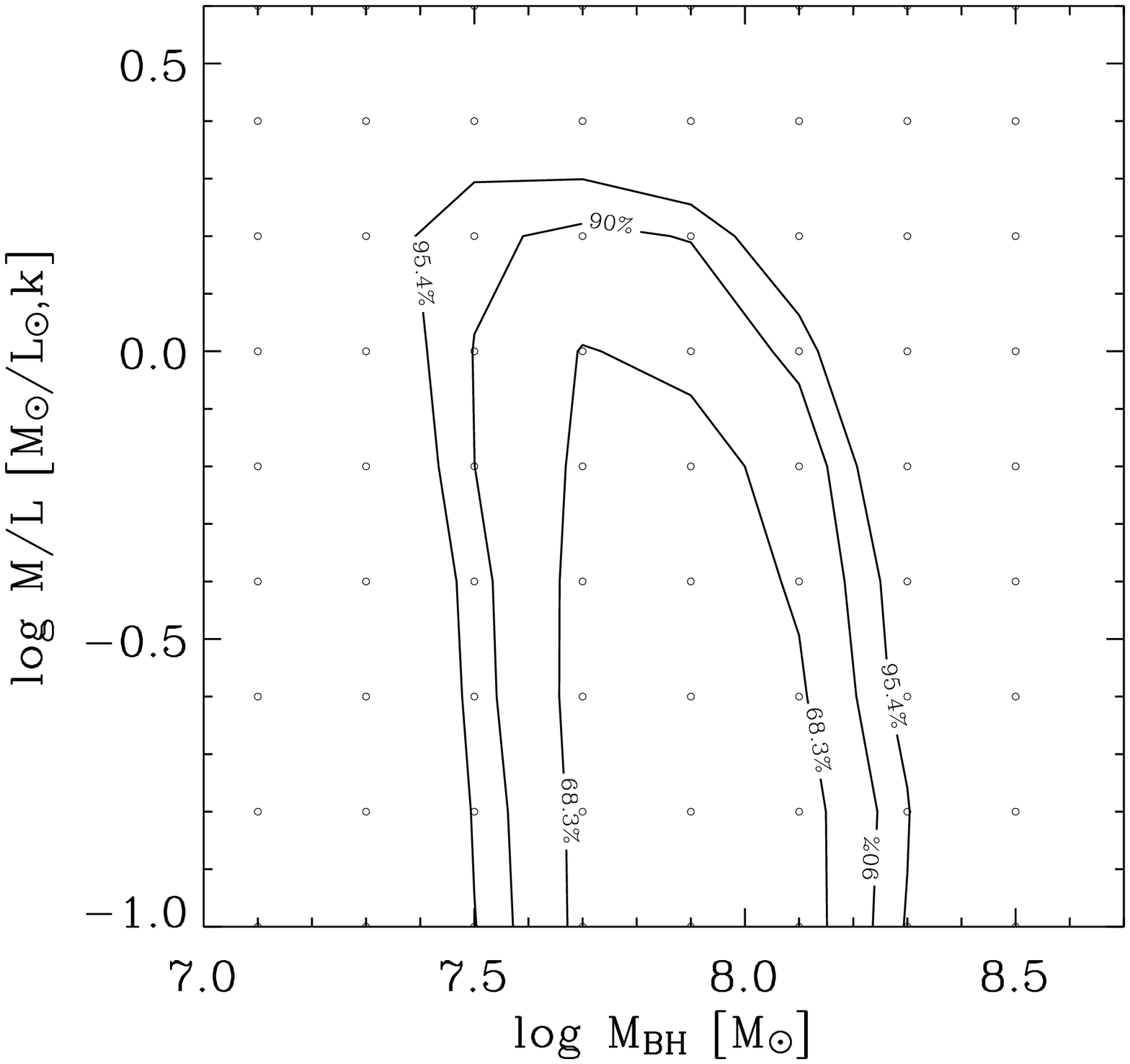, width=0.4\linewidth}}
\caption{\label{4258:incl}\emph{Left panel}: Dependence of $\chi^{2}$ on the adopted disk inclination during NGC4258 fitting only our data. Dotted line indicates the 68.3\% confidence levels (lines indicating respectively 90\% and 95.4\% would have value $1.45$ and $1.34$).\emph{Right panel}: For the same galaxy, $\chi^{2}$ contours for joint variation of $M_{BH}$ and $M/L$ using the best ISBD and $i=60$ deg. Contours levels are for $\chi^{2}=\chi^{2}_{min}+2.3,4.61,6.17$ corresponding to 68.3\%,90\% and 95.4\% confidence levels. }
\end{figure*}

The model with the lowest $\chi^2$ value is the one at $i=60$ deg and, after rescaling, all the models have reduced $\chi^2$ close to 1.  The left panel of Fig.~\ref{4258:incl} shows the dependence of $\chi^{2}$ on the adopted disk inclination (the filled square marks the model with the lowest $\chi^2$). In this figure we did not draw the 90\% and 95\% confidence level because they are larger than all the values of $\chi^{2}$ found by our fitting. This means that its is not possible to place any constraints on the inclination value
but it is interesting to note that the best inclination value is remarkably similar to the inclination of the galactic disk which is $\simeq 64$ deg \citep{cecil92}.

Finally, we present in the right panel of Fig.~\ref{4258:incl} the contours for the joint variation of $M_{BH}$ and $M/L$ computed for the disk inclination of $i=60$ deg. The contour shapes allow to estimate $M_{BH}$, but only an upper limit for the $M/L$ ratio.  At the 68.3\% confidence level, the black hole mass is $M_{BH}=(7.9)^{+6.2}_{-3.5}\times 10^{7}M_{\odot}$ while the 95\% upper limit on $M/L$ is $M/L < 1.8 M_{\odot}/L_{\odot,H}$.

\begin{figure*}
\centerline{\epsfig{file=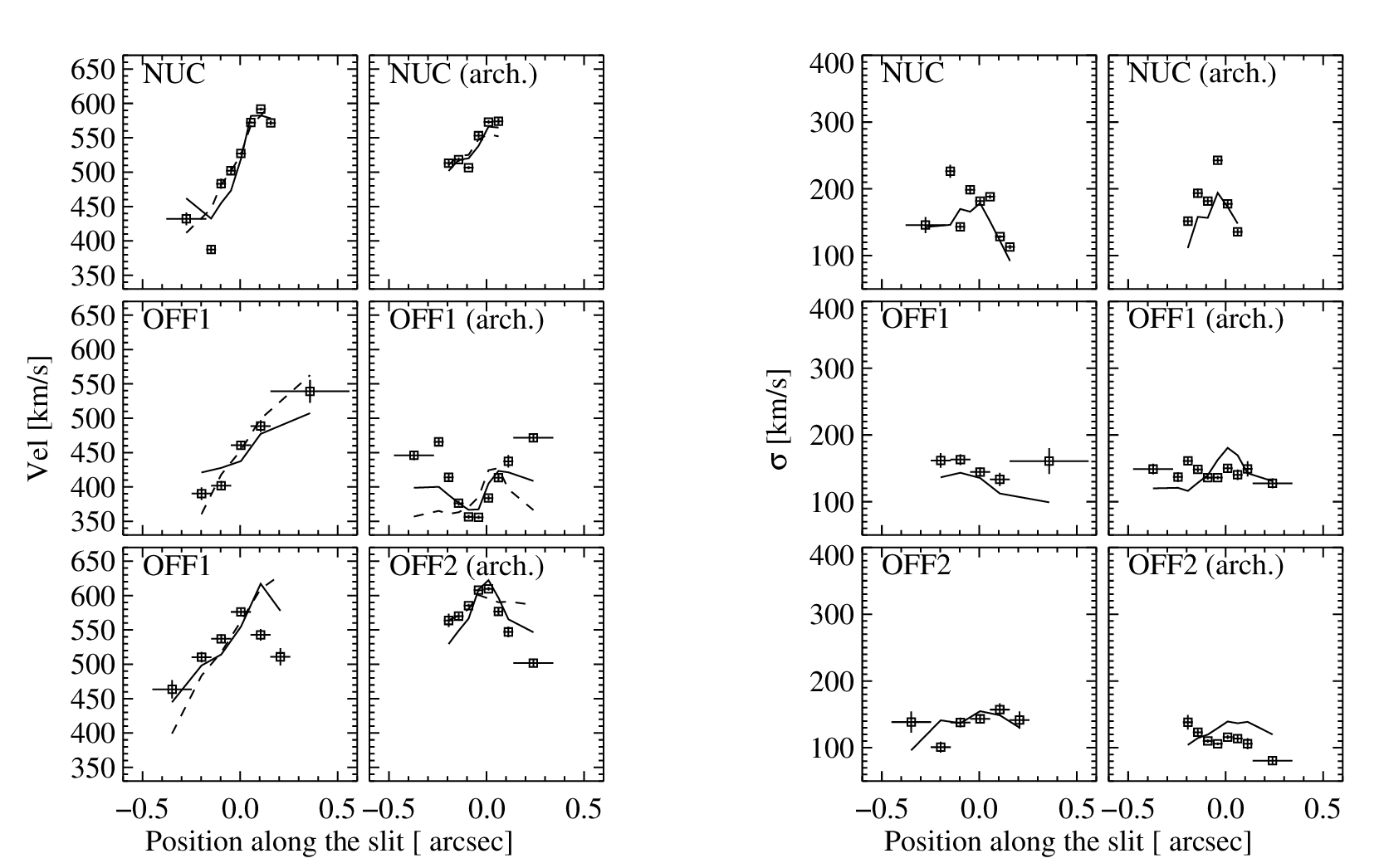,angle=90, width=0.85\linewidth }}
\caption{\label{fig:4258_modf60} NGC 4258:\emph{Left panel}: best-fit model (solid line) computed for $i=60^{\circ}$ fitting simultaneously our and archive data (dots with error bars). From top to bottom the panels refer to NUC, OFF1 and OFF2 respectively. The dashed line is the best fit model obtained without a BH; even if the difference between the two cases does not appear significative, the model without a BH provides an unphysically high value for stellar mass to light ratio which is $M/L=5.08M_{\odot}/L_{\odot,H}$. \emph{Right panel}: expected velocity dispersions for the best fitting rotating disk model with $i=60$ deg (solid line) compared with the observations (dots with error bars).} 
\end{figure*}

In the left panel of Fig.~\ref{fig:4258_modf60} we compare the observed kinematics with the models with (solid line) and without (dashed line) a BH computed at $i=60$ deg. As in the case of NGC 4303, the no--BH fit does not appear much worse than the one with a BH but is statistically unacceptable. After rescaling velocity errors with $\delta v = 32.9 km/s$, as described above, the $68.3\%, 90\%, 95.4\%$ confidence levels are $\chi^{2} = 35.3$, 37.6 and 39.1, respectively while the model without the BH has  $\chi^{2}=57.8$. Moreover, the no-BH model is characterized by a very high value of the mass-to-light ratio $M/L=5.08M_{\odot}/L_{\odot,H}$ which is larger than the maximum predicted by normal stellar populations (see the discussion in the case of NGC 3310 in Sec. 5.1) indicating the presence of a dark mass. Combining these two facts we consider that with our data we have reliably detected the presence of the BH in NGC 4258.

Our black hole mass value is compatible with the $H_2O$-maser estimate by \citet{greenhilla,greenhillb} and \citet{hernest} which is $M_{BH}=(3.9 \pm 0.1)\times 10^{7}M_{\odot}$ at slightly above the $1\sigma$ level. Given the poor match between the data and a simple pure rotating disk model (i.e. no streaming or warps), this must be nonetheless considered an encouraging result.

As in the previous cases, we verified that the statistical errors on $\log M/L$ and $\log M_{BH}$ do not  vary significantly with the adopted inclination. However, in this case we could not place any constraint on the inclination thus, in principle, we could only provide a lower limit on the BH mass value.
The nuclear gas disk we are mapping is spatially intermediate between the large scale galactic disk and the maser disk. The two disks have different inclinations $\sim 60$ and $80$ deg, respectively and it is thus reasonable to assume that the inclination of the nuclear gas disk, connecting the large to the small galactic scales, is limited in the $60-80$ deg range.  With this assumption the black hole mass from gas kinematics is confined in the range $2.5\times 10^{7}$ -- $2.6 \times 10^{8}M_{\odot}$ where the extreme have been computed by taking into account also the 95\% confidence errors on $\log M_{BH}$ (-0.5, +0.4).

In the right panel of Fig.~\ref{fig:4258_modf60} we plot the expected velocity dispersion for the best fitting model at $i=60$ deg which includes intrumental broadening and
the contribution of unresolved rotation (solid line). In contrast with NGC 3310 and NGC 4303, the observed velocity dispersion 
can be well matched with unresolved rotation without the inclusion of any intrinsic dispersion.

The non-perfect agreement between our model and the data suggests that rotation curves might be somehow affected by non circular motions. Indeed a jet-cloud interaction is responsible for the hot X-ray cocoons which give rise to the anomalous arms \citep{wilson}. Their position angle, i.e.~the region where the X-ray cocoons are interacting with the dense interstellar gas, is $\sim -33$ deg, very close to the slit PA of the archival data ($\sim 146=-34$). 
Since our data are at an almost perpendicular slit PA and likely less affected by non-circular motions, we have tried to perform the fit without the archival data. We found no significant change in the black hole mass estimate although a slightly better agreement between model and observed velocities. This indicates that the non-circular motions do not affect the final black hole mass estimate.

\section{Discussion\label{discuss}}

In Sec.~\ref{ris.3310}, \ref{ris.4303} and \ref{ris.4258} we have presented the kinematical modeling of the three Sbc spiral galaxies analyzed in this paper, NGC 3310, NGC 4303 and NGC 4258. We first worked with a fixed value of inclination and then we found the range of possible values for $M_{BH}$ taking into account the allowed values of inclinations. Previous works on spiral galaxies have made the simplified assumption that nuclear gas disks used for $M_{BH}$ detection have the same inclination as the large scale galactic disks. In fact, in contrast to elliptical galaxies, it is not possibile to determine the gas disk morphologies and inclinations from HST images of spiral galaxies. In this paper we have considered the effects of the unknown inclination which results in a range of  $M_{BH}$ values or upper limits.
The $M_{BH}$ values derived for a given disk inclination are characterized by statistical errors as shown in the previous section but the same $M_{BH}$ values vary as $(\sin i)^{-2}$. Thus a range of allowed inclinations results in a range of possible BH masses.

The observed rotation curves of NGC 3310 within the inner 0.6\arcsec\ from the nucleus ($r<50$ pc) agree very well with a circularly rotating disk model. 
The observed velocity dispersions cannot be matched with a simple rotating disk but require a constant intrinsic velocity dispersion of $\sim 55$ km/s. This might indicate the presence of turbulence, non-circular or non-gravitational motions. The former would just contribute to the dispersion of the gas but not to the bulk motions and therefore would not invalidate the gas kinematical method. Conversely, the latter ones could invalidate the measurements depending on their strength compared to the ordered circular rotation (see, e.g., the discussion in \citealt{barth01,verdoes,marconi04}).
However, the good agreement between the observed and model rotation curves suggests that the contribution of non-circular or non-gravitational motions is small and does not invalidate the measurement.
From the analysis of the rotation curves it is not possible to find evidence for the existence of any BH since they can be  accounted for by the stellar mass. The derived mass to light ratio (for the best model at $i=70$ deg) is $M/L=(0.47)^{+0.04}_{-0.07}M_\odot/L_{\odot,H}$ and is consistent with the typical numbers for spiral bulges \citep{moriondo} confirming the validity of the analysis. The spatial resolution of our observations is about 0.07\arcsec (FWHM of the STIS PSF at 6700\AA) and within a sphere with that diameter the stellar mass is $M_{*} \sim 10^{7} M_{\odot}$. This fact, combined with the upper limit  on the BH mass ($M_{BH} < 4.2\times 10^6 M_\odot$, at the 96\%\ level for the lowest inclination allowed), indicates that we were not able to detect the BH because our spatial resolution is not sufficiently high.

The situation for NGC 4303 is different with respect to NGC 3310. The rotation curves require the presence of a BH with mass $M_{BH} = 5.0^{+0.87}_{-2.26}\times 10^6 M_\odot$ ($6.0 \times 10^{5} - 1.6 \times 10^{7} M_{\odot}$ when taking into account $i>40$ deg) but the agreement between observed and model velocities is not as good especially at OFF1. With the inclusion of an intrinsic velocity dispersion of $50$ km/s, the observed velocity dispersions in the offnuclear slits are in fair agreement with the model ones while in the nucleus they show deviations of the order of $10$ km/s. Although the model without a black hole is statistically worse than that with a BH, the combination of the above facts does not allow us to consider the black hole mass estimate in NGC 4303 completely reliable. 

For NGC 4258, the fair agreement between observed and model rotation curves might indicate the presence of non-circular motions but the observed velocity dispersion is completely accounted for by unresolved rotation. These two facts alone would not allow us to
say anything about the reliability of the BH mass estimate. However, NGC 4258 was chosen as a benchmark for the gas kinematical method because it represents the second best case for a BH (after the Milky Way), with an accurate mass estimate ($M_{BH}=(3.9 \pm 0.1) \times 10^{7}M_{\odot}$; \citealt{greenhilla,greenhillb,hernest}). The agreement between our mass estimate ($7.9^{+6.2}_{-3.5} \times 10^{7}M_{\odot}$) ($i=60$ deg) and the $H_2O$-maser value (the difference is just slightly above $1\sigma$) indicates that the gas kinematical method is reliable even if the observed velocities present deviations from the circularly rotating disk model.
\
We now compare the BH mass estimates (or upper limits) in spiral galaxies with the known correlations between $M_{BH}$ and host galaxy structural parameters. Since our work has doubled the number of gas kinematical measurements in late type spiral galaxies it is worthwhile to consider the sample of all late type spirals studied so far. 

\begin{table*}
\caption{\label{tab.corr_log}Comparison between $M_{BH}$ estimates in late type spirals with the expectations from the BH-spheroid scaling relations. The intrinsic scatter of the correlations is $\sim 0.3$ in $\log M_{BH}$ and the uncertainty on the determination of the luminosity is $0.3$ dex (see text).} 
\centering
\begin{tabular}{ l c c c c c c c c r }
\hline 
Galaxy & Type & $D^{a}$  & $ \log M_{BH}^{b}$  & $\log L_K^{c}$ & $\log M_{BH}/M_{BH}(L_K)^{d}$ & $\sigma_c\,^{e}$ & $\log M_{BH}/M_{BH}(\sigma_c)^{f}$ & $\sigma_e\,^{e}$ & $\log M_{BH}/M_{BH}(\sigma_e)^{g}$  \\
\hline
\\
NGC1300  & SB(rs)bc & 18.8  & $7.8 _{-0.2 }^{+0.2 }$& 10.0 &  0.6  &   90  &  1.3   &  87  &  1.1 \\
\\
NGC2748  & SAbc     & 23.2  & $7.6 _{-0.4 }^{+0.2 }$& 9.8  &  0.6  &   79  &  1.4   &  83  &  1.0 \\
\\
NGC3310  & SAB(r)bc & 17.4  & $<7.6 _{    }^{    }$& 9.6  &  $<0.9$  &   101 &  $<0.8$  &  84  &  $<1.0$ \\
\\
NGC4041  & SA(rs)bc & 19.5  & $<7.3 _{    }^{    }$& 9.7 &  $<0.4$  &   92  &  $<0.7$  &  88  &  $<0.6$ \\
\\
NGC4303  & SAB(rs)bc& 16.1  & $6.6 _{-0.2 }^{+0.5 }$& 10.2 & -0.8  &   108 &  -0.3 &  84  & 0.0 \\
\\
\hline
\\
NGC4258  & SAB(s)bc & 7.2   & $7.59_{-0.04}^{+0.04}$ & 10.3 &  0.09  &   120 &  0.45  &  148 & -0.19 \\
\\
Milkyway & SbI-II   & 0.008 & $6.60_{-0.03}^{+0.03}$& 10.2 & -0.9  &   100 &  -0.2 &  100 &  -0.3 \\
\\
M81      & SA(s)ab  & 3.9   & $7.84_{-0.07}^{ +0.1}$& 11.0 & -0.4  &   174 & -0.1  &  165 &  0.0\\
\\
\hline
\hline
\multicolumn{10}{l}{\small{$^{a}$ units of Mpc.            }} \\
\multicolumn{10}{l}{\small{$^{b}$ units of $M_{\odot}$. }}  \\
\multicolumn{10}{l}{\small{$^{c}$ units of $L_{\odot,K}$. }}  \\
\multicolumn{10}{l}{\small{$^{d}$ $M_{BH}(L_K)$ is the value expected from the \citet{MeH} correlation.}} \\
\multicolumn{10}{l}{\small{$^{e}$ units of km/s.}} \\
\multicolumn{10}{l}{\small{$^{f}$ $M_{BH}(\sigma_c)$ is the value expected from the \citet{ferrarese05} correlation.}} \\
\multicolumn{10}{l}{\small{$^{g}$ $M_{BH}(\sigma_e)$ is the value expected from the \citet{trem02} correlation.}} \\
\end{tabular}
\end{table*}

This is shown in Tab.~\ref{tab.corr_log} where for all galaxies studied in our project and those available from the literature we report the measured BH masses or upper limits and the ratio with the expected values from the $M_{BH}-L_{\emph{sph}}$ (\citealt{MeH}; eq.~\ref{eq:ML}) and $M_{BH}/\sigma_{*}$ (\citealt{trem02, ferrarese05};  eq.~\ref{eq:MsG} and \ref{eq:MsF}) correlations:

\begin{equation}
\label{eq:ML}
\log(M_{BH}/M_{\odot})=8.21+1.13 (\log (L_\emph{sph}/L_{\odot,K})  -10.9)
\end{equation}

\begin{equation}
\label{eq:MsG}
\log(M_{BH}/M_{\odot})=8.13+4.02 \log (\sigma_{e}/200\,km/s)
\end{equation}

\begin{equation}
\label{eq:MsF}
\log(M_{BH}/M_{\odot})=8.22+4.86 \log (\sigma_{c}/200\,km/s)
\end{equation}
BH masses are taken from this work (NGC 3310, NGC 4303), from \citealt{atki} (NGC 1300, NGC 2748), from \citealt{4041} (NGC 4041) and from the compilation by \citealt{ferrarese05} (Milky Way, NGC 4258 and M81) to which we refer for the proper references for each galaxy.
K band luminosities of the spheroid are taken from \cite{MeH} (Milky Way, NGC 4258, M81), from \cite{dong} (NGC 2748 and NGC 4041). For the other galaxies (NGC 1300, NGC 3310 and NGC 4303) we took the total K band magnitude from the 2MASS extended source catalogue \citep{jarret} and we applied the near-IR bulge-total correction by \cite{dong}. The correcting factor depends on the galaxy morphological type $T$ and for our galaxies ($T=4.0$), it is $\Delta m=M_{sph}-M_{tot}=2.0$. We then transformed the relative to absolute magnitudes using the distances shown in the table and we obtained the luminosities using the solar K band magnitude $M_{K\odot} = 3.28$. The uncertainty on the determination of $K$-band luminosity of each galaxy is $0.3$ dex and is due to the observed dispersion on the bulge-total correction for a given morphological type \citep{dong}.
The stellar velocity dispersions required by the correlations in Eq.~\ref{eq:MsG} and \ref{eq:MsF} are different:
\citet{trem02} use the luminosity weighted velocity dispersion $\sigma_{e}$ measured within $r_{e}$ i.e.\ within the bulge half-light radius. On the contrary,
\citet{ferrarese05} use the central velocity dispersion $\sigma_{c}$, i.e.~ measured within an aperture of $r = \frac{1}{8}r_{e}$. For all galaxies except the Milky Way and M81 we used the values measured by \citet{dan05}. Otherwise we used the compilations by \citet{trem02} and \citet{ferrarese05}.

When inspecting the logarithms of the ratios between the measured BH masses and the expectations from the correlations one has to bear in mind that the intrinsic dispersion of the correlations is $\sim 0.3$ in $\log M_{BH}$ \citep{MeH}.
The upper limit on the BH mass in NGC 3310 is consistent with all correlations since it is larger than all expected $M_{BH}$ values.
The BH mass measurement in NGC 4303 is less reliable, as discussed above. However, it is comforting that the measured $M_{BH}$ value is in very good agreement with the $M_{BH}-\sigma$ correlations, if their intrinsic dispersion is taken into account. $M_{BH}$ is a factor of 6 smaller than expected from the $M_{BH}-L_K$ correlation however it should be recalled that the bulge luminosity of NGC 4303 has been measured using the scaling relation by \cite{dong} and the observed dispersion for Sbc spiral galaxies is of the order of 0.3 in $\log L_K$.

\begin{figure*}
\centering
\centerline{\epsfig{file=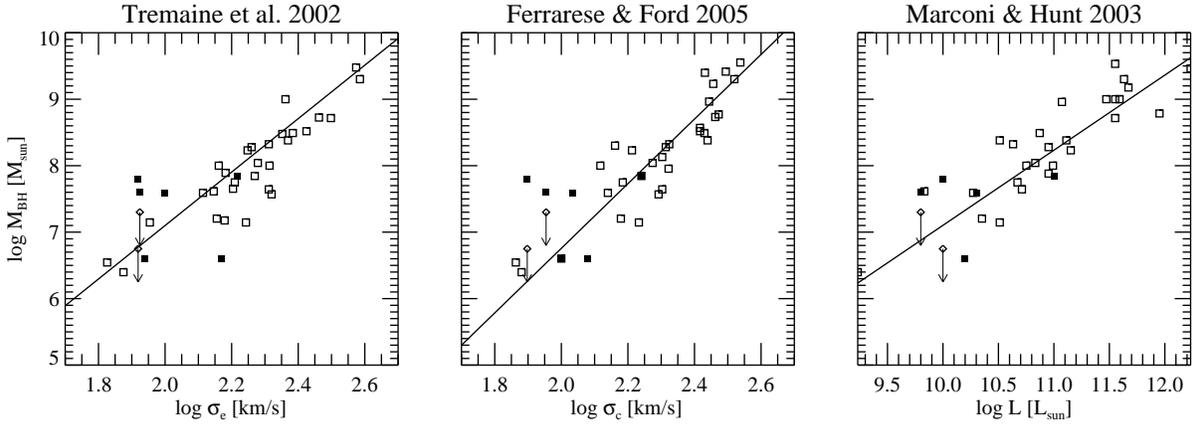, width=0.4\linewidth, angle=90}}
\caption{\label{corr}Correlations between $M_{BH}$ and the host galaxy luminosity  (\emph{right panel}), the central stellar velocity dispersion, $\sigma_{c}$ (\emph{middle panel}) and the effective velocity dispersion $\sigma_{e}$ (\emph{left panel}). The empty squares denote early type galaxies (E/S0) from the literature while the filled ones refer to the spiral galaxies in Tab.\ref{tab.corr_log}.}
\end{figure*}

In Fig.\ref{corr} we compare spiral galaxies (filled squares) with the known $M_{BH}/L_{sph,K}$,$M_{BH}/\sigma_{e}$ and $M_{BH}/\sigma_{c}$ correlations from the compilations by \citet{MeH,ferrarese05,trem02}. The empty squares represent early type galaxies and, for simplicity, we have not shown error bars.

Considering that each correlation has an intrinsic dispersion of $0.3$ in $\log M_{BH}$ we can conclude that spiral galaxies do not show evident deviations from the correlations followed by early type galaxies, apart for a slightly larger scatter.
A statistical analysis on the behavior of spiral galaxies has to wait for a larger number of BH mass detections.

In conclusion, in order to show the difficulty of this kind of measurements, we compute the diameters of the BH spheres of influence ($2\frac{M_{BH}G}{{\sigma_{e}^{2}}}$) for each galaxy and compare them with the available spatial resolution (FWHM $\sim 0.07$ arcsec). These are $<0.08$, 0.08 and $0.44''$ for NGC 3310, NGC 4303 and NGC 4258, respectively. Except for NGC 4258, which is a nearby spiral galaxy, the black hole spheres of influence are at the limit of the HST spatial resolution, indicating how challenging these measurements are.

\section{\label{conclusioni} Conclusions}

We have presented new HST Space Telescope Imaging Spectrograph (STIS) observations of three Sbc galaxies: NGC 3310, NGC 4303 and NGC 4258.
We used $H\alpha$ $\lambda$ $6564 \AA $, [NII] $\lambda$$\lambda$ $6549,6585 \AA$ and [SII] $\lambda$$\lambda$ $ 6718,6732 \AA$ to study the kinematics of nuclear ionized gas disks at $\sim 0.07\arcsec$ spatial resolution.
Our NGC 4258 data were analyzed in conjunction with available HST/STIS archival data.
In all galaxies, we derived the kinematical moments by fitting a Gauss-Hermite expansion to the observed line profiles and,
 in the case of NGC 4258, we also took into account the presence of a broad $H\alpha$ component.
 
We have modeled the observed kinematics by assuming that the gas is confined in a thin circularly rotating disk where hydrodynamical
effects can be neglected. The disk rotates in a gravitational potential which is due to the galaxy stars and to a black hole. 
For each galaxy the stellar mass distribution was derived from HST-NICMOS images by fitting the observed stellar surface brightness.
In computing the model kinematics, we have taken into account the finite spatial resolution of the observations and we have averaged 
over the slit and pixel area, weighting with the intrinsic surface brightness distribution of the emission lines.
The free fit parameters were determined with a $\chi^{2}$ minimization.

In NGC 3310, the observed rotation curves are well reproduced with the circularly rotating disk model. 
However, the spatial resolution of the observations only allowed us to set an upper limit to the BH mass,
$M_{BH} < 5.6 \times 10^{6} M_{\odot}$, for a disk inclination of $i=70$ deg.  When taking into account the allowed disk inclinations, $M_{BH}$ varies in the range $5.0 \times 10^{6} - 4.2 \times 10^{7} M_{\odot}$.  

In NGC 4303 the rotation curves require the presence of a black hole with mass $M_{BH}=(5.0)^{+0.9}_{-2.3} \times 10^{6} M_{\odot}$ ($i=70$ deg) but the agreement with the circularly rotating disk model does not allow to consider this detection completely reliable. If the allowed inclination values are taken into account, $M_{BH}$ varies in the range $6.0 \times 10^{5} - 1.6 \times 10^{7} M_{\odot}$.

Finally, in NGC 4258, the simultaneous modeling of our and archival data constrains the BH mass to  $M_{BH}=(7.9^{+6.2}_{-3.5})\times 10^{7}$ ($i=60$ deg) and, taking into account reasonable limits for the inclination, $M_{BH}$ is in the range $2.5\times 10^{7}$ -- $2.6 \times 10^{8}M_{\odot}$.  The observed rotation curves are not well matched by the model but the good agreement with the accurate $H_2O$-maser estimate ($M_{BH}=(3.9 \pm 0.1) \times 10^{7}\, M_\odot$ \citet{greenhilla,greenhillb,hernest}) is an important indication of the reliability of the gas kinematical method. 

We have combined the BH mass estimates in late type spirals presented in this paper with others from the literature. Although the sample is still small (6 detections and 2 upper limits), spiral galaxies seem to follow the same BH-spheroid scaling relations as early type galaxies.  

The comparison between the BH spheres of influence and the spatial resolution of HST observations underlines the difficulties connected to the study of BH in late type spiral galaxies and indicates that higher spatial resolution is required for a significant step forward.
 
\acknowledgements
We would like to thank the anonymous referee for his/her careful reading of the manuscript and for useful comments which improved the final version of the paper.
\\Funding for L.D was provided by NASA through grant NAG5-8573.



\clearpage



\end{document}